\documentclass[aps,prl,reprint,twocolumn,showpacs,superscriptaddress]{revtex4-2}

\usepackage{amsmath}
\usepackage{graphicx}
\usepackage{lmodern}
\usepackage{amsmath}
\usepackage{color}
\usepackage{amssymb}
\usepackage{bm}
\usepackage{braket}
\usepackage{mathtools}

\newcommand{\br}{\bm{r}}

\newcommand{\bk}{\bm{k}}

\newcommand{\bq}{\bm{q}}
\newcommand{\bQ}{\bm{Q}}
\newcommand{\bP}{\bm{P}}

\usepackage[dvipsnames]{xcolor}
\usepackage[colorlinks=true]{hyperref}
\hypersetup{
    colorlinks=true,
    linkcolor=blue,
    citecolor=blue,
    urlcolor=blue
}
\usepackage{cleveref}
\makeatletter
\def\maketitle{
\@author@finish
\title@column\titleblock@produce
\suppressfloats[t]}
\makeatother

\begin{document}
\title{{Time-reversal symmetry breaking, collective modes, and Raman spectrum in  pair-density-wave states}}

\author{Yi-Ming Wu}
\affiliation{Stanford Institute for Theoretical Physics, Stanford University, Stanford, California 94305, USA}
\author{Andrey V. Chubukov}
\affiliation{Department of Physics, University of Minnesota, Minneapolis, Minnesota 55455, USA}
\author{Yuxuan Wang}
\affiliation{Department of Physics, University of Florida, Gainesville, Florida 32611, USA}
\author{Steven A. Kivelson}
\affiliation{Department of Physics, Stanford University, Stanford, California 94305, USA}

\begin{abstract}
     Inspired by empirical evidence of the existence of pair-density-wave (PDW) order in certain underdoped cuprates, we investigate the collective modes in systems with unidirectional PDW order with momenta $\pm \bm{Q}$ and a $d$-wave form-factor with special focus on the amplitude (Higgs) modes. In the pure PDW state, 
    there are two overdamped Higgs modes. We show that a phase with co-existing PDW and uniform ($d$-wave) superconducting (SC) order, PDW/SC, spontaneously breaks time-reversal symmetry - and thus is distinct from a simpler phase, SC/CDW, with coexisting SC and charge-density-wave (CDW) order.  The PDW/SC phase exhibits three Higgs modes, one of which is sharply peaked and is predominantly a PDW fluctuation, symmetric between $\bm{Q}$ and
    -$\bm{Q}$, whose damping rate is strongly reduced by SC. This sharp mode should be visible in Raman experiments. 
     \end{abstract}
\maketitle

\date{\today}

{\it Introduction.}--A pair density wave (PDW) is an exotic form of superconducting order in which Cooper pairs carry finite center-of-mass momentum\cite{annurev:/content/journals/10.1146/annurev-conmatphys-031119-050711,PhysRevLett.130.126001,PhysRevLett.130.026001,PhysRevLett.131.026601,Setty2023,PhysRevB.107.214504,PhysRevB.110.094515,PhysRevLett.133.176501,PhysRevLett.125.167001,Huang2022,PhysRevB.109.L121101,wang2024pairdensitywavesstrongcoupling,PhysRevX.9.021047,PhysRevB.97.174511,wang2024quantumgeometryfacilitatedpairdensitywave,PhysRevB.107.224516,PhysRevB.105.L100509}. Many recent experiments have reported possible signatures of  PDW order  in the absence of 
a magnetic field
in correlated electronic systems, such as kagome metals\cite{Chen2021,Deng2024,PhysRevB.108.L081117,PhysRevB.110.024501,yao2024selfconsistenttheory2times2pair,PhysRevLett.129.167001}, NbSe$_2$\cite{doi:10.1126/science.abd4607}, UTe$_2$\cite{Gu2023,Aishwarya2023,Aishwarya2024}, EuRbFe$_4$As$_4$\cite{Zhao2023}, SrTa$_2$S$_5$\cite{Devarakonda2024} and rhombohedral graphene\cite{han2024signatureschiralsuperconductivityrhombohedral}.
 Certain La-based underdoped high $T_c$ cuprates, such as La$_{2-x}$Ba$_x$CuO$_4$ (LBCO) and  La$_{2-x-y}$X$_y$Sr$_x$CuO$_4$ with X=Nd (LNSCO)  or X=Eu (LESCO) are the most intensely studied PDW candidate materials, where the bulk superconducting (SC)  $T_c$ has a deep minimum at $x\approx 1/8$, while the ordering temperature for a stripe charge-density  wave (CDW), $T_{\rm cdw}$, is maximal\cite{PhysRevB.38.4596}.  At $T_c<T<T_{\rm cdw}$ transport measurements suggest a dynamical decoupling of the Cu-O layers\cite{PhysRevLett.99.067001,PhysRevLett.99.127003,Berg_2009,PhysRevB.91.115103,PhysRevLett.114.197001,
 PhysRevX.4.031017,RevModPhys.87.457,PhysRevLett.126.167001,PhysRevB.97.134520,Shi2020,PhysRevB.77.214524}, which is plausibly explained by the existence of in-plane  stripe PDW order with twice the period of the CDW. Below $T_c$,  this PDW order most plausibly coexists with a d-wave SC order~\cite{annurevTranquada}.

However, obtaining {\em direct} experimental evidence  of PDW order has proven difficult.
Transport properties can be difficult to interpret uniquely in complex materials. STM is another commonly used technique to provide evidence of PDW order\cite{doi:10.1126/science.aat1773,Du2020,Hamidian2016}. However, STM measurements provide information about surface states, 
and evidence of order can be difficult to disentangle from signatures of quasiparticle interference\cite{gao2023pairbreakingscatteringinterferencemechanism}.
 More fundamentally, it is 
unclear to what extent STM can distinguish a PDW state from a CDW+SC state. Given that a PDW is a ``new phase of matter,'' more direct and unambiguous experimental signatures are needed. An important experimental development in this direction is a recent X-ray study of underdoped LBCO and 
 La$_{2-x}$Sr$_x$Cu$_{1-y}$Fe$_y$O$_4$\cite{lee2023pairdensitywavesignatureobserved} that apparently provides bulk evidence for the coexistence of PDW and uniform SC order in a range of $T$.

In this Letter, we study low energy collective modes in a  system of  unidirectional $d$-wave PDW order~\cite{soto-garrido-wang-fradkin-cooper,PhysRevResearch.2.013034,nagashima2024opticallyactivehiggsleggett} with and without 
coexisting $d$-wave SC order. In the PDW+SC state, we find that the system favors 
spontaneous time-reversal symmetry   breaking 
(TRSB). Moreover, we observe that one of the amplitude (Higgs) modes exhibits a spectral function that is nearly delta-function-like. Such a sharp mode is predominantly from the synchronous motion of the two PDW amplitude modes
with ${\bQ}$ and $-{\bQ}$.
In contrast to the case of a pure PDW state where the same mode is overdamped, 
 the presence of 
uniform SC order can almost completely eliminate the damping for this particular mode.
This mode should be visible in non-resonant Raman scattering measurements, which then can be used as a versatile tool in the search for bulk evidence of PDW order. For comparison we also consider a coexisting CDW and SC state, which has been frequently discussed for La-based cuprates, see e.g. \cite{Miao2021,PhysRevLett.126.167001,Wen2019,PhysRevX.9.031042,PhysRevLett.124.207005,doi:10.1073/pnas.1708549114,PhysRevLett.125.097002}. We argue that Raman experiments can distinguish between CDW +SC and PDW + SC states,  at least when SC order is larger than the order with which it co-exists.

\begin{figure}
    \includegraphics[width=8.5cm]{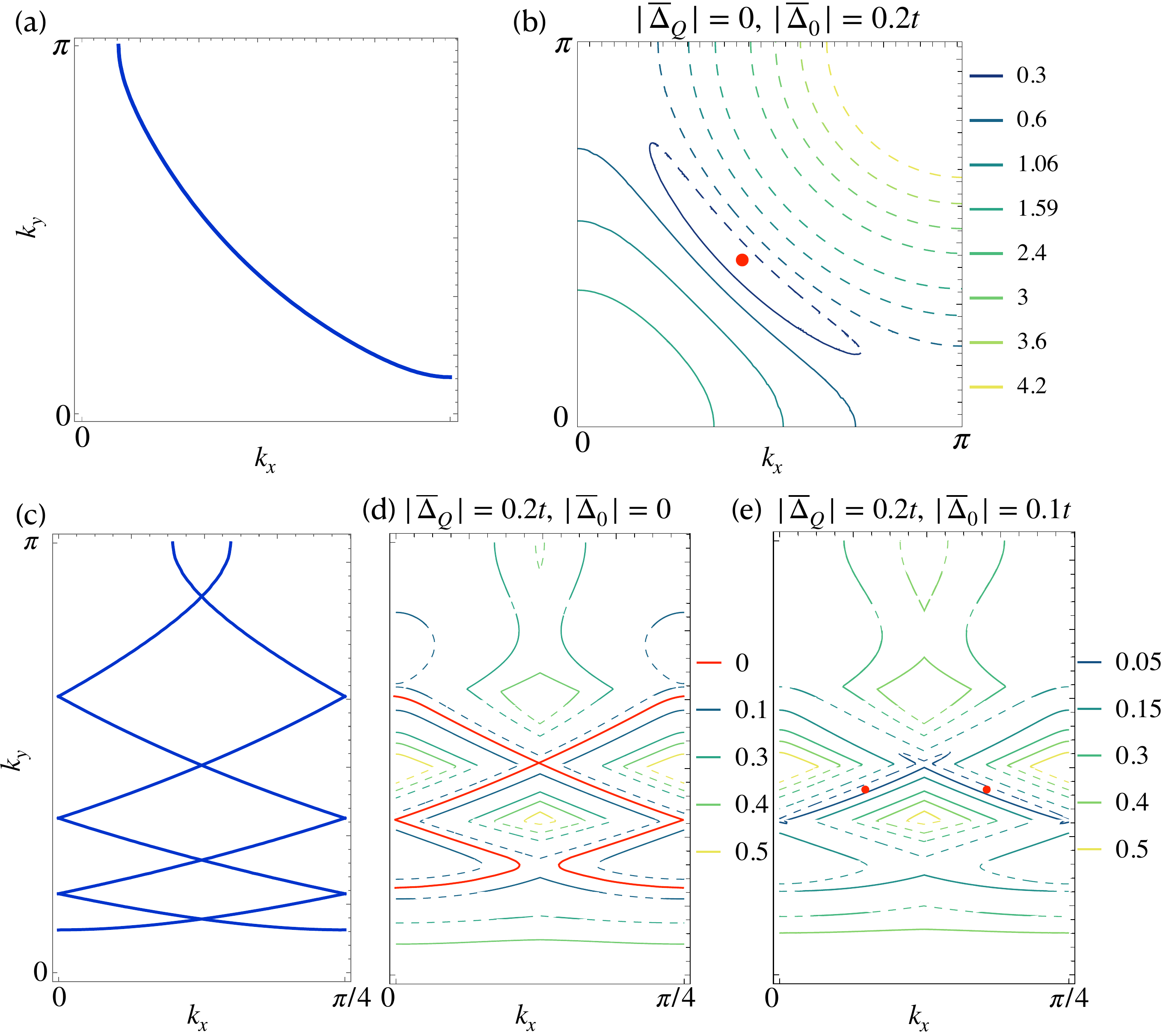}
    \caption{
     Fermi surface (FS) in the original  Brillouin zone (BZ). (b) CEC for a $d$-wave SC in the original BZ  for $|\overline\Delta_0|=0.2t$. The red dot indicates the nodal point. The dashed and solid curves are for electron-like and hole-like dispersions, respectively. (c) FS in the folded BZ. (d) CEC for a pure PDW with momentum $\bQ=(\frac{\pi}{4},0)$ in the folded BZ for $|\overline\Delta_{\pm \bQ}|=0.2t$. The red dashed curves indicate the residual Fermi surfaces. (e) CEC for a PDW+SC case with $\varphi_{\bQ}=\varphi_{-\bQ}$ and $\varphi_{\bm{0}}=\pi/2+\varphi_{\bQ}$ and with $|\overline\Delta_0|=0.1t$ and $|\overline\Delta_{\pm \bQ}|=0.2t$. Different colors correspond to different energies, specified to the right of each panel.}
 \label{fig:PDW_BdG}
\end{figure}

{\it Free energy at mean-field level.}--To enable  explicit calculations, we consider a specific microscopic model of PDW order on a square lattice designed to
represent one Cu-O layer. The fermion dispersion is 
taken to be $ \xi(\bk)=-2t\left[\cos k_x+\cos k_y\right]-4t'\cos k_x\cos k_y-\mu$. We choose 
$t'=-t/4$ and 
 report all energies in units of $t$. The chemical potential $\mu$ is chosen to set $x =1/8$,
resulting in the Fermi surface 
in Fig.\ref{fig:PDW_BdG}(a).
The PDW and SC order parameters 
$\Delta(\bm{R},\tau;\br)$ are 
taken to be of the form
\begin{equation}
    \begin{aligned}
        \Delta(\bm{R},\tau;\br)=&\sum_{\overline{\bq}=0,\pm\bQ}
        \Delta_{\overline{\bq}}(\bm{R},\tau)e^{i\overline{\bq}\cdot\bm{R}} f(\br)
    \end{aligned}\label{eq:orderparameter1}
\end{equation}
where $\tau$ is an imaginary time, $\overline{\bq}=0$ for
the uniform SC order and $\overline{\bq} = \pm \bQ$ for the two PDW orders~\cite{PhysRevB.69.184510,PhysRevB.62.6786,PhysRevB.63.174513}, and $\bm{R}$ and  $\br$ denote the center-of-mass and relative positions of the two fermions in a Cooper pair. The Fourier transform of $f(\br)$
is the pairing form factor, which we take to have
the $d_{x^2-y^2}$ form, $f_{\bk}=\cos k_x - \cos k_y$. We assume a period-8 PDW order so $\bQ=(\frac{\pi}{4},0)$.

At the mean field level, $\Delta_{\overline{\bq}}(\bm{R},\tau)=\overline{\Delta}_{\overline{\bq}} = |\overline\Delta_{\overline{\bq}}|e^{i\varphi_{\overline{\bq}}}$
 does not depend on $\bm{R}$ and $\tau$.   The phase factors $\varphi_{\overline{\bq}}$  have to be determined by minimizing the variational free energy. A conventional coupling of these mean field order parameters to band fermions ($\Delta \psi^\dagger \psi^\dagger + h.c$) yields a set of Bogoliubov quasiparticle bands\cite{SM}. In Fig.\ref{fig:PDW_BdG} (b) we show the constant energy contours (CEC) of the first Bogoliubov band above the Fermi level for a pure $d$-wave SC order. In Fig.\ref{fig:PDW_BdG} (c) we show the FS in the folded BZ, and in Fig.\ref{fig:PDW_BdG} (d,e) we show CEC for the PDW states. For a pure PDW order, there are multiple ``Bogoliubov Fermi surfaces''\cite{Berg_2009,PhysRevX.4.031017,PhysRevB.77.174502,doi:10.1073/pnas.1803009115,PhysRevB.86.104507,PhysRevResearch.3.023199},
 as shown in Fig.\ref{fig:PDW_BdG}(d). These  Fermi surfaces are further gapped if a uniform SC is also present, see Fig.\ref{fig:PDW_BdG}(e).


We obtain the effective Ginzburg-Landau (GL) action in terms of $\Delta_{\overline{\bq}}(\bm{R},\tau)$ by Hubbard-Stratonovich transformation of the underlying model with 4-fermion interactions (see \cite{SM}) 
we obtain
     \begin{align}
    &\mathcal{F}_\text{
    MF}=\alpha_1 (|\overline\Delta_{\bQ}|^2+|\overline\Delta_{-\bQ}|^2)+\alpha_2|\overline\Delta_{0}|^2\nonumber\\
&+\beta_1(|\overline\Delta_{\bQ}|^4+|\overline\Delta_{-\bQ}|^4)+\beta_2|\overline\Delta_{0}|^4\nonumber\\
&+\beta_3|\overline\Delta_{\bQ}|^2|\overline\Delta_{-\bQ}|^2   +\beta_4|\overline\Delta_0|^2\left(|\overline\Delta_{\bQ}|^2+|\overline\Delta_{-\bQ}|^2\right)\nonumber\\     &+\beta_5\left(\overline\Delta_0^2\overline\Delta^*_{-\bQ}\overline\Delta^*_{\bQ}
        + c.c.\right)
       +...
   \label{eq:FGL}
    \end{align}
 where all the coefficients are convolutions of fermionic propagators with the dispersion set by our microscopic model~\cite{SM}. In particular, $\alpha_1 = 1/g_1 - \Pi_Q$, $\alpha_2 = 1/g_2 - \Pi_0$, where $g_1$ and $g_2$ are the interactions responsible for the onset of PDW and SC orders and $\Pi_{\bQ} = \Pi_{-\bQ}$ and $\Pi_0$ are the corresponding polarization bubbles. All $\beta_i$ turn out to be positive,
  at least at $T \ll |{\bar \Delta}_{\bQ}|, |{\bar \Delta}_0|$.
 In this situation, 
 ${\cal F}_{MF}$ is minimized when $|{\overline \Delta}_{\bQ}| =  |{\overline \Delta}_{-\bQ}|$. The relation between the phases of SC and PDW orders is determined by $\beta_5>0$, so
\begin{equation}
    \begin{aligned}
  \varphi_{\bf 0}-\frac{\varphi_{\bQ}+\varphi_{-\bQ}}2
       = \frac \pi 2+n\pi
    \end{aligned}\label{eq:phi1phi2}
\end{equation}
where  $n$ is an integer. Since $\Delta_{\bQ}\to(\Delta_{-\bQ})^*$ under time reversal, Eq.\eqref{eq:phi1phi2} implies that in a mixed PDW + SC state  the system 
favors spontaneous TRS breaking \footnote{
TRSB in the mixed state does not depend on the phase difference $\varphi_{\bQ}-\varphi_{-\bQ}$, 
which is arbitrary to the order described by Eq. (\ref{eq:FGL}). This phase is fixed once we include the 8$^{th}$ order contribution to ${\cal F}_{MF}$:  $-\upsilon[(\overline\Delta_{\bQ}\overline\Delta^*_{-\bQ})^4+c.c.]$, which is allowed 
for the commensurate, period 8 case we have treated. We have computed  the prefactor $v$ and found 
it is  positive. 
Thus, $\varphi_{\bQ}-\varphi_{-\bQ} = m \pi/2$, where $m$ is integer.}. {This may provide a partial explanation of anomalous Kerr and Nernst effects observed in LBCO\cite{PhysRevLett.112.047003,PhysRevLett.107.277001}.}

 {\it Collective modes.}-- We now go beyond mean-field and consider small fluctuations, which in momentum space we parameterize  as~ \cite{PhysRevB.88.214508,PhysRevB.87.144511,PhysRevB.107.134519,PhysRevLett.108.177005,PhysRevB.82.144506,Sharapov2002,PhysRevB.69.184510,PhysRevB.95.214502,nosov2024spatiallyresolveddynamicsamplitudeschmidhiggs,Lee2023}
\begin{equation}
    \Delta_{\overline{\bq}}(q)\approx\overline{\Delta}_{\overline{\bq}}\left[1+\mathcal{A}_{\overline{\bq}}(q)+i\theta_{\overline{\bq}}(q)\right]
\end{equation}
where $q = (\omega_m, {\bQ})$ and $\mathcal{A}_{\overline{\bq}}(q)$ and  $\theta_{\overline{\bq}}(q)$ are the amplitude and phase variations, respectively. They satisfy $\mathcal{A}_{\overline{\bq}}(-q)=\mathcal{A}_{\overline{\bq}}^*(q)$, $\theta_{\overline{\bq}}(-q)=\theta_{\overline{\bq}}^*(q)$ and $\mathcal{A}_{\overline{\bq}}(0)=\theta_{\overline{\bq}}(0)=0$. It is convenient to introduce a vector basis $\zeta(q)=[\mathcal{A}_{\bQ}(q), \mathcal{A}_{-\bQ}(q), \mathcal{A}_{\bm{0}}(q),\theta_{\bQ}(q), \theta_{-\bQ}(q), \theta_{\bm{0}}(q)]^T$. Using this basis and truncating the fluctuating part of the GL action $S_\text{GL}$ at the Gaussian level, we obtain (see \cite{SM} for details)
\begin{equation}
    S^{(2)}_\text{
    GL}=\sum_q {\zeta}^T(q) \hat\Gamma^{-1}(q) \zeta(-q)\label{S2FL}
\end{equation}
 The  matrix  $\hat\Gamma^{-1}(q)$ can be thought of as the inverse matrix Green's function for the fluctuating fields (see \cite{SM} for details). The dispersions of the collective modes along the imaginary Matsubara frequency axis $i\omega_n$ can be found by solving for $\det\hat\Gamma^{-1}(q)=0$. To obtain the dispersions along the real frequency axis $\omega$ and the spectral functions  $B_j(\bq,\omega)$  ($j$  labels the collective modes), we use Pade approximants\cite{PhysRevB.93.075104} to implement the analytic continuation $i\omega_n \to \omega+0^+$. We will be interested in the spectral functions of the amplitude modes in the long-wavelength limit, and define $B_j(\omega)\equiv B_j(\bq=0,\omega)$. 
 The  calculations are done at $T = 0.005t$, which is in all cases that we studied is well below the values of $|{\overline \Delta}_{\bQ}|$, $|{\overline \Delta}_0|$. 

\begin{figure}
    \includegraphics[width=8.5cm]{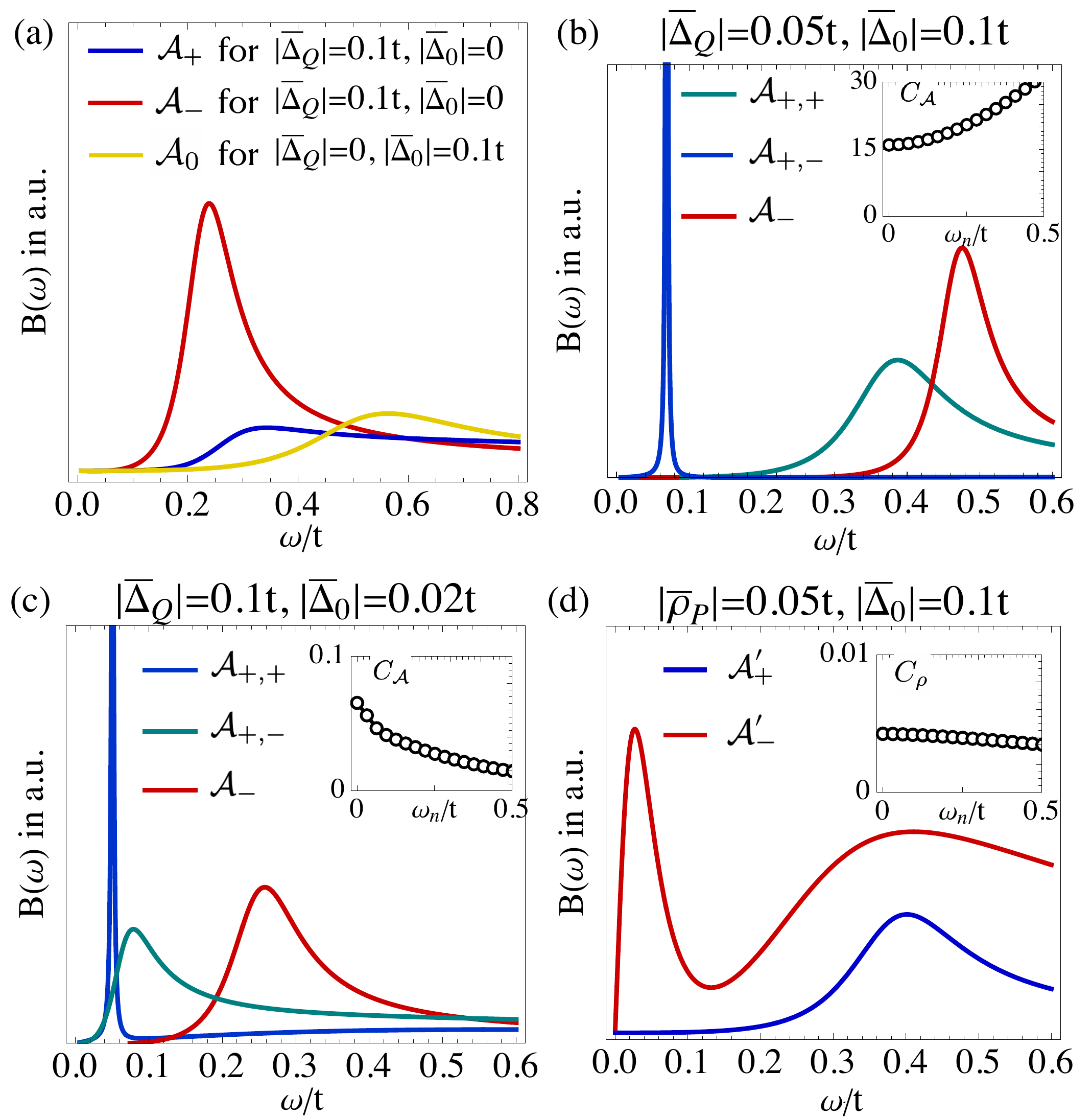}
    \caption{(a) Spectral functions $B(\omega)$ of $\mathcal{A}_{\pm}$ for a pure PDW order and $\mathcal{A}_{\bm{0}}$ for a pure SC order, both with $d$-wave form factor. (b)
     $B(\omega)$ for the three amplitude modes in Eq.\eqref{eq:phase123}. The inset shows $C_{\mathcal{A}}$ as a function of Matsubara frequency $\omega_n$ at $\bq=0$. In this case $C_{\mathcal{A}}\gg1$. (c) is similar to (b) but with parameters such that $C_{\mathcal{A}}\ll1$. (d) $B(\omega)$ for the two amplitude modes for the CDW+SC order.
    The numerical
    calculations are done
     at $T=0.005t$.}
    \label{fig:spectral}
\end{figure}

 {\it Pure PDW and pure SC}--When only $\overline\Delta_{\pm\bQ}$ are present, we find  two phase and two amplitude collective modes
 \begin{equation}
    \begin{aligned}
       &\theta_{\pm}(q)=\frac{\theta_{\bQ}(q)\pm\theta_{-\bQ}(q)}{\sqrt{2}}, ~~\mathcal{A}_{\pm}(q)=\frac{\mathcal{A}_{\bQ}(q)\pm\mathcal{A}_{-\bQ}(q)}{\sqrt{2}}.\\
    \end{aligned}\label{eq:phase12}
\end{equation}
In Fig.\ref{fig:spectral}(a) we plot  the  spectral functions $B(\omega)$ for the two amplitude modes. We see that both modes are strongly overdamped, and the damping is stronger for the $\mathcal{A}_+$ mode. The fact that $B(\omega)$ for the $\mathcal{A}_+$ mode peaks at a higher energy than that for the $\mathcal{A}_-$ mode can be understood analytically within the GL action\cite{soto-garrido-wang-fradkin-cooper,SM}. For comparison we also present $B(\omega)$ for $\mathcal{A}_{\bm{0}}$ when only a $d$-wave SC is present (yellow curve). Unlike in an s-wave SC, where $B(\omega) \propto 1/\sqrt{\omega^2 - 4 |\Delta_0|^2}$,  here for a $d$-wave SC  there is no singularity because of nodal quadiparticles, and the broad maximum is at $\omega/|\overline\Delta_0|\sim6 $.

{\it PDW with SC.}-- When both $\Delta_0$ and $\Delta_{\pm\bQ}$ are present, we find three phase and three amplitude eigen modes
\begin{equation}
    \begin{aligned}
       & \theta_{+,\pm}(q)\propto\frac{\theta_{\bQ}(q)+\theta_{-\bQ}(q)}{\sqrt{2}}\pm\left[C_\theta(q)\right]^{\pm1}\theta_{0}(q),\\
       & \theta_-(q)\propto\theta_{\bQ}(q)-\theta_{-\bQ}(q),\\
       &\mathcal{A}_{+,\pm}(q)\propto\frac{\mathcal{A}_{\bQ}(q)+\mathcal{A}_{-\bQ}(q)}{\sqrt{2}}\pm\left[C_{\mathcal{A}}(q)\right]^{\pm1}\mathcal{A}_{0}(q),\\
       & \mathcal{A}_-(q)\propto\mathcal{A}_{\bQ}(q)-\mathcal{A}_{-\bQ}(q),\\
        \end{aligned}\label{eq:phase123}
\end{equation}
 where $C_{\theta}(q)$ and $C_{\mathcal{A}}(q)$ are two dimensionless numbers. It can be shown that $C_{\theta}(0)=1/\sqrt{2}$, while $C_{\mathcal{A}}$ depends sensitively on the relative gap magnitudes. For small $q$, we find $C_{\mathcal{A}}\gg 1$ for $|\overline\Delta_{0}|\gg|\overline\Delta_{\bQ}|$, and $C_{\mathcal{A}}\ll 1$ for $|\overline\Delta_{0}|\ll|\overline\Delta_{\bQ}|$. We see from Eq.\eqref{eq:phase123} that $\mathcal{A}_-$ is decoupled from $\mathcal{A}_{\bm{0}}$, although its  propagator is affected by the presence of the SC order. Moreover, when $C_{\mathcal{A}}\gg1$, $\mathcal{A}_{+,+}$ and $\mathcal{A}_{+,-}$ become predominantly $\mathcal{A}_0$ and $\mathcal{A}_+$ [as in Eq.\eqref{eq:phase12}] respectively; while in the other limit when $C_{\mathcal{A}}\ll1$, we have $\mathcal{A}_{+,+}\approx \mathcal{A}_+$ and $\mathcal{A}_{+,-}\approx \mathcal{A}_{\bm{0}}$ instead. We show our numerical results of $B(\omega)$ for the  three amplitude modes in  Fig.\ref{fig:spectral} (b) and (c) for  $|{\overline \Delta}_0| > |{\overline \Delta}_{\bQ}|$, $C_{\mathcal{A}}\gg1$ and $|{\overline \Delta}_0| < |{\overline \Delta}_{\bQ}|$, $C_{\mathcal{A}}\ll1$, respectively. We note for both cases there exist an almost delta-function-like peak for the lowest energy mode, which is predominantly $\mathcal{A}_+$.

  A sharp Higgs peak is absent in the pure PDW and pure SC cases and thus appears to be  a unique feature of 
  mixed PDW+SC order. From an analytic 
  perspective, the case of $|\overline\Delta_0|\gg|\overline\Delta_{\bQ}|$, $C_{\mathcal{A}}\gg1$ is relatively easy to understand when $|\overline\Delta_{\bQ}|$ is treated perturbatively. To locate the modes, one has to (i) re-evaluate the frequencies of $\mathcal{A}_{+}$ and $\mathcal{A}_-$ from Eq.\eqref{eq:phase12} in the presence of SC within GL action, (ii) include mode-mode coupling between $\mathcal{A}_+$ and $\mathcal{A}_0$ so that the eigen modes become $\mathcal{A}_{+,\pm}$ and (iii) re-evaluate the damping rates in the presence of SC. We show the calculations in \cite{SM} and here list the results:  On (i), the 
  resulting $\mathcal{A}_{+}$ mode frequency remains comparable to $2 |\overline\Delta_{\bQ}|$,
  as in a pure PDW state, while the frequency of the $\mathcal{A}_-$ mode  increases in the presence of stronger SC and becomes comparable to $2 |\overline\Delta_0|$.  This is
  consistent with Fig.~\ref{fig:spectral}(b), which shows that the peak in the $\mathcal{A}_-$ mode is at a frequency set by $|\overline\Delta_0|$ rather than by $|\overline\Delta_{\bQ}|$. We note in passing that this effect is caused by the same $\beta_5$ term in \eqref{eq:FGL} that is  responsible for TRSB. On (ii), mode-mode coupling (level repulsion) shifts the frequency of the $\mathcal{A}_{+-}$ mode to a smaller frequency,  comparable to $|\overline\Delta_{\bQ}|$.   On (iii),  the $\mathcal{A}_-$ mode is peaked 
  above $2|\overline\Delta_0|$ and its damping is not reduced compared to a pure PDW, but the damping rate of the $\mathcal{A}_{+,-}$ mode, peaked well below $2|\overline\Delta_0|$, is strongly reduced by SC and also by the fact that even in a pure PDW state the damping is very small at $\omega \sim |\overline\Delta_{\bQ}|$.   As
  a consequence, the $\mathcal{A}_{+-}$ mode becomes almost completely propagating and the corresponding $B(\omega)$ displays a near-$\delta$-function 
  peak\footnote{There is a certain similarity between our case and Morr-Pines scenario for the resonance peak in the cuprates~\cite{PhysRevLett.81.1086}. {We also note that the Higgs mode can, in principle, also decay 
  into two quasiparticles of the
  phase mode $\theta_-$.~\cite{soto-garrido-wang-fradkin-cooper,fradkin-book} This is a higher-order process, 
  not included in our analysis}} 

Numerical results~\cite{SM} for a  generic ratio of $|{\overline \Delta}_0|/ |{\overline \Delta}_Q|$ 
and in particular, for the opposite limit  $|{\overline \Delta}_0| \ll |{\overline \Delta}_Q|$ as shown in Fig.\ref{fig:spectral}(c) again show a sharp spectral peak for the lowest energy amplitude mode (note for $C_{\mathcal{A}}\ll1$ as in Fig.\ref{fig:spectral}(c) this undamped mode is $\mathcal{A}_{+,+}$). To understand this analytically one needs to go beyond mode-mode coupling analysis (see \cite{SM}) since even an infinitesimal SC order parameter gaps out the entire Fermi surface (except for the nodal points) thus changing the susceptibilities in a non-perturbative way.

{\it CDW with SC.}--
We now discuss the case of uniform SC coexisting  with
CDW order.
To permit a direct comparison to the PDW +SC state, we 
take the CDW 
ordering vector to be $\bP=2\bQ$.
We also assume the CDW has an electronic origin\cite{PhysRevB.28.4029,Torchinsky2013,PhysRevLett.95.117002,PhysRevLett.96.137003}, and 
neglect the presence of optical phonon modes (for comparison, see \cite{PhysRevLett.31.462,PhysRevB.26.4883,RevModPhys.60.1129,PhysRevB.90.224515,PhysRevB.97.094502,PhysRevB.94.064512}). Thus, we parametrize the CDW fluctuations as
\begin{equation}
    \rho(q)=2|\overline\rho_{\bm{P}}|(1+\mathcal{A}_\rho(q))\cos(\bm{P}\cdot\br+
    \theta_{\rho}(q)),
\end{equation}
where $\mathcal{A}_\rho$ and $\theta_\rho$ are the CDW amplitude and phase modes (amplitudon and phason). As before, we introduce the attractive interactions in the CDW and SC channels and obtain an effective action for CDW and SC orders. At the mean-field level, we find that the phase of a SC order parameter can be arbitrary, i.e., TRS is not broken. 
 {{\em Importantly, this means that the phase with coexisting SC and PDW order is thermodynamically distinct from the phase with coexisting CDW and SC order!}} The part of the action describing fluctuations around mean-field is formally the same as Eq.\eqref{S2FL} in a new basis $\zeta'=[\mathcal{A}_{\rho}(q),\mathcal{A}_{\bm{0}}(q),\theta_{\rho}(q),\theta_0(q)]^T$. There are two phase and two amplitude eigenmodes
 \begin{equation}
    \begin{aligned}
        & \theta_{\rho}(q), ~~\theta_0(q), ~~\mathcal{A}'_{\pm}\propto \mathcal{A}_{\bm{0}}\pm [C_{\rho}(q)]^{\pm1}\mathcal{A}_{\rho}. \\
        \end{aligned}\label{eq:amplitude23}
\end{equation}
In Fig.\ref{fig:spectral}(d) we show the spectral functions for the two amplitude modes $\mathcal{A}'_{\pm}$. First we note $C_\rho\ll1$ for the chosen parameters. In fact, $C_\rho$ remains small even with larger $|\overline\rho_{\bP}|$ or $|\overline\Delta_0|$, meaning $\mathcal{A}'_+\approx\mathcal{A}_{\bm{0}}$ and $\mathcal{A}'_-\approx\mathcal{A}_{\rho}$. 
As in the PDW + SC case, the mode, for which $B(\omega)$ displays a visible peak, largely describes fluctuations of a non-SC order (here, CDW). We see, however, that the peak is substantially broader than in the PDW+SC case.

{\it Raman spectrum.}--
We next check whether the sharp mode for the PDW+SC case and a more broadened $\mathcal{A}'_{-}$ mode for the CDW+SC case can be detected in Raman scattering.  For this, we compute
 dressed Raman susceptibilities defined as $ \chi_R(\omega_n)=\int d\tau e^{i\omega_n\tau}\lim_{\bq\to0}\braket{\mathcal{T}_\tau \tilde \rho (\bq,\tau)\tilde \rho (-\bq,0)}$ where $\mathcal{T}_\tau$ is the time ordering and  $\tilde \rho (\bq,\tau)=\sum_{\bk,\sigma}\gamma(\bk)\psi^\dagger_\sigma(\bk+\bq/2,\tau)\psi_\sigma(\bk-\bq/2,\tau)$ is the Raman 
density with the Raman vertex  $\gamma_{\bk}$. Applying linear response theory~\cite{SM,PhysRevB.29.4976,PhysRevB.82.014507,RevModPhys.79.175,PhysRevB.52.9760,PhysRevB.51.16336,PhysRevLett.72.396}, we obtain
\begin{equation}
    \chi_R(\omega_n)=
    K(\omega_n)-\Lambda^T(-\omega_n)\hat\Gamma(\omega_n)\Lambda(\omega_n).\label{eq:Ramansus}
\end{equation}
Here $K=\chi_{\tilde\rho,\tilde\rho}$ is the bare Raman susceptibility, $\Lambda=\chi_{\zeta,\tilde\rho}$ is the coupling between the Raman density and a collective mode and $\hat \Gamma$ is the susceptibility of 
the collective mode.

In the presence of either PDW or CDW order, the original four-fold rotational symmetry of a square lattice is broken down to two-fold. {It is 
also 
likely that the stripe order
 breaks the $z\to-z$ mirror symmetry, in which case} 
the relevant  symmetry group becomes $D_2$.
The Raman vertices $\gamma(\bk)$ from 2D irreducible representations of $D_2$ are $A: \cos k_x \text{ or } \cos k_y$ and $B_1: \sin k_x\sin k_y$. For 
{pure stripe PDW order}
in either $x$ or $y$ direction,  
the $B_1$ {Raman} channel is inactive {since it is odd under $x\to-x$ {and} $y\to-y$,} 
leaving only the $A$-channel, 
{where} only modes that are even under $\bQ\to -\bQ$ are visible. 
For the PDW+SC case, we find by directly computing $\chi_R$ that the  $\mathcal{A}_{++}$ and $\mathcal{A}_{+-}$ modes are Raman active, while the
$\mathcal{A}_-$ mode is Raman inactive. For the CDW+SC case, both $\mathcal{A}'_{\pm}$ are Raman active.

\begin{figure}
    \includegraphics[width=8.5cm]{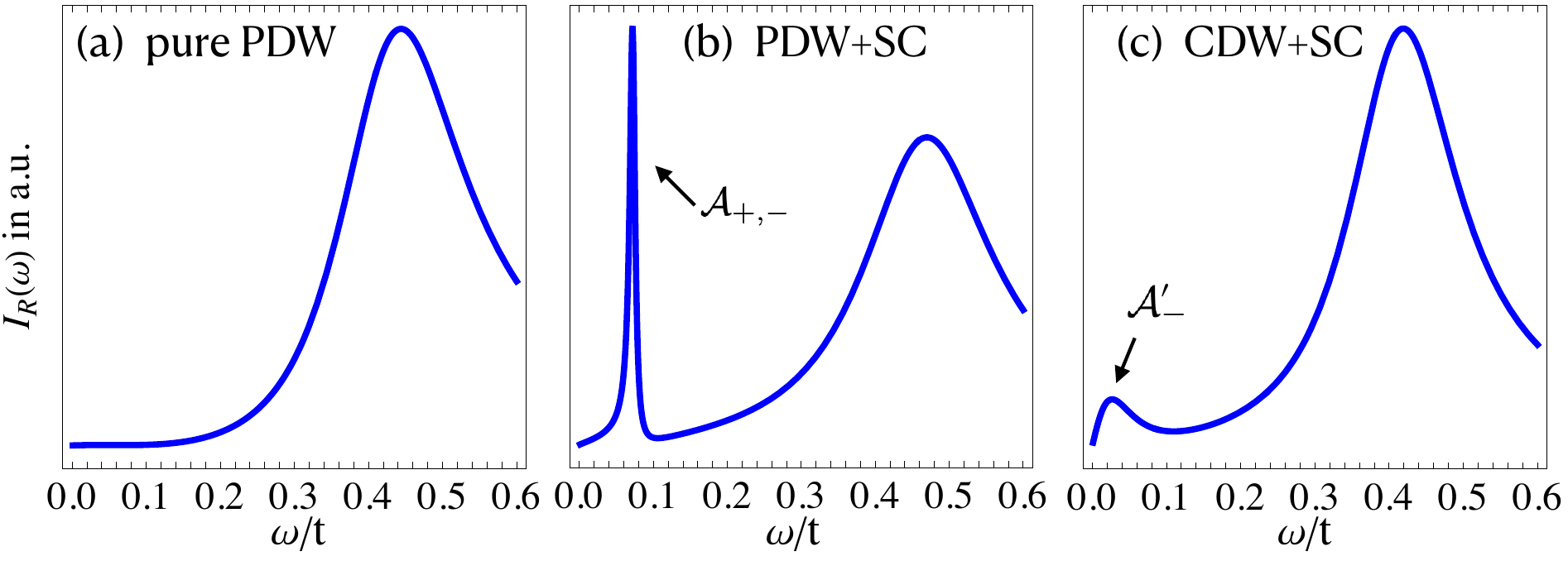}
    \caption{
    Raman intensities in the $A$-channel. The  parameters are the same as those in Fig.\ref{fig:spectral} (a), (b) and (d) respectively. Collective modes, which chiefly contribute to the peaks in $I_R(\omega)$ are marked.
    }\label{fig:Raman}
\end{figure}

In Fig.~\ref{fig:Raman} we show the calculated Raman intensity
$I_R(\omega)=\frac{-1}{\pi}\text{Im}\chi_R(i\omega_n\to\omega+i0^+)$ for all the three cases discussed. For a pure PDW [panel (a)],  the Raman response is featureless at small frequencies because the $\mathcal{A}_-$ mode, which could potentially give rise to a peak in $I_R(\omega)$~\cite{soto-garrido-wang-fradkin-cooper},  is Raman inactive. For PDW+SC order [panel (b)], $I_R (\omega)$ reproduces the  sharp peak in the spectral function of the $\mathcal{A}_{+-}$ mode. For the CDW + SC state, $I_R (\omega)$ reproduces the peak in the spectral function of the $\mathcal{A}'_{-}$ mode, but the peak is far smaller because of 
the finite width of the peak in $B(\omega)$ for this mode. Based on these results, we argue that Raman scattering can distinguish among these three different phases. 

{\it Effects of weak disorder:} 
In the presence of weak disorder, the existence of 2D long-range CDW or PDW order is precluded (for incommensurate order this is true even in 3D), and we expect that PDW and CDW order only exists with finite (possibly large) correlation length. The nature of the remaining (vestigial) orders when a pure PDW state is disrupted by weak disorder is not entirely clear\cite{PhysRevB.86.115138,PhysRevX.5.031008}. 
However, a PDW+SC state remains distinct from a CDW+SC state
since the TRSB characteristic of the PDW+SC phase should survive as vestigial order for a finite range of disorder strengths.

Another feature of disorder is that it induces a new form of coupling between the uniform SC and PDW orders when they coexist. Such a coupling is realized by a possible $1\bQ$ CDW order
\begin{align}
    \rho_{\pm {\bQ}}({\bf r})  \sim\left[\Delta_0^*({\bf r})\Delta_{\pm {\bQ}}({\bf r}) + \Delta_{\mp {\bQ}}^*({\bf r})\Delta_0({\bf r})\right].
\end{align}
In the absence of disorder, TRSB implies that this 1$\bQ$ order vanishes.
However, in the presence of a disorder $V(\br)$, there is an additive contribution to the effective action of the form ${\cal F}_{dis} = V({\bf r)}\left[\rho_{ {\bQ}}({\bf r})e^{i{\bQ}\cdot{\bf r}}+c.c. \right]$. Including this contribution in the PDW+SC case and assuming $V(\br)$ is short-range correlated we find\cite{SM}:
1)  At least to lowest order in the disorder strength, the disorder coupling always favors spontaneous TRSB, enforcing the tendency already derived in the clean limit\footnote{The physics behind this is the same as that which underlies the disorder driven breaking of TRS in junctions in which the leading order Josephson coupling is frustrated.\cite{PhysRevB.61.5902,PhysRevB.108.L100505}};
2) It induces local $1{\bQ}$ CDW order which is weak (small amplitude) but can have a relatively long correlation length that diverges when $\overline{V^2} \to 0$.  This sort of disorder-stabilized density-wave correlations is reminiscent of Zn doping stabilized spin-stripe order in LSCO\cite{HIROTA200161} and YBCO\cite{PhysRevLett.105.037207}.

It is also important to mention that in materials which exhibit signatures of the coexistence of uniform SC and PDW or uniform SC and CDW orders, it is always questionable  whether they coexist uniformly, or if instead they occur in distinct mesoscopic regions of the material, and only truly coexist at the interfaces between regions\cite{annurevTranquada}. 
Such coexistence could exhibit features quite different from anything we have analyzed.

{\it Concluding remarks.}--The purpose of this paper was to identify experimentally accessible features which, if detected, could serve as clear evidence of the existence of a PDW order. We have found that there is a 
clear distinction between a state with coexisting uniform SC and PDW order and a state with coexisting uniform SC and CDW order. 
First, the former state is expected to spontaneously break TRS.
Second, we studied amplitude collective modes for  PDW +SC and CDW +SC states and found that for a PDW+SC state, 
the spectral function of one collective mode displays a near-$\delta$-function peak. We found that this mode is Raman active and argued that a sharp peak should be visible in the Raman intensity. There is no such sharp peak in a CDW +SC state, i.e., the peak is  
 apparently a unique feature of the PDW+SC order.

\noindent{\bf Acknowledgements}:  We thank Eduardo Fradkin, Rudi Hackl, Tom Devereaux, Suk Bum Chung and Saurabh Maiti for helpful discussions. Y.-M.W. acknowledges support from the Gordon and Betty Moore Foundation’s EPiQS Initiative through GBMF8686. S.A.K. was supported in part by the Department of Energy, Office of Basic Energy Sciences, under contract No.~DEAC02-76SF00515.  A.V.C. was supported by the NSF-DMR Grant No.2325357. 
 Y. Wang was supported by NSF-DMR Grant No.2045781.
Y.Wang and A.V.C acknowledge support by grant NSF PHY-1748958 to the Kavli Institute for Theoretical Physics (YW and MY), where this work was partly performed.

\bibliography{Ramanv2}

\begin{thebibliography}{104}%
\makeatletter
\providecommand \@ifxundefined [1]{%
 \@ifx{#1\undefined}
}%
\providecommand \@ifnum [1]{%
 \ifnum #1\expandafter \@firstoftwo
 \else \expandafter \@secondoftwo
 \fi
}%
\providecommand \@ifx [1]{%
 \ifx #1\expandafter \@firstoftwo
 \else \expandafter \@secondoftwo
 \fi
}%
\providecommand \natexlab [1]{#1}%
\providecommand \enquote  [1]{``#1''}%
\providecommand \bibnamefont  [1]{#1}%
\providecommand \bibfnamefont [1]{#1}%
\providecommand \citenamefont [1]{#1}%
\providecommand \href@noop [0]{\@secondoftwo}%
\providecommand \href [0]{\begingroup \@sanitize@url \@href}%
\providecommand \@href[1]{\@@startlink{#1}\@@href}%
\providecommand \@@href[1]{\endgroup#1\@@endlink}%
\providecommand \@sanitize@url [0]{\catcode `\\12\catcode `\$12\catcode
  `\&12\catcode `\#12\catcode `\^12\catcode `\_12\catcode `\%12\relax}%
\providecommand \@@startlink[1]{}%
\providecommand \@@endlink[0]{}%
\providecommand \url  [0]{\begingroup\@sanitize@url \@url }%
\providecommand \@url [1]{\endgroup\@href {#1}{\urlprefix }}%
\providecommand \urlprefix  [0]{URL }%
\providecommand \Eprint [0]{\href }%
\providecommand \doibase [0]{https://doi.org/}%
\providecommand \selectlanguage [0]{\@gobble}%
\providecommand \bibinfo  [0]{\@secondoftwo}%
\providecommand \bibfield  [0]{\@secondoftwo}%
\providecommand \translation [1]{[#1]}%
\providecommand \BibitemOpen [0]{}%
\providecommand \bibitemStop [0]{}%
\providecommand \bibitemNoStop [0]{.\EOS\space}%
\providecommand \EOS [0]{\spacefactor3000\relax}%
\providecommand \BibitemShut  [1]{\csname bibitem#1\endcsname}%
\let\auto@bib@innerbib\@empty
\bibitem [{\citenamefont {Agterberg}\ \emph {et~al.}(2020)\citenamefont
  {Agterberg}, \citenamefont {Davis}, \citenamefont {Edkins}, \citenamefont
  {Fradkin}, \citenamefont {Van~Harlingen}, \citenamefont {Kivelson},
  \citenamefont {Lee}, \citenamefont {Radzihovsky}, \citenamefont {Tranquada},\
  and\ \citenamefont
  {Wang}}]{annurev:/content/journals/10.1146/annurev-conmatphys-031119-050711}%
  \BibitemOpen
  \bibfield  {author} {\bibinfo {author} {\bibfnamefont {D.~F.}\ \bibnamefont
  {Agterberg}}, \bibinfo {author} {\bibfnamefont {J.~S.}\ \bibnamefont
  {Davis}}, \bibinfo {author} {\bibfnamefont {S.~D.}\ \bibnamefont {Edkins}},
  \bibinfo {author} {\bibfnamefont {E.}~\bibnamefont {Fradkin}}, \bibinfo
  {author} {\bibfnamefont {D.~J.}\ \bibnamefont {Van~Harlingen}}, \bibinfo
  {author} {\bibfnamefont {S.~A.}\ \bibnamefont {Kivelson}}, \bibinfo {author}
  {\bibfnamefont {P.~A.}\ \bibnamefont {Lee}}, \bibinfo {author} {\bibfnamefont
  {L.}~\bibnamefont {Radzihovsky}}, \bibinfo {author} {\bibfnamefont {J.~M.}\
  \bibnamefont {Tranquada}},\ and\ \bibinfo {author} {\bibfnamefont
  {Y.}~\bibnamefont {Wang}},\ }\bibfield  {title} {\bibinfo {title} {The
  physics of pair-density waves: Cuprate superconductors and beyond},\ }\href
  {https://doi.org/https://doi.org/10.1146/annurev-conmatphys-031119-050711}
  {\bibfield  {journal} {\bibinfo  {journal} {Annual Review of Condensed Matter
  Physics}\ }\textbf {\bibinfo {volume} {11}},\ \bibinfo {pages} {231}
  (\bibinfo {year} {2020})}\BibitemShut {NoStop}%
\bibitem [{\citenamefont {Wu}\ \emph {et~al.}(2023{\natexlab{a}})\citenamefont
  {Wu}, \citenamefont {Wu},\ and\ \citenamefont
  {Yao}}]{PhysRevLett.130.126001}%
  \BibitemOpen
  \bibfield  {author} {\bibinfo {author} {\bibfnamefont {Y.-M.}\ \bibnamefont
  {Wu}}, \bibinfo {author} {\bibfnamefont {Z.}~\bibnamefont {Wu}},\ and\
  \bibinfo {author} {\bibfnamefont {H.}~\bibnamefont {Yao}},\ }\bibfield
  {title} {\bibinfo {title} {Pair-density-wave and chiral superconductivity in
  twisted bilayer transition metal dichalcogenides},\ }\href
  {https://doi.org/10.1103/PhysRevLett.130.126001} {\bibfield  {journal}
  {\bibinfo  {journal} {Phys. Rev. Lett.}\ }\textbf {\bibinfo {volume} {130}},\
  \bibinfo {pages} {126001} (\bibinfo {year} {2023}{\natexlab{a}})}\BibitemShut
  {NoStop}%
\bibitem [{\citenamefont {Wu}\ \emph {et~al.}(2023{\natexlab{b}})\citenamefont
  {Wu}, \citenamefont {Nosov}, \citenamefont {Patel},\ and\ \citenamefont
  {Raghu}}]{PhysRevLett.130.026001}%
  \BibitemOpen
  \bibfield  {author} {\bibinfo {author} {\bibfnamefont {Y.-M.}\ \bibnamefont
  {Wu}}, \bibinfo {author} {\bibfnamefont {P.~A.}\ \bibnamefont {Nosov}},
  \bibinfo {author} {\bibfnamefont {A.~A.}\ \bibnamefont {Patel}},\ and\
  \bibinfo {author} {\bibfnamefont {S.}~\bibnamefont {Raghu}},\ }\bibfield
  {title} {\bibinfo {title} {Pair density wave order from electron repulsion},\
  }\href {https://doi.org/10.1103/PhysRevLett.130.026001} {\bibfield  {journal}
  {\bibinfo  {journal} {Phys. Rev. Lett.}\ }\textbf {\bibinfo {volume} {130}},\
  \bibinfo {pages} {026001} (\bibinfo {year} {2023}{\natexlab{b}})}\BibitemShut
  {NoStop}%
\bibitem [{\citenamefont {Castro}\ \emph {et~al.}(2023)\citenamefont {Castro},
  \citenamefont {Shaffer}, \citenamefont {Wu},\ and\ \citenamefont
  {Santos}}]{PhysRevLett.131.026601}%
  \BibitemOpen
  \bibfield  {author} {\bibinfo {author} {\bibfnamefont {P.}~\bibnamefont
  {Castro}}, \bibinfo {author} {\bibfnamefont {D.}~\bibnamefont {Shaffer}},
  \bibinfo {author} {\bibfnamefont {Y.-M.}\ \bibnamefont {Wu}},\ and\ \bibinfo
  {author} {\bibfnamefont {L.~H.}\ \bibnamefont {Santos}},\ }\bibfield  {title}
  {\bibinfo {title} {Emergence of the chern supermetal and pair-density wave
  through higher-order van hove singularities in the haldane-hubbard model},\
  }\href {https://doi.org/10.1103/PhysRevLett.131.026601} {\bibfield  {journal}
  {\bibinfo  {journal} {Phys. Rev. Lett.}\ }\textbf {\bibinfo {volume} {131}},\
  \bibinfo {pages} {026601} (\bibinfo {year} {2023})}\BibitemShut {NoStop}%
\bibitem [{\citenamefont {Setty}\ \emph {et~al.}(2023)\citenamefont {Setty},
  \citenamefont {Fanfarillo},\ and\ \citenamefont {Hirschfeld}}]{Setty2023}%
  \BibitemOpen
  \bibfield  {author} {\bibinfo {author} {\bibfnamefont {C.}~\bibnamefont
  {Setty}}, \bibinfo {author} {\bibfnamefont {L.}~\bibnamefont {Fanfarillo}},\
  and\ \bibinfo {author} {\bibfnamefont {P.~J.}\ \bibnamefont {Hirschfeld}},\
  }\bibfield  {title} {\bibinfo {title} {Mechanism for fluctuating pair density
  wave},\ }\href {https://doi.org/10.1038/s41467-023-38956-x} {\bibfield
  {journal} {\bibinfo  {journal} {Nature Communications}\ }\textbf {\bibinfo
  {volume} {14}},\ \bibinfo {pages} {3181} (\bibinfo {year}
  {2023})}\BibitemShut {NoStop}%
\bibitem [{\citenamefont {Jiang}(2023)}]{PhysRevB.107.214504}%
  \BibitemOpen
  \bibfield  {author} {\bibinfo {author} {\bibfnamefont {H.-C.}\ \bibnamefont
  {Jiang}},\ }\bibfield  {title} {\bibinfo {title} {Pair density wave in the
  doped three-band hubbard model on two-leg square cylinders},\ }\href
  {https://doi.org/10.1103/PhysRevB.107.214504} {\bibfield  {journal} {\bibinfo
   {journal} {Phys. Rev. B}\ }\textbf {\bibinfo {volume} {107}},\ \bibinfo
  {pages} {214504} (\bibinfo {year} {2023})}\BibitemShut {NoStop}%
\bibitem [{\citenamefont {Ticea}\ \emph {et~al.}(2024)\citenamefont {Ticea},
  \citenamefont {Raghu},\ and\ \citenamefont {Wu}}]{PhysRevB.110.094515}%
  \BibitemOpen
  \bibfield  {author} {\bibinfo {author} {\bibfnamefont {N.~S.}\ \bibnamefont
  {Ticea}}, \bibinfo {author} {\bibfnamefont {S.}~\bibnamefont {Raghu}},\ and\
  \bibinfo {author} {\bibfnamefont {Y.-M.}\ \bibnamefont {Wu}},\ }\bibfield
  {title} {\bibinfo {title} {Pair density wave order in multiband systems},\
  }\href {https://doi.org/10.1103/PhysRevB.110.094515} {\bibfield  {journal}
  {\bibinfo  {journal} {Phys. Rev. B}\ }\textbf {\bibinfo {volume} {110}},\
  \bibinfo {pages} {094515} (\bibinfo {year} {2024})}\BibitemShut {NoStop}%
\bibitem [{\citenamefont {Jiang}\ and\ \citenamefont
  {Yao}(2024)}]{PhysRevLett.133.176501}%
  \BibitemOpen
  \bibfield  {author} {\bibinfo {author} {\bibfnamefont {Y.-F.}\ \bibnamefont
  {Jiang}}\ and\ \bibinfo {author} {\bibfnamefont {H.}~\bibnamefont {Yao}},\
  }\bibfield  {title} {\bibinfo {title} {Pair-density-wave superconductivity: A
  microscopic model on the 2d honeycomb lattice},\ }\href
  {https://doi.org/10.1103/PhysRevLett.133.176501} {\bibfield  {journal}
  {\bibinfo  {journal} {Phys. Rev. Lett.}\ }\textbf {\bibinfo {volume} {133}},\
  \bibinfo {pages} {176501} (\bibinfo {year} {2024})}\BibitemShut {NoStop}%
\bibitem [{\citenamefont {Han}\ \emph {et~al.}(2020)\citenamefont {Han},
  \citenamefont {Kivelson},\ and\ \citenamefont
  {Yao}}]{PhysRevLett.125.167001}%
  \BibitemOpen
  \bibfield  {author} {\bibinfo {author} {\bibfnamefont {Z.}~\bibnamefont
  {Han}}, \bibinfo {author} {\bibfnamefont {S.~A.}\ \bibnamefont {Kivelson}},\
  and\ \bibinfo {author} {\bibfnamefont {H.}~\bibnamefont {Yao}},\ }\bibfield
  {title} {\bibinfo {title} {Strong coupling limit of the holstein-hubbard
  model},\ }\href {https://doi.org/10.1103/PhysRevLett.125.167001} {\bibfield
  {journal} {\bibinfo  {journal} {Phys. Rev. Lett.}\ }\textbf {\bibinfo
  {volume} {125}},\ \bibinfo {pages} {167001} (\bibinfo {year}
  {2020})}\BibitemShut {NoStop}%
\bibitem [{\citenamefont {Huang}\ \emph {et~al.}(2022)\citenamefont {Huang},
  \citenamefont {Han}, \citenamefont {Kivelson},\ and\ \citenamefont
  {Yao}}]{Huang2022}%
  \BibitemOpen
  \bibfield  {author} {\bibinfo {author} {\bibfnamefont {K.~S.}\ \bibnamefont
  {Huang}}, \bibinfo {author} {\bibfnamefont {Z.}~\bibnamefont {Han}}, \bibinfo
  {author} {\bibfnamefont {S.~A.}\ \bibnamefont {Kivelson}},\ and\ \bibinfo
  {author} {\bibfnamefont {H.}~\bibnamefont {Yao}},\ }\bibfield  {title}
  {\bibinfo {title} {Pair-density-wave in the strong coupling limit of the
  holstein-hubbard model},\ }\href {https://doi.org/10.1038/s41535-022-00426-w}
  {\bibfield  {journal} {\bibinfo  {journal} {npj Quantum Materials}\ }\textbf
  {\bibinfo {volume} {7}},\ \bibinfo {pages} {17} (\bibinfo {year}
  {2022})}\BibitemShut {NoStop}%
\bibitem [{\citenamefont {Liu}\ and\ \citenamefont
  {Han}(2024)}]{PhysRevB.109.L121101}%
  \BibitemOpen
  \bibfield  {author} {\bibinfo {author} {\bibfnamefont {F.}~\bibnamefont
  {Liu}}\ and\ \bibinfo {author} {\bibfnamefont {Z.}~\bibnamefont {Han}},\
  }\bibfield  {title} {\bibinfo {title} {Pair density wave and
  $\mathit{s}\ifmmode\pm\else\textpm\fi{}\mathit{id}$ superconductivity in a
  strongly coupled lightly doped kondo insulator},\ }\href
  {https://doi.org/10.1103/PhysRevB.109.L121101} {\bibfield  {journal}
  {\bibinfo  {journal} {Phys. Rev. B}\ }\textbf {\bibinfo {volume} {109}},\
  \bibinfo {pages} {L121101} (\bibinfo {year} {2024})}\BibitemShut {NoStop}%
\bibitem [{\citenamefont {Wang}\ \emph {et~al.}(2024)\citenamefont {Wang},
  \citenamefont {Sun}, \citenamefont {Wang}, \citenamefont {Han}, \citenamefont
  {Kivelson},\ and\ \citenamefont
  {Yao}}]{wang2024pairdensitywavesstrongcoupling}%
  \BibitemOpen
  \bibfield  {author} {\bibinfo {author} {\bibfnamefont {J.}~\bibnamefont
  {Wang}}, \bibinfo {author} {\bibfnamefont {W.}~\bibnamefont {Sun}}, \bibinfo
  {author} {\bibfnamefont {H.-X.}\ \bibnamefont {Wang}}, \bibinfo {author}
  {\bibfnamefont {Z.}~\bibnamefont {Han}}, \bibinfo {author} {\bibfnamefont
  {S.~A.}\ \bibnamefont {Kivelson}},\ and\ \bibinfo {author} {\bibfnamefont
  {H.}~\bibnamefont {Yao}},\ }\href {https://arxiv.org/abs/2404.11950}
  {\bibinfo {title} {Pair density waves in the strong-coupling two-dimensional
  holstein-hubbard model: a variational monte carlo study}} (\bibinfo {year}
  {2024}),\ \Eprint {https://arxiv.org/abs/2404.11950} {arXiv:2404.11950
  [cond-mat.str-el]} \BibitemShut {NoStop}%
\bibitem [{\citenamefont {Santos}\ \emph {et~al.}(2019)\citenamefont {Santos},
  \citenamefont {Wang},\ and\ \citenamefont {Fradkin}}]{PhysRevX.9.021047}%
  \BibitemOpen
  \bibfield  {author} {\bibinfo {author} {\bibfnamefont {L.~H.}\ \bibnamefont
  {Santos}}, \bibinfo {author} {\bibfnamefont {Y.}~\bibnamefont {Wang}},\ and\
  \bibinfo {author} {\bibfnamefont {E.}~\bibnamefont {Fradkin}},\ }\bibfield
  {title} {\bibinfo {title} {Pair-density-wave order and paired fractional
  quantum hall fluids},\ }\href {https://doi.org/10.1103/PhysRevX.9.021047}
  {\bibfield  {journal} {\bibinfo  {journal} {Phys. Rev. X}\ }\textbf {\bibinfo
  {volume} {9}},\ \bibinfo {pages} {021047} (\bibinfo {year}
  {2019})}\BibitemShut {NoStop}%
\bibitem [{\citenamefont {Dai}\ \emph {et~al.}(2018)\citenamefont {Dai},
  \citenamefont {Zhang}, \citenamefont {Senthil},\ and\ \citenamefont
  {Lee}}]{PhysRevB.97.174511}%
  \BibitemOpen
  \bibfield  {author} {\bibinfo {author} {\bibfnamefont {Z.}~\bibnamefont
  {Dai}}, \bibinfo {author} {\bibfnamefont {Y.-H.}\ \bibnamefont {Zhang}},
  \bibinfo {author} {\bibfnamefont {T.}~\bibnamefont {Senthil}},\ and\ \bibinfo
  {author} {\bibfnamefont {P.~A.}\ \bibnamefont {Lee}},\ }\bibfield  {title}
  {\bibinfo {title} {Pair-density waves, charge-density waves, and vortices in
  high-${T}_{c}$ cuprates},\ }\href
  {https://doi.org/10.1103/PhysRevB.97.174511} {\bibfield  {journal} {\bibinfo
  {journal} {Phys. Rev. B}\ }\textbf {\bibinfo {volume} {97}},\ \bibinfo
  {pages} {174511} (\bibinfo {year} {2018})}\BibitemShut {NoStop}%
\bibitem [{\citenamefont {Wang}\ and\ \citenamefont
  {Huang}(2024)}]{wang2024quantumgeometryfacilitatedpairdensitywave}%
  \BibitemOpen
  \bibfield  {author} {\bibinfo {author} {\bibfnamefont {H.-X.}\ \bibnamefont
  {Wang}}\ and\ \bibinfo {author} {\bibfnamefont {W.}~\bibnamefont {Huang}},\
  }\href {https://arxiv.org/abs/2406.17187} {\bibinfo {title}
  {Quantum-geometry-facilitated pair density wave order: Density matrix
  renormalization group study}} (\bibinfo {year} {2024}),\ \Eprint
  {https://arxiv.org/abs/2406.17187} {arXiv:2406.17187 [cond-mat.str-el]}
  \BibitemShut {NoStop}%
\bibitem [{\citenamefont {Shaffer}\ \emph {et~al.}(2023)\citenamefont
  {Shaffer}, \citenamefont {Burnell},\ and\ \citenamefont
  {Fernandes}}]{PhysRevB.107.224516}%
  \BibitemOpen
  \bibfield  {author} {\bibinfo {author} {\bibfnamefont {D.}~\bibnamefont
  {Shaffer}}, \bibinfo {author} {\bibfnamefont {F.~J.}\ \bibnamefont
  {Burnell}},\ and\ \bibinfo {author} {\bibfnamefont {R.~M.}\ \bibnamefont
  {Fernandes}},\ }\bibfield  {title} {\bibinfo {title} {Weak-coupling theory of
  pair density wave instabilities in transition metal dichalcogenides},\ }\href
  {https://doi.org/10.1103/PhysRevB.107.224516} {\bibfield  {journal} {\bibinfo
   {journal} {Phys. Rev. B}\ }\textbf {\bibinfo {volume} {107}},\ \bibinfo
  {pages} {224516} (\bibinfo {year} {2023})}\BibitemShut {NoStop}%
\bibitem [{\citenamefont {Han}\ and\ \citenamefont
  {Kivelson}(2022)}]{PhysRevB.105.L100509}%
  \BibitemOpen
  \bibfield  {author} {\bibinfo {author} {\bibfnamefont {Z.}~\bibnamefont
  {Han}}\ and\ \bibinfo {author} {\bibfnamefont {S.~A.}\ \bibnamefont
  {Kivelson}},\ }\bibfield  {title} {\bibinfo {title} {Pair density wave and
  reentrant superconducting tendencies originating from valley polarization},\
  }\href {https://doi.org/10.1103/PhysRevB.105.L100509} {\bibfield  {journal}
  {\bibinfo  {journal} {Phys. Rev. B}\ }\textbf {\bibinfo {volume} {105}},\
  \bibinfo {pages} {L100509} (\bibinfo {year} {2022})}\BibitemShut {NoStop}%
\bibitem [{\citenamefont {Chen}\ \emph {et~al.}(2021)\citenamefont {Chen},
  \citenamefont {Yang}, \citenamefont {Hu}, \citenamefont {Zhao}, \citenamefont
  {Yuan}, \citenamefont {Xing}, \citenamefont {Qian}, \citenamefont {Huang},
  \citenamefont {Li}, \citenamefont {Ye}, \citenamefont {Ma}, \citenamefont
  {Ni}, \citenamefont {Zhang}, \citenamefont {Yin}, \citenamefont {Gong},
  \citenamefont {Tu}, \citenamefont {Lei}, \citenamefont {Tan}, \citenamefont
  {Zhou}, \citenamefont {Shen}, \citenamefont {Dong}, \citenamefont {Yan},
  \citenamefont {Wang},\ and\ \citenamefont {Gao}}]{Chen2021}%
  \BibitemOpen
  \bibfield  {author} {\bibinfo {author} {\bibfnamefont {H.}~\bibnamefont
  {Chen}}, \bibinfo {author} {\bibfnamefont {H.}~\bibnamefont {Yang}}, \bibinfo
  {author} {\bibfnamefont {B.}~\bibnamefont {Hu}}, \bibinfo {author}
  {\bibfnamefont {Z.}~\bibnamefont {Zhao}}, \bibinfo {author} {\bibfnamefont
  {J.}~\bibnamefont {Yuan}}, \bibinfo {author} {\bibfnamefont {Y.}~\bibnamefont
  {Xing}}, \bibinfo {author} {\bibfnamefont {G.}~\bibnamefont {Qian}}, \bibinfo
  {author} {\bibfnamefont {Z.}~\bibnamefont {Huang}}, \bibinfo {author}
  {\bibfnamefont {G.}~\bibnamefont {Li}}, \bibinfo {author} {\bibfnamefont
  {Y.}~\bibnamefont {Ye}}, \bibinfo {author} {\bibfnamefont {S.}~\bibnamefont
  {Ma}}, \bibinfo {author} {\bibfnamefont {S.}~\bibnamefont {Ni}}, \bibinfo
  {author} {\bibfnamefont {H.}~\bibnamefont {Zhang}}, \bibinfo {author}
  {\bibfnamefont {Q.}~\bibnamefont {Yin}}, \bibinfo {author} {\bibfnamefont
  {C.}~\bibnamefont {Gong}}, \bibinfo {author} {\bibfnamefont {Z.}~\bibnamefont
  {Tu}}, \bibinfo {author} {\bibfnamefont {H.}~\bibnamefont {Lei}}, \bibinfo
  {author} {\bibfnamefont {H.}~\bibnamefont {Tan}}, \bibinfo {author}
  {\bibfnamefont {S.}~\bibnamefont {Zhou}}, \bibinfo {author} {\bibfnamefont
  {C.}~\bibnamefont {Shen}}, \bibinfo {author} {\bibfnamefont {X.}~\bibnamefont
  {Dong}}, \bibinfo {author} {\bibfnamefont {B.}~\bibnamefont {Yan}}, \bibinfo
  {author} {\bibfnamefont {Z.}~\bibnamefont {Wang}},\ and\ \bibinfo {author}
  {\bibfnamefont {H.-J.}\ \bibnamefont {Gao}},\ }\bibfield  {title} {\bibinfo
  {title} {Roton pair density wave in a strong-coupling kagome
  superconductor},\ }\href {https://doi.org/10.1038/s41586-021-03983-5}
  {\bibfield  {journal} {\bibinfo  {journal} {Nature}\ }\textbf {\bibinfo
  {volume} {599}},\ \bibinfo {pages} {222} (\bibinfo {year}
  {2021})}\BibitemShut {NoStop}%
\bibitem [{\citenamefont {Deng}\ \emph {et~al.}(2024)\citenamefont {Deng},
  \citenamefont {Qin}, \citenamefont {Liu}, \citenamefont {Yang}, \citenamefont
  {Fu}, \citenamefont {Zhang}, \citenamefont {Wu}, \citenamefont {Wang},
  \citenamefont {Shi}, \citenamefont {Liu}, \citenamefont {Liu}, \citenamefont
  {Yan}, \citenamefont {Song}, \citenamefont {Xu}, \citenamefont {Zhao},
  \citenamefont {Yi}, \citenamefont {Xu}, \citenamefont {Hohmann},
  \citenamefont {Holb{\ae}k}, \citenamefont {D{\"u}rrnagel}, \citenamefont
  {Zhou}, \citenamefont {Chang}, \citenamefont {Yao}, \citenamefont {Wang},
  \citenamefont {Guguchia}, \citenamefont {Neupert}, \citenamefont {Thomale},
  \citenamefont {Fischer},\ and\ \citenamefont {Yin}}]{Deng2024}%
  \BibitemOpen
  \bibfield  {author} {\bibinfo {author} {\bibfnamefont {H.}~\bibnamefont
  {Deng}}, \bibinfo {author} {\bibfnamefont {H.}~\bibnamefont {Qin}}, \bibinfo
  {author} {\bibfnamefont {G.}~\bibnamefont {Liu}}, \bibinfo {author}
  {\bibfnamefont {T.}~\bibnamefont {Yang}}, \bibinfo {author} {\bibfnamefont
  {R.}~\bibnamefont {Fu}}, \bibinfo {author} {\bibfnamefont {Z.}~\bibnamefont
  {Zhang}}, \bibinfo {author} {\bibfnamefont {X.}~\bibnamefont {Wu}}, \bibinfo
  {author} {\bibfnamefont {Z.}~\bibnamefont {Wang}}, \bibinfo {author}
  {\bibfnamefont {Y.}~\bibnamefont {Shi}}, \bibinfo {author} {\bibfnamefont
  {J.}~\bibnamefont {Liu}}, \bibinfo {author} {\bibfnamefont {H.}~\bibnamefont
  {Liu}}, \bibinfo {author} {\bibfnamefont {X.-Y.}\ \bibnamefont {Yan}},
  \bibinfo {author} {\bibfnamefont {W.}~\bibnamefont {Song}}, \bibinfo {author}
  {\bibfnamefont {X.}~\bibnamefont {Xu}}, \bibinfo {author} {\bibfnamefont
  {Y.}~\bibnamefont {Zhao}}, \bibinfo {author} {\bibfnamefont {M.}~\bibnamefont
  {Yi}}, \bibinfo {author} {\bibfnamefont {G.}~\bibnamefont {Xu}}, \bibinfo
  {author} {\bibfnamefont {H.}~\bibnamefont {Hohmann}}, \bibinfo {author}
  {\bibfnamefont {S.~C.}\ \bibnamefont {Holb{\ae}k}}, \bibinfo {author}
  {\bibfnamefont {M.}~\bibnamefont {D{\"u}rrnagel}}, \bibinfo {author}
  {\bibfnamefont {S.}~\bibnamefont {Zhou}}, \bibinfo {author} {\bibfnamefont
  {G.}~\bibnamefont {Chang}}, \bibinfo {author} {\bibfnamefont
  {Y.}~\bibnamefont {Yao}}, \bibinfo {author} {\bibfnamefont {Q.}~\bibnamefont
  {Wang}}, \bibinfo {author} {\bibfnamefont {Z.}~\bibnamefont {Guguchia}},
  \bibinfo {author} {\bibfnamefont {T.}~\bibnamefont {Neupert}}, \bibinfo
  {author} {\bibfnamefont {R.}~\bibnamefont {Thomale}}, \bibinfo {author}
  {\bibfnamefont {M.~H.}\ \bibnamefont {Fischer}},\ and\ \bibinfo {author}
  {\bibfnamefont {J.-X.}\ \bibnamefont {Yin}},\ }\bibfield  {title} {\bibinfo
  {title} {Chiral kagome superconductivity modulations with residual fermi
  arcs},\ }\href {https://doi.org/10.1038/s41586-024-07798-y} {\bibfield
  {journal} {\bibinfo  {journal} {Nature}\ }\textbf {\bibinfo {volume} {632}},\
  \bibinfo {pages} {775} (\bibinfo {year} {2024})}\BibitemShut {NoStop}%
\bibitem [{\citenamefont {Wu}\ \emph {et~al.}(2023{\natexlab{c}})\citenamefont
  {Wu}, \citenamefont {Thomale},\ and\ \citenamefont
  {Raghu}}]{PhysRevB.108.L081117}%
  \BibitemOpen
  \bibfield  {author} {\bibinfo {author} {\bibfnamefont {Y.-M.}\ \bibnamefont
  {Wu}}, \bibinfo {author} {\bibfnamefont {R.}~\bibnamefont {Thomale}},\ and\
  \bibinfo {author} {\bibfnamefont {S.}~\bibnamefont {Raghu}},\ }\bibfield
  {title} {\bibinfo {title} {Sublattice interference promotes pair density wave
  order in kagome metals},\ }\href
  {https://doi.org/10.1103/PhysRevB.108.L081117} {\bibfield  {journal}
  {\bibinfo  {journal} {Phys. Rev. B}\ }\textbf {\bibinfo {volume} {108}},\
  \bibinfo {pages} {L081117} (\bibinfo {year}
  {2023}{\natexlab{c}})}\BibitemShut {NoStop}%
\bibitem [{\citenamefont {Schwemmer}\ \emph {et~al.}(2024)\citenamefont
  {Schwemmer}, \citenamefont {Hohmann}, \citenamefont {D\"urrnagel},
  \citenamefont {Potten}, \citenamefont {Beyer}, \citenamefont {Rachel},
  \citenamefont {Wu}, \citenamefont {Raghu}, \citenamefont {M\"uller},
  \citenamefont {Hanke},\ and\ \citenamefont {Thomale}}]{PhysRevB.110.024501}%
  \BibitemOpen
  \bibfield  {author} {\bibinfo {author} {\bibfnamefont {T.}~\bibnamefont
  {Schwemmer}}, \bibinfo {author} {\bibfnamefont {H.}~\bibnamefont {Hohmann}},
  \bibinfo {author} {\bibfnamefont {M.}~\bibnamefont {D\"urrnagel}}, \bibinfo
  {author} {\bibfnamefont {J.}~\bibnamefont {Potten}}, \bibinfo {author}
  {\bibfnamefont {J.}~\bibnamefont {Beyer}}, \bibinfo {author} {\bibfnamefont
  {S.}~\bibnamefont {Rachel}}, \bibinfo {author} {\bibfnamefont {Y.-M.}\
  \bibnamefont {Wu}}, \bibinfo {author} {\bibfnamefont {S.}~\bibnamefont
  {Raghu}}, \bibinfo {author} {\bibfnamefont {T.}~\bibnamefont {M\"uller}},
  \bibinfo {author} {\bibfnamefont {W.}~\bibnamefont {Hanke}},\ and\ \bibinfo
  {author} {\bibfnamefont {R.}~\bibnamefont {Thomale}},\ }\bibfield  {title}
  {\bibinfo {title} {Sublattice modulated superconductivity in the kagome
  hubbard model},\ }\href {https://doi.org/10.1103/PhysRevB.110.024501}
  {\bibfield  {journal} {\bibinfo  {journal} {Phys. Rev. B}\ }\textbf {\bibinfo
  {volume} {110}},\ \bibinfo {pages} {024501} (\bibinfo {year}
  {2024})}\BibitemShut {NoStop}%
\bibitem [{\citenamefont {Yao}\ \emph {et~al.}(2024)\citenamefont {Yao},
  \citenamefont {Wang}, \citenamefont {Wang}, \citenamefont {Yin},\ and\
  \citenamefont {Wang}}]{yao2024selfconsistenttheory2times2pair}%
  \BibitemOpen
  \bibfield  {author} {\bibinfo {author} {\bibfnamefont {M.}~\bibnamefont
  {Yao}}, \bibinfo {author} {\bibfnamefont {Y.}~\bibnamefont {Wang}}, \bibinfo
  {author} {\bibfnamefont {D.}~\bibnamefont {Wang}}, \bibinfo {author}
  {\bibfnamefont {J.-X.}\ \bibnamefont {Yin}},\ and\ \bibinfo {author}
  {\bibfnamefont {Q.-H.}\ \bibnamefont {Wang}},\ }\href
  {https://arxiv.org/abs/2408.03056} {\bibinfo {title} {Self-consistent theory
  of $2\times2$ pair density waves in kagome superconductors}} (\bibinfo {year}
  {2024}),\ \Eprint {https://arxiv.org/abs/2408.03056} {arXiv:2408.03056
  [cond-mat.supr-con]} \BibitemShut {NoStop}%
\bibitem [{\citenamefont {Jin}\ \emph {et~al.}(2022)\citenamefont {Jin},
  \citenamefont {Jiang}, \citenamefont {Yao},\ and\ \citenamefont
  {Zhou}}]{PhysRevLett.129.167001}%
  \BibitemOpen
  \bibfield  {author} {\bibinfo {author} {\bibfnamefont {J.-T.}\ \bibnamefont
  {Jin}}, \bibinfo {author} {\bibfnamefont {K.}~\bibnamefont {Jiang}}, \bibinfo
  {author} {\bibfnamefont {H.}~\bibnamefont {Yao}},\ and\ \bibinfo {author}
  {\bibfnamefont {Y.}~\bibnamefont {Zhou}},\ }\bibfield  {title} {\bibinfo
  {title} {Interplay between pair density wave and a nested fermi surface},\
  }\href {https://doi.org/10.1103/PhysRevLett.129.167001} {\bibfield  {journal}
  {\bibinfo  {journal} {Phys. Rev. Lett.}\ }\textbf {\bibinfo {volume} {129}},\
  \bibinfo {pages} {167001} (\bibinfo {year} {2022})}\BibitemShut {NoStop}%
\bibitem [{\citenamefont {Liu}\ \emph {et~al.}(2021)\citenamefont {Liu},
  \citenamefont {Chong}, \citenamefont {Sharma},\ and\ \citenamefont
  {Davis}}]{doi:10.1126/science.abd4607}%
  \BibitemOpen
  \bibfield  {author} {\bibinfo {author} {\bibfnamefont {X.}~\bibnamefont
  {Liu}}, \bibinfo {author} {\bibfnamefont {Y.~X.}\ \bibnamefont {Chong}},
  \bibinfo {author} {\bibfnamefont {R.}~\bibnamefont {Sharma}},\ and\ \bibinfo
  {author} {\bibfnamefont {J.~C.~S.}\ \bibnamefont {Davis}},\ }\bibfield
  {title} {\bibinfo {title} {Discovery of a cooper-pair density wave state in a
  transition-metal dichalcogenide},\ }\href
  {https://doi.org/10.1126/science.abd4607} {\bibfield  {journal} {\bibinfo
  {journal} {Science}\ }\textbf {\bibinfo {volume} {372}},\ \bibinfo {pages}
  {1447} (\bibinfo {year} {2021})}\BibitemShut {NoStop}%
\bibitem [{\citenamefont {Gu}\ \emph {et~al.}(2023)\citenamefont {Gu},
  \citenamefont {Carroll}, \citenamefont {Wang}, \citenamefont {Ran},
  \citenamefont {Broyles}, \citenamefont {Siddiquee}, \citenamefont {Butch},
  \citenamefont {Saha}, \citenamefont {Paglione}, \citenamefont {Davis},\ and\
  \citenamefont {Liu}}]{Gu2023}%
  \BibitemOpen
  \bibfield  {author} {\bibinfo {author} {\bibfnamefont {Q.}~\bibnamefont
  {Gu}}, \bibinfo {author} {\bibfnamefont {J.~P.}\ \bibnamefont {Carroll}},
  \bibinfo {author} {\bibfnamefont {S.}~\bibnamefont {Wang}}, \bibinfo {author}
  {\bibfnamefont {S.}~\bibnamefont {Ran}}, \bibinfo {author} {\bibfnamefont
  {C.}~\bibnamefont {Broyles}}, \bibinfo {author} {\bibfnamefont
  {H.}~\bibnamefont {Siddiquee}}, \bibinfo {author} {\bibfnamefont {N.~P.}\
  \bibnamefont {Butch}}, \bibinfo {author} {\bibfnamefont {S.~R.}\ \bibnamefont
  {Saha}}, \bibinfo {author} {\bibfnamefont {J.}~\bibnamefont {Paglione}},
  \bibinfo {author} {\bibfnamefont {J.~C.~S.}\ \bibnamefont {Davis}},\ and\
  \bibinfo {author} {\bibfnamefont {X.}~\bibnamefont {Liu}},\ }\bibfield
  {title} {\bibinfo {title} {Detection of a pair density wave state in ute2},\
  }\href {https://doi.org/10.1038/s41586-023-05919-7} {\bibfield  {journal}
  {\bibinfo  {journal} {Nature}\ }\textbf {\bibinfo {volume} {618}},\ \bibinfo
  {pages} {921} (\bibinfo {year} {2023})}\BibitemShut {NoStop}%
\bibitem [{\citenamefont {Aishwarya}\ \emph {et~al.}(2023)\citenamefont
  {Aishwarya}, \citenamefont {May-Mann}, \citenamefont {Raghavan},
  \citenamefont {Nie}, \citenamefont {Romanelli}, \citenamefont {Ran},
  \citenamefont {Saha}, \citenamefont {Paglione}, \citenamefont {Butch},
  \citenamefont {Fradkin},\ and\ \citenamefont {Madhavan}}]{Aishwarya2023}%
  \BibitemOpen
  \bibfield  {author} {\bibinfo {author} {\bibfnamefont {A.}~\bibnamefont
  {Aishwarya}}, \bibinfo {author} {\bibfnamefont {J.}~\bibnamefont {May-Mann}},
  \bibinfo {author} {\bibfnamefont {A.}~\bibnamefont {Raghavan}}, \bibinfo
  {author} {\bibfnamefont {L.}~\bibnamefont {Nie}}, \bibinfo {author}
  {\bibfnamefont {M.}~\bibnamefont {Romanelli}}, \bibinfo {author}
  {\bibfnamefont {S.}~\bibnamefont {Ran}}, \bibinfo {author} {\bibfnamefont
  {S.~R.}\ \bibnamefont {Saha}}, \bibinfo {author} {\bibfnamefont
  {J.}~\bibnamefont {Paglione}}, \bibinfo {author} {\bibfnamefont {N.~P.}\
  \bibnamefont {Butch}}, \bibinfo {author} {\bibfnamefont {E.}~\bibnamefont
  {Fradkin}},\ and\ \bibinfo {author} {\bibfnamefont {V.}~\bibnamefont
  {Madhavan}},\ }\bibfield  {title} {\bibinfo {title} {Magnetic-field-sensitive
  charge density waves in the superconductor ute2},\ }\href
  {https://doi.org/10.1038/s41586-023-06005-8} {\bibfield  {journal} {\bibinfo
  {journal} {Nature}\ }\textbf {\bibinfo {volume} {618}},\ \bibinfo {pages}
  {928} (\bibinfo {year} {2023})}\BibitemShut {NoStop}%
\bibitem [{\citenamefont {Aishwarya}\ \emph {et~al.}(2024)\citenamefont
  {Aishwarya}, \citenamefont {May-Mann}, \citenamefont {Almoalem},
  \citenamefont {Ran}, \citenamefont {Saha}, \citenamefont {Paglione},
  \citenamefont {Butch}, \citenamefont {Fradkin},\ and\ \citenamefont
  {Madhavan}}]{Aishwarya2024}%
  \BibitemOpen
  \bibfield  {author} {\bibinfo {author} {\bibfnamefont {A.}~\bibnamefont
  {Aishwarya}}, \bibinfo {author} {\bibfnamefont {J.}~\bibnamefont {May-Mann}},
  \bibinfo {author} {\bibfnamefont {A.}~\bibnamefont {Almoalem}}, \bibinfo
  {author} {\bibfnamefont {S.}~\bibnamefont {Ran}}, \bibinfo {author}
  {\bibfnamefont {S.~R.}\ \bibnamefont {Saha}}, \bibinfo {author}
  {\bibfnamefont {J.}~\bibnamefont {Paglione}}, \bibinfo {author}
  {\bibfnamefont {N.~P.}\ \bibnamefont {Butch}}, \bibinfo {author}
  {\bibfnamefont {E.}~\bibnamefont {Fradkin}},\ and\ \bibinfo {author}
  {\bibfnamefont {V.}~\bibnamefont {Madhavan}},\ }\bibfield  {title} {\bibinfo
  {title} {Melting of the charge density wave by generation of pairs of
  topological defects in ute2},\ }\href
  {https://doi.org/10.1038/s41567-024-02429-9} {\bibfield  {journal} {\bibinfo
  {journal} {Nature Physics}\ }\textbf {\bibinfo {volume} {20}},\ \bibinfo
  {pages} {964} (\bibinfo {year} {2024})}\BibitemShut {NoStop}%
\bibitem [{\citenamefont {Zhao}\ \emph {et~al.}(2023)\citenamefont {Zhao},
  \citenamefont {Blackwell}, \citenamefont {Thinel}, \citenamefont {Handa},
  \citenamefont {Ishida}, \citenamefont {Zhu}, \citenamefont {Iyo},
  \citenamefont {Eisaki}, \citenamefont {Pasupathy},\ and\ \citenamefont
  {Fujita}}]{Zhao2023}%
  \BibitemOpen
  \bibfield  {author} {\bibinfo {author} {\bibfnamefont {H.}~\bibnamefont
  {Zhao}}, \bibinfo {author} {\bibfnamefont {R.}~\bibnamefont {Blackwell}},
  \bibinfo {author} {\bibfnamefont {M.}~\bibnamefont {Thinel}}, \bibinfo
  {author} {\bibfnamefont {T.}~\bibnamefont {Handa}}, \bibinfo {author}
  {\bibfnamefont {S.}~\bibnamefont {Ishida}}, \bibinfo {author} {\bibfnamefont
  {X.}~\bibnamefont {Zhu}}, \bibinfo {author} {\bibfnamefont {A.}~\bibnamefont
  {Iyo}}, \bibinfo {author} {\bibfnamefont {H.}~\bibnamefont {Eisaki}},
  \bibinfo {author} {\bibfnamefont {A.~N.}\ \bibnamefont {Pasupathy}},\ and\
  \bibinfo {author} {\bibfnamefont {K.}~\bibnamefont {Fujita}},\ }\bibfield
  {title} {\bibinfo {title} {Smectic pair-density-wave order in eurbfe4as4},\
  }\href {https://doi.org/10.1038/s41586-023-06103-7} {\bibfield  {journal}
  {\bibinfo  {journal} {Nature}\ }\textbf {\bibinfo {volume} {618}},\ \bibinfo
  {pages} {940} (\bibinfo {year} {2023})}\BibitemShut {NoStop}%
\bibitem [{\citenamefont {Devarakonda}\ \emph {et~al.}(2024)\citenamefont
  {Devarakonda}, \citenamefont {Chen}, \citenamefont {Fang}, \citenamefont
  {Graf}, \citenamefont {Kriener}, \citenamefont {Akey}, \citenamefont {Bell},
  \citenamefont {Suzuki},\ and\ \citenamefont {Checkelsky}}]{Devarakonda2024}%
  \BibitemOpen
  \bibfield  {author} {\bibinfo {author} {\bibfnamefont {A.}~\bibnamefont
  {Devarakonda}}, \bibinfo {author} {\bibfnamefont {A.}~\bibnamefont {Chen}},
  \bibinfo {author} {\bibfnamefont {S.}~\bibnamefont {Fang}}, \bibinfo {author}
  {\bibfnamefont {D.}~\bibnamefont {Graf}}, \bibinfo {author} {\bibfnamefont
  {M.}~\bibnamefont {Kriener}}, \bibinfo {author} {\bibfnamefont {A.~J.}\
  \bibnamefont {Akey}}, \bibinfo {author} {\bibfnamefont {D.~C.}\ \bibnamefont
  {Bell}}, \bibinfo {author} {\bibfnamefont {T.}~\bibnamefont {Suzuki}},\ and\
  \bibinfo {author} {\bibfnamefont {J.~G.}\ \bibnamefont {Checkelsky}},\
  }\bibfield  {title} {\bibinfo {title} {Evidence of striped electronic phases
  in a structurally modulated superlattice},\ }\href
  {https://doi.org/10.1038/s41586-024-07589-5} {\bibfield  {journal} {\bibinfo
  {journal} {Nature}\ }\textbf {\bibinfo {volume} {631}},\ \bibinfo {pages}
  {526} (\bibinfo {year} {2024})}\BibitemShut {NoStop}%
\bibitem [{\citenamefont {Han}\ \emph {et~al.}(2024)\citenamefont {Han},
  \citenamefont {Lu}, \citenamefont {Yao}, \citenamefont {Shi}, \citenamefont
  {Yang}, \citenamefont {Seo}, \citenamefont {Ye}, \citenamefont {Wu},
  \citenamefont {Zhou}, \citenamefont {Liu}, \citenamefont {Shi}, \citenamefont
  {Hua}, \citenamefont {Watanabe}, \citenamefont {Taniguchi}, \citenamefont
  {Xiong}, \citenamefont {Fu},\ and\ \citenamefont
  {Ju}}]{han2024signatureschiralsuperconductivityrhombohedral}%
  \BibitemOpen
  \bibfield  {author} {\bibinfo {author} {\bibfnamefont {T.}~\bibnamefont
  {Han}}, \bibinfo {author} {\bibfnamefont {Z.}~\bibnamefont {Lu}}, \bibinfo
  {author} {\bibfnamefont {Y.}~\bibnamefont {Yao}}, \bibinfo {author}
  {\bibfnamefont {L.}~\bibnamefont {Shi}}, \bibinfo {author} {\bibfnamefont
  {J.}~\bibnamefont {Yang}}, \bibinfo {author} {\bibfnamefont {J.}~\bibnamefont
  {Seo}}, \bibinfo {author} {\bibfnamefont {S.}~\bibnamefont {Ye}}, \bibinfo
  {author} {\bibfnamefont {Z.}~\bibnamefont {Wu}}, \bibinfo {author}
  {\bibfnamefont {M.}~\bibnamefont {Zhou}}, \bibinfo {author} {\bibfnamefont
  {H.}~\bibnamefont {Liu}}, \bibinfo {author} {\bibfnamefont {G.}~\bibnamefont
  {Shi}}, \bibinfo {author} {\bibfnamefont {Z.}~\bibnamefont {Hua}}, \bibinfo
  {author} {\bibfnamefont {K.}~\bibnamefont {Watanabe}}, \bibinfo {author}
  {\bibfnamefont {T.}~\bibnamefont {Taniguchi}}, \bibinfo {author}
  {\bibfnamefont {P.}~\bibnamefont {Xiong}}, \bibinfo {author} {\bibfnamefont
  {L.}~\bibnamefont {Fu}},\ and\ \bibinfo {author} {\bibfnamefont
  {L.}~\bibnamefont {Ju}},\ }\href {https://arxiv.org/abs/2408.15233} {\bibinfo
  {title} {Signatures of chiral superconductivity in rhombohedral graphene}}
  (\bibinfo {year} {2024}),\ \Eprint {https://arxiv.org/abs/2408.15233}
  {arXiv:2408.15233 [cond-mat.mes-hall]} \BibitemShut {NoStop}%
\bibitem [{\citenamefont {Moodenbaugh}\ \emph {et~al.}(1988)\citenamefont
  {Moodenbaugh}, \citenamefont {Xu}, \citenamefont {Suenaga}, \citenamefont
  {Folkerts},\ and\ \citenamefont {Shelton}}]{PhysRevB.38.4596}%
  \BibitemOpen
  \bibfield  {author} {\bibinfo {author} {\bibfnamefont {A.~R.}\ \bibnamefont
  {Moodenbaugh}}, \bibinfo {author} {\bibfnamefont {Y.}~\bibnamefont {Xu}},
  \bibinfo {author} {\bibfnamefont {M.}~\bibnamefont {Suenaga}}, \bibinfo
  {author} {\bibfnamefont {T.~J.}\ \bibnamefont {Folkerts}},\ and\ \bibinfo
  {author} {\bibfnamefont {R.~N.}\ \bibnamefont {Shelton}},\ }\bibfield
  {title} {\bibinfo {title} {Superconducting properties of
  ${\mathrm{la}}_{2\mathrm{\ensuremath{-}}\mathrm{x}}$${\mathrm{ba}}_{\mathrm{x}}$${\mathrm{cuo}}_{4}$},\
  }\href {https://doi.org/10.1103/PhysRevB.38.4596} {\bibfield  {journal}
  {\bibinfo  {journal} {Phys. Rev. B}\ }\textbf {\bibinfo {volume} {38}},\
  \bibinfo {pages} {4596} (\bibinfo {year} {1988})}\BibitemShut {NoStop}%
\bibitem [{\citenamefont {Li}\ \emph {et~al.}(2007)\citenamefont {Li},
  \citenamefont {H\"ucker}, \citenamefont {Gu}, \citenamefont {Tsvelik},\ and\
  \citenamefont {Tranquada}}]{PhysRevLett.99.067001}%
  \BibitemOpen
  \bibfield  {author} {\bibinfo {author} {\bibfnamefont {Q.}~\bibnamefont
  {Li}}, \bibinfo {author} {\bibfnamefont {M.}~\bibnamefont {H\"ucker}},
  \bibinfo {author} {\bibfnamefont {G.~D.}\ \bibnamefont {Gu}}, \bibinfo
  {author} {\bibfnamefont {A.~M.}\ \bibnamefont {Tsvelik}},\ and\ \bibinfo
  {author} {\bibfnamefont {J.~M.}\ \bibnamefont {Tranquada}},\ }\bibfield
  {title} {\bibinfo {title} {Two-dimensional superconducting fluctuations in
  stripe-ordered
  ${\mathrm{la}}_{1.875}{\mathrm{ba}}_{0.125}{\mathrm{cuo}}_{4}$},\ }\href
  {https://doi.org/10.1103/PhysRevLett.99.067001} {\bibfield  {journal}
  {\bibinfo  {journal} {Phys. Rev. Lett.}\ }\textbf {\bibinfo {volume} {99}},\
  \bibinfo {pages} {067001} (\bibinfo {year} {2007})}\BibitemShut {NoStop}%
\bibitem [{\citenamefont {Berg}\ \emph {et~al.}(2007)\citenamefont {Berg},
  \citenamefont {Fradkin}, \citenamefont {Kim}, \citenamefont {Kivelson},
  \citenamefont {Oganesyan}, \citenamefont {Tranquada},\ and\ \citenamefont
  {Zhang}}]{PhysRevLett.99.127003}%
  \BibitemOpen
  \bibfield  {author} {\bibinfo {author} {\bibfnamefont {E.}~\bibnamefont
  {Berg}}, \bibinfo {author} {\bibfnamefont {E.}~\bibnamefont {Fradkin}},
  \bibinfo {author} {\bibfnamefont {E.-A.}\ \bibnamefont {Kim}}, \bibinfo
  {author} {\bibfnamefont {S.~A.}\ \bibnamefont {Kivelson}}, \bibinfo {author}
  {\bibfnamefont {V.}~\bibnamefont {Oganesyan}}, \bibinfo {author}
  {\bibfnamefont {J.~M.}\ \bibnamefont {Tranquada}},\ and\ \bibinfo {author}
  {\bibfnamefont {S.~C.}\ \bibnamefont {Zhang}},\ }\bibfield  {title} {\bibinfo
  {title} {Dynamical layer decoupling in a stripe-ordered high-${T}_{c}$
  superconductor},\ }\href {https://doi.org/10.1103/PhysRevLett.99.127003}
  {\bibfield  {journal} {\bibinfo  {journal} {Phys. Rev. Lett.}\ }\textbf
  {\bibinfo {volume} {99}},\ \bibinfo {pages} {127003} (\bibinfo {year}
  {2007})}\BibitemShut {NoStop}%
\bibitem [{\citenamefont {Berg}\ \emph {et~al.}(2009)\citenamefont {Berg},
  \citenamefont {Fradkin}, \citenamefont {Kivelson},\ and\ \citenamefont
  {Tranquada}}]{Berg_2009}%
  \BibitemOpen
  \bibfield  {author} {\bibinfo {author} {\bibfnamefont {E.}~\bibnamefont
  {Berg}}, \bibinfo {author} {\bibfnamefont {E.}~\bibnamefont {Fradkin}},
  \bibinfo {author} {\bibfnamefont {S.~A.}\ \bibnamefont {Kivelson}},\ and\
  \bibinfo {author} {\bibfnamefont {J.~M.}\ \bibnamefont {Tranquada}},\
  }\bibfield  {title} {\bibinfo {title} {Striped superconductors: how spin,
  charge and superconducting orders intertwine in the cuprates},\ }\href
  {https://doi.org/10.1088/1367-2630/11/11/115004} {\bibfield  {journal}
  {\bibinfo  {journal} {New Journal of Physics}\ }\textbf {\bibinfo {volume}
  {11}},\ \bibinfo {pages} {115004} (\bibinfo {year} {2009})}\BibitemShut
  {NoStop}%
\bibitem [{\citenamefont {Wang}\ \emph
  {et~al.}(2015{\natexlab{a}})\citenamefont {Wang}, \citenamefont {Agterberg},\
  and\ \citenamefont {Chubukov}}]{PhysRevB.91.115103}%
  \BibitemOpen
  \bibfield  {author} {\bibinfo {author} {\bibfnamefont {Y.}~\bibnamefont
  {Wang}}, \bibinfo {author} {\bibfnamefont {D.~F.}\ \bibnamefont
  {Agterberg}},\ and\ \bibinfo {author} {\bibfnamefont {A.}~\bibnamefont
  {Chubukov}},\ }\bibfield  {title} {\bibinfo {title} {Interplay between pair-
  and charge-density-wave orders in underdoped cuprates},\ }\href
  {https://doi.org/10.1103/PhysRevB.91.115103} {\bibfield  {journal} {\bibinfo
  {journal} {Phys. Rev. B}\ }\textbf {\bibinfo {volume} {91}},\ \bibinfo
  {pages} {115103} (\bibinfo {year} {2015}{\natexlab{a}})}\BibitemShut
  {NoStop}%
\bibitem [{\citenamefont {Wang}\ \emph
  {et~al.}(2015{\natexlab{b}})\citenamefont {Wang}, \citenamefont {Agterberg},\
  and\ \citenamefont {Chubukov}}]{PhysRevLett.114.197001}%
  \BibitemOpen
  \bibfield  {author} {\bibinfo {author} {\bibfnamefont {Y.}~\bibnamefont
  {Wang}}, \bibinfo {author} {\bibfnamefont {D.~F.}\ \bibnamefont
  {Agterberg}},\ and\ \bibinfo {author} {\bibfnamefont {A.}~\bibnamefont
  {Chubukov}},\ }\bibfield  {title} {\bibinfo {title} {Coexistence of
  charge-density-wave and pair-density-wave orders in underdoped cuprates},\
  }\href {https://doi.org/10.1103/PhysRevLett.114.197001} {\bibfield  {journal}
  {\bibinfo  {journal} {Phys. Rev. Lett.}\ }\textbf {\bibinfo {volume} {114}},\
  \bibinfo {pages} {197001} (\bibinfo {year} {2015}{\natexlab{b}})}\BibitemShut
  {NoStop}%
\bibitem [{\citenamefont {Lee}(2014)}]{PhysRevX.4.031017}%
  \BibitemOpen
  \bibfield  {author} {\bibinfo {author} {\bibfnamefont {P.~A.}\ \bibnamefont
  {Lee}},\ }\bibfield  {title} {\bibinfo {title} {Amperean pairing and the
  pseudogap phase of cuprate superconductors},\ }\href
  {https://doi.org/10.1103/PhysRevX.4.031017} {\bibfield  {journal} {\bibinfo
  {journal} {Phys. Rev. X}\ }\textbf {\bibinfo {volume} {4}},\ \bibinfo {pages}
  {031017} (\bibinfo {year} {2014})}\BibitemShut {NoStop}%
\bibitem [{\citenamefont {Fradkin}\ \emph {et~al.}(2015)\citenamefont
  {Fradkin}, \citenamefont {Kivelson},\ and\ \citenamefont
  {Tranquada}}]{RevModPhys.87.457}%
  \BibitemOpen
  \bibfield  {author} {\bibinfo {author} {\bibfnamefont {E.}~\bibnamefont
  {Fradkin}}, \bibinfo {author} {\bibfnamefont {S.~A.}\ \bibnamefont
  {Kivelson}},\ and\ \bibinfo {author} {\bibfnamefont {J.~M.}\ \bibnamefont
  {Tranquada}},\ }\bibfield  {title} {\bibinfo {title} {Colloquium: Theory of
  intertwined orders in high temperature superconductors},\ }\href
  {https://doi.org/10.1103/RevModPhys.87.457} {\bibfield  {journal} {\bibinfo
  {journal} {Rev. Mod. Phys.}\ }\textbf {\bibinfo {volume} {87}},\ \bibinfo
  {pages} {457} (\bibinfo {year} {2015})}\BibitemShut {NoStop}%
\bibitem [{\citenamefont {Huang}\ \emph {et~al.}(2021)\citenamefont {Huang},
  \citenamefont {Lee}, \citenamefont {Ikeda}, \citenamefont {Taniguchi},
  \citenamefont {Takahama}, \citenamefont {Kao}, \citenamefont {Fujita},\ and\
  \citenamefont {Lee}}]{PhysRevLett.126.167001}%
  \BibitemOpen
  \bibfield  {author} {\bibinfo {author} {\bibfnamefont {H.}~\bibnamefont
  {Huang}}, \bibinfo {author} {\bibfnamefont {S.-J.}\ \bibnamefont {Lee}},
  \bibinfo {author} {\bibfnamefont {Y.}~\bibnamefont {Ikeda}}, \bibinfo
  {author} {\bibfnamefont {T.}~\bibnamefont {Taniguchi}}, \bibinfo {author}
  {\bibfnamefont {M.}~\bibnamefont {Takahama}}, \bibinfo {author}
  {\bibfnamefont {C.-C.}\ \bibnamefont {Kao}}, \bibinfo {author} {\bibfnamefont
  {M.}~\bibnamefont {Fujita}},\ and\ \bibinfo {author} {\bibfnamefont {J.-S.}\
  \bibnamefont {Lee}},\ }\bibfield  {title} {\bibinfo {title} {Two-dimensional
  superconducting fluctuations associated with charge-density-wave stripes in
  ${\mathrm{la}}_{1.87}{\mathrm{sr}}_{0.13}{\mathrm{cu}}_{0.99}{\mathrm{fe}}_{0.01}{\mathrm{o}}_{4}$},\
  }\href {https://doi.org/10.1103/PhysRevLett.126.167001} {\bibfield  {journal}
  {\bibinfo  {journal} {Phys. Rev. Lett.}\ }\textbf {\bibinfo {volume} {126}},\
  \bibinfo {pages} {167001} (\bibinfo {year} {2021})}\BibitemShut {NoStop}%
\bibitem [{\citenamefont {Zhong}\ \emph {et~al.}(2018)\citenamefont {Zhong},
  \citenamefont {Schneeloch}, \citenamefont {Chi}, \citenamefont {Li},
  \citenamefont {Gu},\ and\ \citenamefont {Tranquada}}]{PhysRevB.97.134520}%
  \BibitemOpen
  \bibfield  {author} {\bibinfo {author} {\bibfnamefont {R.}~\bibnamefont
  {Zhong}}, \bibinfo {author} {\bibfnamefont {J.~A.}\ \bibnamefont
  {Schneeloch}}, \bibinfo {author} {\bibfnamefont {H.}~\bibnamefont {Chi}},
  \bibinfo {author} {\bibfnamefont {Q.}~\bibnamefont {Li}}, \bibinfo {author}
  {\bibfnamefont {G.}~\bibnamefont {Gu}},\ and\ \bibinfo {author}
  {\bibfnamefont {J.~M.}\ \bibnamefont {Tranquada}},\ }\bibfield  {title}
  {\bibinfo {title} {Evidence for magnetic-field-induced decoupling of
  superconducting bilayers in
  ${\mathrm{la}}_{2\ensuremath{-}x}{\mathrm{ca}}_{1+x}{\mathrm{cu}}_{2}{\mathrm{o}}_{6}$},\
  }\href {https://doi.org/10.1103/PhysRevB.97.134520} {\bibfield  {journal}
  {\bibinfo  {journal} {Phys. Rev. B}\ }\textbf {\bibinfo {volume} {97}},\
  \bibinfo {pages} {134520} (\bibinfo {year} {2018})}\BibitemShut {NoStop}%
\bibitem [{\citenamefont {Shi}\ \emph {et~al.}(2020)\citenamefont {Shi},
  \citenamefont {Baity}, \citenamefont {Terzic}, \citenamefont {Sasagawa},\
  and\ \citenamefont {Popovi{\'{c}}}}]{Shi2020}%
  \BibitemOpen
  \bibfield  {author} {\bibinfo {author} {\bibfnamefont {Z.}~\bibnamefont
  {Shi}}, \bibinfo {author} {\bibfnamefont {P.~G.}\ \bibnamefont {Baity}},
  \bibinfo {author} {\bibfnamefont {J.}~\bibnamefont {Terzic}}, \bibinfo
  {author} {\bibfnamefont {T.}~\bibnamefont {Sasagawa}},\ and\ \bibinfo
  {author} {\bibfnamefont {D.}~\bibnamefont {Popovi{\'{c}}}},\ }\bibfield
  {title} {\bibinfo {title} {Pair density wave at high magnetic fields in
  cuprates with charge and spin orders},\ }\href
  {https://doi.org/10.1038/s41467-020-17138-z} {\bibfield  {journal} {\bibinfo
  {journal} {Nature Communications}\ }\textbf {\bibinfo {volume} {11}},\
  \bibinfo {pages} {3323} (\bibinfo {year} {2020})}\BibitemShut {NoStop}%
\bibitem [{\citenamefont {Ding}\ \emph {et~al.}(2008)\citenamefont {Ding},
  \citenamefont {Xiang}, \citenamefont {Zhang}, \citenamefont {Liu},\ and\
  \citenamefont {Li}}]{PhysRevB.77.214524}%
  \BibitemOpen
  \bibfield  {author} {\bibinfo {author} {\bibfnamefont {J.~F.}\ \bibnamefont
  {Ding}}, \bibinfo {author} {\bibfnamefont {X.~Q.}\ \bibnamefont {Xiang}},
  \bibinfo {author} {\bibfnamefont {Y.~Q.}\ \bibnamefont {Zhang}}, \bibinfo
  {author} {\bibfnamefont {H.}~\bibnamefont {Liu}},\ and\ \bibinfo {author}
  {\bibfnamefont {X.~G.}\ \bibnamefont {Li}},\ }\bibfield  {title} {\bibinfo
  {title} {Two-dimensional superconductivity in stripe-ordered
  ${\text{la}}_{1.6\ensuremath{-}x}{\text{nd}}_{0.4}{\text{sr}}_{x}{\text{cuo}}_{4}$
  single crystals},\ }\href {https://doi.org/10.1103/PhysRevB.77.214524}
  {\bibfield  {journal} {\bibinfo  {journal} {Phys. Rev. B}\ }\textbf {\bibinfo
  {volume} {77}},\ \bibinfo {pages} {214524} (\bibinfo {year}
  {2008})}\BibitemShut {NoStop}%
\bibitem [{\citenamefont {Hayden}\ and\ \citenamefont
  {Tranquada}(2024)}]{annurevTranquada}%
  \BibitemOpen
  \bibfield  {author} {\bibinfo {author} {\bibfnamefont {S.~M.}\ \bibnamefont
  {Hayden}}\ and\ \bibinfo {author} {\bibfnamefont {J.~M.}\ \bibnamefont
  {Tranquada}},\ }\bibfield  {title} {\bibinfo {title} {Charge correlations in
  cuprate superconductors},\ }\href
  {https://doi.org/https://doi.org/10.1146/annurev-conmatphys-032922-094430}
  {\bibfield  {journal} {\bibinfo  {journal} {Annual Review of Condensed Matter
  Physics}\ }\textbf {\bibinfo {volume} {15}},\ \bibinfo {pages} {215}
  (\bibinfo {year} {2024})}\BibitemShut {NoStop}%
\bibitem [{\citenamefont {Edkins}\ \emph {et~al.}(2019)\citenamefont {Edkins},
  \citenamefont {Kostin}, \citenamefont {Fujita}, \citenamefont {Mackenzie},
  \citenamefont {Eisaki}, \citenamefont {Uchida}, \citenamefont {Sachdev},
  \citenamefont {Lawler}, \citenamefont {Kim}, \citenamefont {Davis},\ and\
  \citenamefont {Hamidian}}]{doi:10.1126/science.aat1773}%
  \BibitemOpen
  \bibfield  {author} {\bibinfo {author} {\bibfnamefont {S.~D.}\ \bibnamefont
  {Edkins}}, \bibinfo {author} {\bibfnamefont {A.}~\bibnamefont {Kostin}},
  \bibinfo {author} {\bibfnamefont {K.}~\bibnamefont {Fujita}}, \bibinfo
  {author} {\bibfnamefont {A.~P.}\ \bibnamefont {Mackenzie}}, \bibinfo {author}
  {\bibfnamefont {H.}~\bibnamefont {Eisaki}}, \bibinfo {author} {\bibfnamefont
  {S.}~\bibnamefont {Uchida}}, \bibinfo {author} {\bibfnamefont
  {S.}~\bibnamefont {Sachdev}}, \bibinfo {author} {\bibfnamefont {M.~J.}\
  \bibnamefont {Lawler}}, \bibinfo {author} {\bibfnamefont {E.-A.}\
  \bibnamefont {Kim}}, \bibinfo {author} {\bibfnamefont {J.~C.~S.}\
  \bibnamefont {Davis}},\ and\ \bibinfo {author} {\bibfnamefont {M.~H.}\
  \bibnamefont {Hamidian}},\ }\bibfield  {title} {\bibinfo {title} {Magnetic
  field–induced pair density wave state in the cuprate vortex halo},\ }\href
  {https://doi.org/10.1126/science.aat1773} {\bibfield  {journal} {\bibinfo
  {journal} {Science}\ }\textbf {\bibinfo {volume} {364}},\ \bibinfo {pages}
  {976} (\bibinfo {year} {2019})}\BibitemShut {NoStop}%
\bibitem [{\citenamefont {Du}\ \emph {et~al.}(2020)\citenamefont {Du},
  \citenamefont {Li}, \citenamefont {Joo}, \citenamefont {Donoway},
  \citenamefont {Lee}, \citenamefont {Davis}, \citenamefont {Gu}, \citenamefont
  {Johnson},\ and\ \citenamefont {Fujita}}]{Du2020}%
  \BibitemOpen
  \bibfield  {author} {\bibinfo {author} {\bibfnamefont {Z.}~\bibnamefont
  {Du}}, \bibinfo {author} {\bibfnamefont {H.}~\bibnamefont {Li}}, \bibinfo
  {author} {\bibfnamefont {S.~H.}\ \bibnamefont {Joo}}, \bibinfo {author}
  {\bibfnamefont {E.~P.}\ \bibnamefont {Donoway}}, \bibinfo {author}
  {\bibfnamefont {J.}~\bibnamefont {Lee}}, \bibinfo {author} {\bibfnamefont
  {J.~C.~S.}\ \bibnamefont {Davis}}, \bibinfo {author} {\bibfnamefont
  {G.}~\bibnamefont {Gu}}, \bibinfo {author} {\bibfnamefont {P.~D.}\
  \bibnamefont {Johnson}},\ and\ \bibinfo {author} {\bibfnamefont
  {K.}~\bibnamefont {Fujita}},\ }\bibfield  {title} {\bibinfo {title} {Imaging
  the energy gap modulations of the cuprate pair-density-wave state},\ }\href
  {https://doi.org/10.1038/s41586-020-2143-x} {\bibfield  {journal} {\bibinfo
  {journal} {Nature}\ }\textbf {\bibinfo {volume} {580}},\ \bibinfo {pages}
  {65} (\bibinfo {year} {2020})}\BibitemShut {NoStop}%
\bibitem [{\citenamefont {Hamidian}\ \emph {et~al.}(2016)\citenamefont
  {Hamidian}, \citenamefont {Edkins}, \citenamefont {Joo}, \citenamefont
  {Kostin}, \citenamefont {Eisaki}, \citenamefont {Uchida}, \citenamefont
  {Lawler}, \citenamefont {Kim}, \citenamefont {Mackenzie}, \citenamefont
  {Fujita}, \citenamefont {Lee},\ and\ \citenamefont {Davis}}]{Hamidian2016}%
  \BibitemOpen
  \bibfield  {author} {\bibinfo {author} {\bibfnamefont {M.~H.}\ \bibnamefont
  {Hamidian}}, \bibinfo {author} {\bibfnamefont {S.~D.}\ \bibnamefont
  {Edkins}}, \bibinfo {author} {\bibfnamefont {S.~H.}\ \bibnamefont {Joo}},
  \bibinfo {author} {\bibfnamefont {A.}~\bibnamefont {Kostin}}, \bibinfo
  {author} {\bibfnamefont {H.}~\bibnamefont {Eisaki}}, \bibinfo {author}
  {\bibfnamefont {S.}~\bibnamefont {Uchida}}, \bibinfo {author} {\bibfnamefont
  {M.~J.}\ \bibnamefont {Lawler}}, \bibinfo {author} {\bibfnamefont {E.-A.}\
  \bibnamefont {Kim}}, \bibinfo {author} {\bibfnamefont {A.~P.}\ \bibnamefont
  {Mackenzie}}, \bibinfo {author} {\bibfnamefont {K.}~\bibnamefont {Fujita}},
  \bibinfo {author} {\bibfnamefont {J.}~\bibnamefont {Lee}},\ and\ \bibinfo
  {author} {\bibfnamefont {J.~C.~S.}\ \bibnamefont {Davis}},\ }\bibfield
  {title} {\bibinfo {title} {Detection of a cooper-pair density wave in
  bi2sr2cacu2o8+x},\ }\href {https://doi.org/10.1038/nature17411} {\bibfield
  {journal} {\bibinfo  {journal} {Nature}\ }\textbf {\bibinfo {volume} {532}},\
  \bibinfo {pages} {343} (\bibinfo {year} {2016})}\BibitemShut {NoStop}%
\bibitem [{\citenamefont {Gao}\ \emph {et~al.}(2023)\citenamefont {Gao},
  \citenamefont {Lin},\ and\ \citenamefont
  {Lee}}]{gao2023pairbreakingscatteringinterferencemechanism}%
  \BibitemOpen
  \bibfield  {author} {\bibinfo {author} {\bibfnamefont {Z.-Q.}\ \bibnamefont
  {Gao}}, \bibinfo {author} {\bibfnamefont {Y.-P.}\ \bibnamefont {Lin}},\ and\
  \bibinfo {author} {\bibfnamefont {D.-H.}\ \bibnamefont {Lee}},\ }\href
  {https://arxiv.org/abs/2310.06024} {\bibinfo {title} {Pair-breaking
  scattering interference as a mechanism for superconducting gap modulation}}
  (\bibinfo {year} {2023}),\ \Eprint {https://arxiv.org/abs/2310.06024}
  {arXiv:2310.06024 [cond-mat.supr-con]} \BibitemShut {NoStop}%
\bibitem [{\citenamefont {Lee}\ \emph {et~al.}(2023)\citenamefont {Lee},
  \citenamefont {Kivelson}, \citenamefont {Wang}, \citenamefont {Ikeda},
  \citenamefont {Taniguchi}, \citenamefont {Fujita},\ and\ \citenamefont
  {Kao}}]{lee2023pairdensitywavesignatureobserved}%
  \BibitemOpen
  \bibfield  {author} {\bibinfo {author} {\bibfnamefont {J.-S.}\ \bibnamefont
  {Lee}}, \bibinfo {author} {\bibfnamefont {S.~A.}\ \bibnamefont {Kivelson}},
  \bibinfo {author} {\bibfnamefont {T.}~\bibnamefont {Wang}}, \bibinfo {author}
  {\bibfnamefont {Y.}~\bibnamefont {Ikeda}}, \bibinfo {author} {\bibfnamefont
  {T.}~\bibnamefont {Taniguchi}}, \bibinfo {author} {\bibfnamefont
  {M.}~\bibnamefont {Fujita}},\ and\ \bibinfo {author} {\bibfnamefont {C.-C.}\
  \bibnamefont {Kao}},\ }\href {https://arxiv.org/abs/2310.19907} {\bibinfo
  {title} {Pair-density wave signature observed by x-ray scattering in la-based
  high-$t_{\rm c}$ cuprates}} (\bibinfo {year} {2023}),\ \Eprint
  {https://arxiv.org/abs/2310.19907} {arXiv:2310.19907 [cond-mat.supr-con]}
  \BibitemShut {NoStop}%
\bibitem [{\citenamefont {Soto-Garrido}\ \emph
  {et~al.}(2017{\natexlab{a}})\citenamefont {Soto-Garrido}, \citenamefont
  {Wang}, \citenamefont {Fradkin},\ and\ \citenamefont
  {Cooper}}]{soto-garrido-wang-fradkin-cooper}%
  \BibitemOpen
  \bibfield  {author} {\bibinfo {author} {\bibfnamefont {R.}~\bibnamefont
  {Soto-Garrido}}, \bibinfo {author} {\bibfnamefont {Y.}~\bibnamefont {Wang}},
  \bibinfo {author} {\bibfnamefont {E.}~\bibnamefont {Fradkin}},\ and\ \bibinfo
  {author} {\bibfnamefont {S.~L.}\ \bibnamefont {Cooper}},\ }\bibfield  {title}
  {\bibinfo {title} {Higgs modes in the pair density wave superconducting
  state},\ }\href {https://doi.org/10.1103/PhysRevB.95.214502} {\bibfield
  {journal} {\bibinfo  {journal} {Phys. Rev. B}\ }\textbf {\bibinfo {volume}
  {95}},\ \bibinfo {pages} {214502} (\bibinfo {year}
  {2017}{\natexlab{a}})}\BibitemShut {NoStop}%
\bibitem [{\citenamefont {Jian}\ \emph {et~al.}(2020)\citenamefont {Jian},
  \citenamefont {Scherer},\ and\ \citenamefont
  {Yao}}]{PhysRevResearch.2.013034}%
  \BibitemOpen
  \bibfield  {author} {\bibinfo {author} {\bibfnamefont {S.-K.}\ \bibnamefont
  {Jian}}, \bibinfo {author} {\bibfnamefont {M.~M.}\ \bibnamefont {Scherer}},\
  and\ \bibinfo {author} {\bibfnamefont {H.}~\bibnamefont {Yao}},\ }\bibfield
  {title} {\bibinfo {title} {Mass hierarchy in collective modes of
  pair-density-wave superconductors},\ }\href
  {https://doi.org/10.1103/PhysRevResearch.2.013034} {\bibfield  {journal}
  {\bibinfo  {journal} {Phys. Rev. Res.}\ }\textbf {\bibinfo {volume} {2}},\
  \bibinfo {pages} {013034} (\bibinfo {year} {2020})}\BibitemShut {NoStop}%
\bibitem [{\citenamefont {Nagashima}\ \emph {et~al.}(2024)\citenamefont
  {Nagashima}, \citenamefont {Mouilleron},\ and\ \citenamefont
  {Tsuji}}]{nagashima2024opticallyactivehiggsleggett}%
  \BibitemOpen
  \bibfield  {author} {\bibinfo {author} {\bibfnamefont {R.}~\bibnamefont
  {Nagashima}}, \bibinfo {author} {\bibfnamefont {T.}~\bibnamefont
  {Mouilleron}},\ and\ \bibinfo {author} {\bibfnamefont {N.}~\bibnamefont
  {Tsuji}},\ }\href {https://arxiv.org/abs/2410.18438} {\bibinfo {title}
  {Optically active higgs and leggett modes in multiband pair-density-wave
  superconductors with lifshitz invariant}} (\bibinfo {year} {2024}),\ \Eprint
  {https://arxiv.org/abs/2410.18438} {arXiv:2410.18438 [cond-mat.supr-con]}
  \BibitemShut {NoStop}%
\bibitem [{\citenamefont {Miao}\ \emph {et~al.}(2021)\citenamefont {Miao},
  \citenamefont {Fabbris}, \citenamefont {Koch}, \citenamefont {Mazzone},
  \citenamefont {Nelson}, \citenamefont {Acevedo-Esteves}, \citenamefont {Gu},
  \citenamefont {Li}, \citenamefont {Yilimaz}, \citenamefont {Kaznatcheev},
  \citenamefont {Vescovo}, \citenamefont {Oda}, \citenamefont {Kurosawa},
  \citenamefont {Momono}, \citenamefont {Assefa}, \citenamefont {Robinson},
  \citenamefont {Bozin}, \citenamefont {Tranquada}, \citenamefont {Johnson},\
  and\ \citenamefont {Dean}}]{Miao2021}%
  \BibitemOpen
  \bibfield  {author} {\bibinfo {author} {\bibfnamefont {H.}~\bibnamefont
  {Miao}}, \bibinfo {author} {\bibfnamefont {G.}~\bibnamefont {Fabbris}},
  \bibinfo {author} {\bibfnamefont {R.~J.}\ \bibnamefont {Koch}}, \bibinfo
  {author} {\bibfnamefont {D.~G.}\ \bibnamefont {Mazzone}}, \bibinfo {author}
  {\bibfnamefont {C.~S.}\ \bibnamefont {Nelson}}, \bibinfo {author}
  {\bibfnamefont {R.}~\bibnamefont {Acevedo-Esteves}}, \bibinfo {author}
  {\bibfnamefont {G.~D.}\ \bibnamefont {Gu}}, \bibinfo {author} {\bibfnamefont
  {Y.}~\bibnamefont {Li}}, \bibinfo {author} {\bibfnamefont {T.}~\bibnamefont
  {Yilimaz}}, \bibinfo {author} {\bibfnamefont {K.}~\bibnamefont
  {Kaznatcheev}}, \bibinfo {author} {\bibfnamefont {E.}~\bibnamefont
  {Vescovo}}, \bibinfo {author} {\bibfnamefont {M.}~\bibnamefont {Oda}},
  \bibinfo {author} {\bibfnamefont {T.}~\bibnamefont {Kurosawa}}, \bibinfo
  {author} {\bibfnamefont {N.}~\bibnamefont {Momono}}, \bibinfo {author}
  {\bibfnamefont {T.}~\bibnamefont {Assefa}}, \bibinfo {author} {\bibfnamefont
  {I.~K.}\ \bibnamefont {Robinson}}, \bibinfo {author} {\bibfnamefont {E.~S.}\
  \bibnamefont {Bozin}}, \bibinfo {author} {\bibfnamefont {J.~M.}\ \bibnamefont
  {Tranquada}}, \bibinfo {author} {\bibfnamefont {P.~D.}\ \bibnamefont
  {Johnson}},\ and\ \bibinfo {author} {\bibfnamefont {M.~P.~M.}\ \bibnamefont
  {Dean}},\ }\bibfield  {title} {\bibinfo {title} {Charge density waves in
  cuprate superconductors beyond the critical doping},\ }\href
  {https://doi.org/10.1038/s41535-021-00327-4} {\bibfield  {journal} {\bibinfo
  {journal} {npj Quantum Materials}\ }\textbf {\bibinfo {volume} {6}},\
  \bibinfo {pages} {31} (\bibinfo {year} {2021})}\BibitemShut {NoStop}%
\bibitem [{\citenamefont {Wen}\ \emph {et~al.}(2019)\citenamefont {Wen},
  \citenamefont {Huang}, \citenamefont {Lee}, \citenamefont {Jang},
  \citenamefont {Knight}, \citenamefont {Lee}, \citenamefont {Fujita},
  \citenamefont {Suzuki}, \citenamefont {Asano}, \citenamefont {Kivelson},
  \citenamefont {Kao},\ and\ \citenamefont {Lee}}]{Wen2019}%
  \BibitemOpen
  \bibfield  {author} {\bibinfo {author} {\bibfnamefont {J.-J.}\ \bibnamefont
  {Wen}}, \bibinfo {author} {\bibfnamefont {H.}~\bibnamefont {Huang}}, \bibinfo
  {author} {\bibfnamefont {S.-J.}\ \bibnamefont {Lee}}, \bibinfo {author}
  {\bibfnamefont {H.}~\bibnamefont {Jang}}, \bibinfo {author} {\bibfnamefont
  {J.}~\bibnamefont {Knight}}, \bibinfo {author} {\bibfnamefont {Y.~S.}\
  \bibnamefont {Lee}}, \bibinfo {author} {\bibfnamefont {M.}~\bibnamefont
  {Fujita}}, \bibinfo {author} {\bibfnamefont {K.~M.}\ \bibnamefont {Suzuki}},
  \bibinfo {author} {\bibfnamefont {S.}~\bibnamefont {Asano}}, \bibinfo
  {author} {\bibfnamefont {S.~A.}\ \bibnamefont {Kivelson}}, \bibinfo {author}
  {\bibfnamefont {C.-C.}\ \bibnamefont {Kao}},\ and\ \bibinfo {author}
  {\bibfnamefont {J.-S.}\ \bibnamefont {Lee}},\ }\bibfield  {title} {\bibinfo
  {title} {Observation of two types of charge-density-wave orders in
  superconducting la2-xsrxcuo4},\ }\href
  {https://doi.org/10.1038/s41467-019-11167-z} {\bibfield  {journal} {\bibinfo
  {journal} {Nature Communications}\ }\textbf {\bibinfo {volume} {10}},\
  \bibinfo {pages} {3269} (\bibinfo {year} {2019})}\BibitemShut {NoStop}%
\bibitem [{\citenamefont {Miao}\ \emph {et~al.}(2019)\citenamefont {Miao},
  \citenamefont {Fumagalli}, \citenamefont {Rossi}, \citenamefont {Lorenzana},
  \citenamefont {Seibold}, \citenamefont {Yakhou-Harris}, \citenamefont
  {Kummer}, \citenamefont {Brookes}, \citenamefont {Gu}, \citenamefont
  {Braicovich}, \citenamefont {Ghiringhelli},\ and\ \citenamefont
  {Dean}}]{PhysRevX.9.031042}%
  \BibitemOpen
  \bibfield  {author} {\bibinfo {author} {\bibfnamefont {H.}~\bibnamefont
  {Miao}}, \bibinfo {author} {\bibfnamefont {R.}~\bibnamefont {Fumagalli}},
  \bibinfo {author} {\bibfnamefont {M.}~\bibnamefont {Rossi}}, \bibinfo
  {author} {\bibfnamefont {J.}~\bibnamefont {Lorenzana}}, \bibinfo {author}
  {\bibfnamefont {G.}~\bibnamefont {Seibold}}, \bibinfo {author} {\bibfnamefont
  {F.}~\bibnamefont {Yakhou-Harris}}, \bibinfo {author} {\bibfnamefont
  {K.}~\bibnamefont {Kummer}}, \bibinfo {author} {\bibfnamefont {N.~B.}\
  \bibnamefont {Brookes}}, \bibinfo {author} {\bibfnamefont {G.~D.}\
  \bibnamefont {Gu}}, \bibinfo {author} {\bibfnamefont {L.}~\bibnamefont
  {Braicovich}}, \bibinfo {author} {\bibfnamefont {G.}~\bibnamefont
  {Ghiringhelli}},\ and\ \bibinfo {author} {\bibfnamefont {M.~P.~M.}\
  \bibnamefont {Dean}},\ }\bibfield  {title} {\bibinfo {title} {Formation of
  incommensurate charge density waves in cuprates},\ }\href
  {https://doi.org/10.1103/PhysRevX.9.031042} {\bibfield  {journal} {\bibinfo
  {journal} {Phys. Rev. X}\ }\textbf {\bibinfo {volume} {9}},\ \bibinfo {pages}
  {031042} (\bibinfo {year} {2019})}\BibitemShut {NoStop}%
\bibitem [{\citenamefont {Lin}\ \emph {et~al.}(2020)\citenamefont {Lin},
  \citenamefont {Miao}, \citenamefont {Mazzone}, \citenamefont {Gu},
  \citenamefont {Nag}, \citenamefont {Walters}, \citenamefont
  {Garc\'{\i}a-Fern\'andez}, \citenamefont {Barbour}, \citenamefont
  {Pelliciari}, \citenamefont {Jarrige}, \citenamefont {Oda}, \citenamefont
  {Kurosawa}, \citenamefont {Momono}, \citenamefont {Zhou}, \citenamefont
  {Bisogni}, \citenamefont {Liu},\ and\ \citenamefont
  {Dean}}]{PhysRevLett.124.207005}%
  \BibitemOpen
  \bibfield  {author} {\bibinfo {author} {\bibfnamefont {J.~Q.}\ \bibnamefont
  {Lin}}, \bibinfo {author} {\bibfnamefont {H.}~\bibnamefont {Miao}}, \bibinfo
  {author} {\bibfnamefont {D.~G.}\ \bibnamefont {Mazzone}}, \bibinfo {author}
  {\bibfnamefont {G.~D.}\ \bibnamefont {Gu}}, \bibinfo {author} {\bibfnamefont
  {A.}~\bibnamefont {Nag}}, \bibinfo {author} {\bibfnamefont {A.~C.}\
  \bibnamefont {Walters}}, \bibinfo {author} {\bibfnamefont {M.}~\bibnamefont
  {Garc\'{\i}a-Fern\'andez}}, \bibinfo {author} {\bibfnamefont
  {A.}~\bibnamefont {Barbour}}, \bibinfo {author} {\bibfnamefont
  {J.}~\bibnamefont {Pelliciari}}, \bibinfo {author} {\bibfnamefont
  {I.}~\bibnamefont {Jarrige}}, \bibinfo {author} {\bibfnamefont
  {M.}~\bibnamefont {Oda}}, \bibinfo {author} {\bibfnamefont {K.}~\bibnamefont
  {Kurosawa}}, \bibinfo {author} {\bibfnamefont {N.}~\bibnamefont {Momono}},
  \bibinfo {author} {\bibfnamefont {K.-J.}\ \bibnamefont {Zhou}}, \bibinfo
  {author} {\bibfnamefont {V.}~\bibnamefont {Bisogni}}, \bibinfo {author}
  {\bibfnamefont {X.}~\bibnamefont {Liu}},\ and\ \bibinfo {author}
  {\bibfnamefont {M.~P.~M.}\ \bibnamefont {Dean}},\ }\bibfield  {title}
  {\bibinfo {title} {Strongly correlated charge density wave in
  ${\mathrm{la}}_{2\ensuremath{-}x}{\mathrm{sr}}_{x}{\mathrm{cuo}}_{4}$
  evidenced by doping-dependent phonon anomaly},\ }\href
  {https://doi.org/10.1103/PhysRevLett.124.207005} {\bibfield  {journal}
  {\bibinfo  {journal} {Phys. Rev. Lett.}\ }\textbf {\bibinfo {volume} {124}},\
  \bibinfo {pages} {207005} (\bibinfo {year} {2020})}\BibitemShut {NoStop}%
\bibitem [{\citenamefont {Miao}\ \emph {et~al.}(2017)\citenamefont {Miao},
  \citenamefont {Lorenzana}, \citenamefont {Seibold}, \citenamefont {Peng},
  \citenamefont {Amorese}, \citenamefont {Yakhou-Harris}, \citenamefont
  {Kummer}, \citenamefont {Brookes}, \citenamefont {Konik}, \citenamefont
  {Thampy}, \citenamefont {Gu}, \citenamefont {Ghiringhelli}, \citenamefont
  {Braicovich},\ and\ \citenamefont {Dean}}]{doi:10.1073/pnas.1708549114}%
  \BibitemOpen
  \bibfield  {author} {\bibinfo {author} {\bibfnamefont {H.}~\bibnamefont
  {Miao}}, \bibinfo {author} {\bibfnamefont {J.}~\bibnamefont {Lorenzana}},
  \bibinfo {author} {\bibfnamefont {G.}~\bibnamefont {Seibold}}, \bibinfo
  {author} {\bibfnamefont {Y.~Y.}\ \bibnamefont {Peng}}, \bibinfo {author}
  {\bibfnamefont {A.}~\bibnamefont {Amorese}}, \bibinfo {author} {\bibfnamefont
  {F.}~\bibnamefont {Yakhou-Harris}}, \bibinfo {author} {\bibfnamefont
  {K.}~\bibnamefont {Kummer}}, \bibinfo {author} {\bibfnamefont {N.~B.}\
  \bibnamefont {Brookes}}, \bibinfo {author} {\bibfnamefont {R.~M.}\
  \bibnamefont {Konik}}, \bibinfo {author} {\bibfnamefont {V.}~\bibnamefont
  {Thampy}}, \bibinfo {author} {\bibfnamefont {G.~D.}\ \bibnamefont {Gu}},
  \bibinfo {author} {\bibfnamefont {G.}~\bibnamefont {Ghiringhelli}}, \bibinfo
  {author} {\bibfnamefont {L.}~\bibnamefont {Braicovich}},\ and\ \bibinfo
  {author} {\bibfnamefont {M.~P.~M.}\ \bibnamefont {Dean}},\ }\bibfield
  {title} {\bibinfo {title} {High-temperature charge density wave correlations
  in la<sub>1.875</sub>ba<sub>0.125</sub>cuo<sub>4</sub> without
  spin\&\#x2013;charge locking},\ }\href
  {https://doi.org/10.1073/pnas.1708549114} {\bibfield  {journal} {\bibinfo
  {journal} {Proceedings of the National Academy of Sciences}\ }\textbf
  {\bibinfo {volume} {114}},\ \bibinfo {pages} {12430} (\bibinfo {year}
  {2017})}\BibitemShut {NoStop}%
\bibitem [{\citenamefont {Peng}\ \emph {et~al.}(2020)\citenamefont {Peng},
  \citenamefont {Husain}, \citenamefont {Mitrano}, \citenamefont {Sun},
  \citenamefont {Johnson}, \citenamefont {Zakrzewski}, \citenamefont
  {MacDougall}, \citenamefont {Barbour}, \citenamefont {Jarrige}, \citenamefont
  {Bisogni},\ and\ \citenamefont {Abbamonte}}]{PhysRevLett.125.097002}%
  \BibitemOpen
  \bibfield  {author} {\bibinfo {author} {\bibfnamefont {Y.~Y.}\ \bibnamefont
  {Peng}}, \bibinfo {author} {\bibfnamefont {A.~A.}\ \bibnamefont {Husain}},
  \bibinfo {author} {\bibfnamefont {M.}~\bibnamefont {Mitrano}}, \bibinfo
  {author} {\bibfnamefont {S.~X.-L.}\ \bibnamefont {Sun}}, \bibinfo {author}
  {\bibfnamefont {T.~A.}\ \bibnamefont {Johnson}}, \bibinfo {author}
  {\bibfnamefont {A.~V.}\ \bibnamefont {Zakrzewski}}, \bibinfo {author}
  {\bibfnamefont {G.~J.}\ \bibnamefont {MacDougall}}, \bibinfo {author}
  {\bibfnamefont {A.}~\bibnamefont {Barbour}}, \bibinfo {author} {\bibfnamefont
  {I.}~\bibnamefont {Jarrige}}, \bibinfo {author} {\bibfnamefont
  {V.}~\bibnamefont {Bisogni}},\ and\ \bibinfo {author} {\bibfnamefont
  {P.}~\bibnamefont {Abbamonte}},\ }\bibfield  {title} {\bibinfo {title}
  {Enhanced electron-phonon coupling for charge-density-wave formation in
  ${\mathrm{la}}_{1.8\ensuremath{-}x}{\mathrm{eu}}_{0.2}{\mathrm{sr}}_{x}{\mathrm{cuo}}_{4+\ensuremath{\delta}}$},\
  }\href {https://doi.org/10.1103/PhysRevLett.125.097002} {\bibfield  {journal}
  {\bibinfo  {journal} {Phys. Rev. Lett.}\ }\textbf {\bibinfo {volume} {125}},\
  \bibinfo {pages} {097002} (\bibinfo {year} {2020})}\BibitemShut {NoStop}%
\bibitem [{\citenamefont {Benfatto}\ \emph {et~al.}(2004)\citenamefont
  {Benfatto}, \citenamefont {Toschi},\ and\ \citenamefont
  {Caprara}}]{PhysRevB.69.184510}%
  \BibitemOpen
  \bibfield  {author} {\bibinfo {author} {\bibfnamefont {L.}~\bibnamefont
  {Benfatto}}, \bibinfo {author} {\bibfnamefont {A.}~\bibnamefont {Toschi}},\
  and\ \bibinfo {author} {\bibfnamefont {S.}~\bibnamefont {Caprara}},\
  }\bibfield  {title} {\bibinfo {title} {Low-energy phase-only action in a
  superconductor: A comparison with the $\mathrm{XY}$ model},\ }\href
  {https://doi.org/10.1103/PhysRevB.69.184510} {\bibfield  {journal} {\bibinfo
  {journal} {Phys. Rev. B}\ }\textbf {\bibinfo {volume} {69}},\ \bibinfo
  {pages} {184510} (\bibinfo {year} {2004})}\BibitemShut {NoStop}%
\bibitem [{\citenamefont {Paramekanti}\ \emph {et~al.}(2000)\citenamefont
  {Paramekanti}, \citenamefont {Randeria}, \citenamefont {Ramakrishnan},\ and\
  \citenamefont {Mandal}}]{PhysRevB.62.6786}%
  \BibitemOpen
  \bibfield  {author} {\bibinfo {author} {\bibfnamefont {A.}~\bibnamefont
  {Paramekanti}}, \bibinfo {author} {\bibfnamefont {M.}~\bibnamefont
  {Randeria}}, \bibinfo {author} {\bibfnamefont {T.~V.}\ \bibnamefont
  {Ramakrishnan}},\ and\ \bibinfo {author} {\bibfnamefont {S.~S.}\ \bibnamefont
  {Mandal}},\ }\bibfield  {title} {\bibinfo {title} {Effective actions and
  phase fluctuations in d-wave superconductors},\ }\href
  {https://doi.org/10.1103/PhysRevB.62.6786} {\bibfield  {journal} {\bibinfo
  {journal} {Phys. Rev. B}\ }\textbf {\bibinfo {volume} {62}},\ \bibinfo
  {pages} {6786} (\bibinfo {year} {2000})}\BibitemShut {NoStop}%
\bibitem [{\citenamefont {Benfatto}\ \emph {et~al.}(2001)\citenamefont
  {Benfatto}, \citenamefont {Caprara}, \citenamefont {Castellani},
  \citenamefont {Paramekanti},\ and\ \citenamefont
  {Randeria}}]{PhysRevB.63.174513}%
  \BibitemOpen
  \bibfield  {author} {\bibinfo {author} {\bibfnamefont {L.}~\bibnamefont
  {Benfatto}}, \bibinfo {author} {\bibfnamefont {S.}~\bibnamefont {Caprara}},
  \bibinfo {author} {\bibfnamefont {C.}~\bibnamefont {Castellani}}, \bibinfo
  {author} {\bibfnamefont {A.}~\bibnamefont {Paramekanti}},\ and\ \bibinfo
  {author} {\bibfnamefont {M.}~\bibnamefont {Randeria}},\ }\bibfield  {title}
  {\bibinfo {title} {Phase fluctuations, dissipation, and superfluid stiffness
  in d-wave superconductors},\ }\href
  {https://doi.org/10.1103/PhysRevB.63.174513} {\bibfield  {journal} {\bibinfo
  {journal} {Phys. Rev. B}\ }\textbf {\bibinfo {volume} {63}},\ \bibinfo
  {pages} {174513} (\bibinfo {year} {2001})}\BibitemShut {NoStop}%
\bibitem [{SM()}]{SM}%
  \BibitemOpen
  \href@noop {} {\bibinfo {title} {See the supplementary material for detailed
  information about i) the bogoliubov quasiparticle band structure, ii)
  mean-field gap equation and the identification of trsb, iii) obtaining the
  fluctuating part $s_\text{FL}$, iv) discussion about energy hierarchy in
  amplitude modes of the pdw+sc case, v) derivation of the raman susceptibility
  and vi) role of disorder.}}\BibitemShut {Stop}%
\bibitem [{\citenamefont {Baruch}\ and\ \citenamefont
  {Orgad}(2008)}]{PhysRevB.77.174502}%
  \BibitemOpen
  \bibfield  {author} {\bibinfo {author} {\bibfnamefont {S.}~\bibnamefont
  {Baruch}}\ and\ \bibinfo {author} {\bibfnamefont {D.}~\bibnamefont {Orgad}},\
  }\bibfield  {title} {\bibinfo {title} {Spectral signatures of modulated
  $d$-wave superconducting phases},\ }\href
  {https://doi.org/10.1103/PhysRevB.77.174502} {\bibfield  {journal} {\bibinfo
  {journal} {Phys. Rev. B}\ }\textbf {\bibinfo {volume} {77}},\ \bibinfo
  {pages} {174502} (\bibinfo {year} {2008})}\BibitemShut {NoStop}%
\bibitem [{\citenamefont {Norman}\ and\ \citenamefont
  {Davis}(2018)}]{doi:10.1073/pnas.1803009115}%
  \BibitemOpen
  \bibfield  {author} {\bibinfo {author} {\bibfnamefont {M.~R.}\ \bibnamefont
  {Norman}}\ and\ \bibinfo {author} {\bibfnamefont {J.~C.~S.}\ \bibnamefont
  {Davis}},\ }\bibfield  {title} {\bibinfo {title} {Quantum oscillations in a
  biaxial pair density wave state},\ }\href
  {https://doi.org/10.1073/pnas.1803009115} {\bibfield  {journal} {\bibinfo
  {journal} {Proceedings of the National Academy of Sciences}\ }\textbf
  {\bibinfo {volume} {115}},\ \bibinfo {pages} {5389} (\bibinfo {year}
  {2018})}\BibitemShut {NoStop}%
\bibitem [{\citenamefont {Zelli}\ \emph {et~al.}(2012)\citenamefont {Zelli},
  \citenamefont {Kallin},\ and\ \citenamefont
  {Berlinsky}}]{PhysRevB.86.104507}%
  \BibitemOpen
  \bibfield  {author} {\bibinfo {author} {\bibfnamefont {M.}~\bibnamefont
  {Zelli}}, \bibinfo {author} {\bibfnamefont {C.}~\bibnamefont {Kallin}},\ and\
  \bibinfo {author} {\bibfnamefont {A.~J.}\ \bibnamefont {Berlinsky}},\
  }\bibfield  {title} {\bibinfo {title} {Quantum oscillations in a
  $\ensuremath{\pi}$-striped superconductor},\ }\href
  {https://doi.org/10.1103/PhysRevB.86.104507} {\bibfield  {journal} {\bibinfo
  {journal} {Phys. Rev. B}\ }\textbf {\bibinfo {volume} {86}},\ \bibinfo
  {pages} {104507} (\bibinfo {year} {2012})}\BibitemShut {NoStop}%
\bibitem [{\citenamefont {Caplan}\ and\ \citenamefont
  {Orgad}(2021)}]{PhysRevResearch.3.023199}%
  \BibitemOpen
  \bibfield  {author} {\bibinfo {author} {\bibfnamefont {Y.}~\bibnamefont
  {Caplan}}\ and\ \bibinfo {author} {\bibfnamefont {D.}~\bibnamefont {Orgad}},\
  }\bibfield  {title} {\bibinfo {title} {Quantum oscillations from a
  pair-density wave},\ }\href
  {https://doi.org/10.1103/PhysRevResearch.3.023199} {\bibfield  {journal}
  {\bibinfo  {journal} {Phys. Rev. Res.}\ }\textbf {\bibinfo {volume} {3}},\
  \bibinfo {pages} {023199} (\bibinfo {year} {2021})}\BibitemShut {NoStop}%
\bibitem [{Note1()}]{Note1}%
  \BibitemOpen
  \bibinfo {note} {TRSB in the mixed state does not depend on the phase
  difference $\varphi _{\protect \bm {Q}}-\varphi _{-\protect \bm {Q}}$, which
  is arbitrary to the order described by Eq. (\ref {eq:FGL}). This phase is
  fixed once we include the 8$^{th}$ order contribution to ${\protect \cal
  F}_{MF}$: $-\upsilon [(\protect \overline \Delta _{\protect \bm {Q}}\protect
  \overline \Delta ^*_{-\protect \bm {Q}})^4+c.c.]$, which is allowed for the
  commensurate, period 8 case we have treated. We have computed the prefactor
  $v$ and found it is positive. Thus, $\varphi _{\protect \bm {Q}}-\varphi
  _{-\protect \bm {Q}} = m \pi /2$, where $m$ is integer.}\BibitemShut {Stop}%
\bibitem [{\citenamefont {Karapetyan}\ \emph {et~al.}(2014)\citenamefont
  {Karapetyan}, \citenamefont {Xia}, \citenamefont {H\"ucker}, \citenamefont
  {Gu}, \citenamefont {Tranquada}, \citenamefont {Fejer},\ and\ \citenamefont
  {Kapitulnik}}]{PhysRevLett.112.047003}%
  \BibitemOpen
  \bibfield  {author} {\bibinfo {author} {\bibfnamefont {H.}~\bibnamefont
  {Karapetyan}}, \bibinfo {author} {\bibfnamefont {J.}~\bibnamefont {Xia}},
  \bibinfo {author} {\bibfnamefont {M.}~\bibnamefont {H\"ucker}}, \bibinfo
  {author} {\bibfnamefont {G.~D.}\ \bibnamefont {Gu}}, \bibinfo {author}
  {\bibfnamefont {J.~M.}\ \bibnamefont {Tranquada}}, \bibinfo {author}
  {\bibfnamefont {M.~M.}\ \bibnamefont {Fejer}},\ and\ \bibinfo {author}
  {\bibfnamefont {A.}~\bibnamefont {Kapitulnik}},\ }\bibfield  {title}
  {\bibinfo {title} {Evidence of chiral order in the charge-ordered phase of
  superconducting
  ${\mathrm{la}}_{1.875}{\mathrm{ba}}_{0.125}{\mathrm{cuo}}_{4}$ single
  crystals using polar kerr-effect measurements},\ }\href
  {https://doi.org/10.1103/PhysRevLett.112.047003} {\bibfield  {journal}
  {\bibinfo  {journal} {Phys. Rev. Lett.}\ }\textbf {\bibinfo {volume} {112}},\
  \bibinfo {pages} {047003} (\bibinfo {year} {2014})}\BibitemShut {NoStop}%
\bibitem [{\citenamefont {Li}\ \emph {et~al.}(2011)\citenamefont {Li},
  \citenamefont {Alidoust}, \citenamefont {Tranquada}, \citenamefont {Gu},\
  and\ \citenamefont {Ong}}]{PhysRevLett.107.277001}%
  \BibitemOpen
  \bibfield  {author} {\bibinfo {author} {\bibfnamefont {L.}~\bibnamefont
  {Li}}, \bibinfo {author} {\bibfnamefont {N.}~\bibnamefont {Alidoust}},
  \bibinfo {author} {\bibfnamefont {J.~M.}\ \bibnamefont {Tranquada}}, \bibinfo
  {author} {\bibfnamefont {G.~D.}\ \bibnamefont {Gu}},\ and\ \bibinfo {author}
  {\bibfnamefont {N.~P.}\ \bibnamefont {Ong}},\ }\bibfield  {title} {\bibinfo
  {title} {Unusual nernst effect suggesting time-reversal violation in the
  striped cuprate superconductor
  ${\mathrm{la}}_{2\ensuremath{-}x}{\mathrm{ba}}_{x}{\mathrm{cuo}}_{4}$},\
  }\href {https://doi.org/10.1103/PhysRevLett.107.277001} {\bibfield  {journal}
  {\bibinfo  {journal} {Phys. Rev. Lett.}\ }\textbf {\bibinfo {volume} {107}},\
  \bibinfo {pages} {277001} (\bibinfo {year} {2011})}\BibitemShut {NoStop}%
\bibitem [{\citenamefont {Marciani}\ \emph {et~al.}(2013)\citenamefont
  {Marciani}, \citenamefont {Fanfarillo}, \citenamefont {Castellani},\ and\
  \citenamefont {Benfatto}}]{PhysRevB.88.214508}%
  \BibitemOpen
  \bibfield  {author} {\bibinfo {author} {\bibfnamefont {M.}~\bibnamefont
  {Marciani}}, \bibinfo {author} {\bibfnamefont {L.}~\bibnamefont
  {Fanfarillo}}, \bibinfo {author} {\bibfnamefont {C.}~\bibnamefont
  {Castellani}},\ and\ \bibinfo {author} {\bibfnamefont {L.}~\bibnamefont
  {Benfatto}},\ }\bibfield  {title} {\bibinfo {title} {Leggett modes in
  iron-based superconductors as a probe of time-reversal symmetry breaking},\
  }\href {https://doi.org/10.1103/PhysRevB.88.214508} {\bibfield  {journal}
  {\bibinfo  {journal} {Phys. Rev. B}\ }\textbf {\bibinfo {volume} {88}},\
  \bibinfo {pages} {214508} (\bibinfo {year} {2013})}\BibitemShut {NoStop}%
\bibitem [{\citenamefont {Maiti}\ and\ \citenamefont
  {Chubukov}(2013)}]{PhysRevB.87.144511}%
  \BibitemOpen
  \bibfield  {author} {\bibinfo {author} {\bibfnamefont {S.}~\bibnamefont
  {Maiti}}\ and\ \bibinfo {author} {\bibfnamefont {A.~V.}\ \bibnamefont
  {Chubukov}},\ }\bibfield  {title} {\bibinfo {title} {$s+is$ state with broken
  time-reversal symmetry in fe-based superconductors},\ }\href
  {https://doi.org/10.1103/PhysRevB.87.144511} {\bibfield  {journal} {\bibinfo
  {journal} {Phys. Rev. B}\ }\textbf {\bibinfo {volume} {87}},\ \bibinfo
  {pages} {144511} (\bibinfo {year} {2013})}\BibitemShut {NoStop}%
\bibitem [{\citenamefont {Phan}\ and\ \citenamefont
  {Chubukov}(2023)}]{PhysRevB.107.134519}%
  \BibitemOpen
  \bibfield  {author} {\bibinfo {author} {\bibfnamefont {D.}~\bibnamefont
  {Phan}}\ and\ \bibinfo {author} {\bibfnamefont {A.~V.}\ \bibnamefont
  {Chubukov}},\ }\bibfield  {title} {\bibinfo {title} {Following the higgs mode
  across the bcs-bec crossover in two dimensions},\ }\href
  {https://doi.org/10.1103/PhysRevB.107.134519} {\bibfield  {journal} {\bibinfo
   {journal} {Phys. Rev. B}\ }\textbf {\bibinfo {volume} {107}},\ \bibinfo
  {pages} {134519} (\bibinfo {year} {2023})}\BibitemShut {NoStop}%
\bibitem [{\citenamefont {Lin}\ and\ \citenamefont
  {Hu}(2012)}]{PhysRevLett.108.177005}%
  \BibitemOpen
  \bibfield  {author} {\bibinfo {author} {\bibfnamefont {S.-Z.}\ \bibnamefont
  {Lin}}\ and\ \bibinfo {author} {\bibfnamefont {X.}~\bibnamefont {Hu}},\
  }\bibfield  {title} {\bibinfo {title} {Massless leggett mode in three-band
  superconductors with time-reversal-symmetry breaking},\ }\href
  {https://doi.org/10.1103/PhysRevLett.108.177005} {\bibfield  {journal}
  {\bibinfo  {journal} {Phys. Rev. Lett.}\ }\textbf {\bibinfo {volume} {108}},\
  \bibinfo {pages} {177005} (\bibinfo {year} {2012})}\BibitemShut {NoStop}%
\bibitem [{\citenamefont {Burnell}\ \emph {et~al.}(2010)\citenamefont
  {Burnell}, \citenamefont {Hu}, \citenamefont {Parish},\ and\ \citenamefont
  {Bernevig}}]{PhysRevB.82.144506}%
  \BibitemOpen
  \bibfield  {author} {\bibinfo {author} {\bibfnamefont {F.~J.}\ \bibnamefont
  {Burnell}}, \bibinfo {author} {\bibfnamefont {J.}~\bibnamefont {Hu}},
  \bibinfo {author} {\bibfnamefont {M.~M.}\ \bibnamefont {Parish}},\ and\
  \bibinfo {author} {\bibfnamefont {B.~A.}\ \bibnamefont {Bernevig}},\
  }\bibfield  {title} {\bibinfo {title} {Leggett mode in a strong-coupling
  model of iron arsenide superconductors},\ }\href
  {https://doi.org/10.1103/PhysRevB.82.144506} {\bibfield  {journal} {\bibinfo
  {journal} {Phys. Rev. B}\ }\textbf {\bibinfo {volume} {82}},\ \bibinfo
  {pages} {144506} (\bibinfo {year} {2010})}\BibitemShut {NoStop}%
\bibitem [{\citenamefont {Sharapov}\ \emph {et~al.}(2002)\citenamefont
  {Sharapov}, \citenamefont {Gusynin},\ and\ \citenamefont
  {Beck}}]{Sharapov2002}%
  \BibitemOpen
  \bibfield  {author} {\bibinfo {author} {\bibfnamefont {S.~G.}\ \bibnamefont
  {Sharapov}}, \bibinfo {author} {\bibfnamefont {V.~P.}\ \bibnamefont
  {Gusynin}},\ and\ \bibinfo {author} {\bibfnamefont {H.}~\bibnamefont
  {Beck}},\ }\bibfield  {title} {\bibinfo {title} {Effective action approach to
  the leggett's mode in two-band superconductors},\ }\href
  {https://doi.org/10.1140/epjb/e2002-00356-9} {\bibfield  {journal} {\bibinfo
  {journal} {The European Physical Journal B - Condensed Matter and Complex
  Systems}\ }\textbf {\bibinfo {volume} {30}},\ \bibinfo {pages} {45} (\bibinfo
  {year} {2002})}\BibitemShut {NoStop}%
\bibitem [{\citenamefont {Soto-Garrido}\ \emph
  {et~al.}(2017{\natexlab{b}})\citenamefont {Soto-Garrido}, \citenamefont
  {Wang}, \citenamefont {Fradkin},\ and\ \citenamefont
  {Cooper}}]{PhysRevB.95.214502}%
  \BibitemOpen
  \bibfield  {author} {\bibinfo {author} {\bibfnamefont {R.}~\bibnamefont
  {Soto-Garrido}}, \bibinfo {author} {\bibfnamefont {Y.}~\bibnamefont {Wang}},
  \bibinfo {author} {\bibfnamefont {E.}~\bibnamefont {Fradkin}},\ and\ \bibinfo
  {author} {\bibfnamefont {S.~L.}\ \bibnamefont {Cooper}},\ }\bibfield  {title}
  {\bibinfo {title} {Higgs modes in the pair density wave superconducting
  state},\ }\href {https://doi.org/10.1103/PhysRevB.95.214502} {\bibfield
  {journal} {\bibinfo  {journal} {Phys. Rev. B}\ }\textbf {\bibinfo {volume}
  {95}},\ \bibinfo {pages} {214502} (\bibinfo {year}
  {2017}{\natexlab{b}})}\BibitemShut {NoStop}%
\bibitem [{\citenamefont {Nosov}\ \emph {et~al.}(2024)\citenamefont {Nosov},
  \citenamefont {Andriyakhina},\ and\ \citenamefont
  {Burmistrov}}]{nosov2024spatiallyresolveddynamicsamplitudeschmidhiggs}%
  \BibitemOpen
  \bibfield  {author} {\bibinfo {author} {\bibfnamefont {P.~A.}\ \bibnamefont
  {Nosov}}, \bibinfo {author} {\bibfnamefont {E.~S.}\ \bibnamefont
  {Andriyakhina}},\ and\ \bibinfo {author} {\bibfnamefont {I.~S.}\ \bibnamefont
  {Burmistrov}},\ }\href {https://arxiv.org/abs/2409.11647} {\bibinfo {title}
  {Spatially-resolved dynamics of the amplitude schmid-higgs mode in disordered
  superconductors}} (\bibinfo {year} {2024}),\ \Eprint
  {https://arxiv.org/abs/2409.11647} {arXiv:2409.11647 [cond-mat.supr-con]}
  \BibitemShut {NoStop}%
\bibitem [{\citenamefont {Lee}\ and\ \citenamefont {Chung}(2023)}]{Lee2023}%
  \BibitemOpen
  \bibfield  {author} {\bibinfo {author} {\bibfnamefont {C.}~\bibnamefont
  {Lee}}\ and\ \bibinfo {author} {\bibfnamefont {S.~B.}\ \bibnamefont
  {Chung}},\ }\bibfield  {title} {\bibinfo {title} {Linear optical response
  from the odd-parity bardasis-schrieffer mode in locally non-centrosymmetric
  superconductors},\ }\href {https://doi.org/10.1038/s42005-023-01421-8}
  {\bibfield  {journal} {\bibinfo  {journal} {Communications Physics}\ }\textbf
  {\bibinfo {volume} {6}},\ \bibinfo {pages} {307} (\bibinfo {year}
  {2023})}\BibitemShut {NoStop}%
\bibitem [{\citenamefont {Sch\"ott}\ \emph {et~al.}(2016)\citenamefont
  {Sch\"ott}, \citenamefont {Locht}, \citenamefont {Lundin}, \citenamefont
  {Gr\aa{}n\"as}, \citenamefont {Eriksson},\ and\ \citenamefont
  {Di~Marco}}]{PhysRevB.93.075104}%
  \BibitemOpen
  \bibfield  {author} {\bibinfo {author} {\bibfnamefont {J.}~\bibnamefont
  {Sch\"ott}}, \bibinfo {author} {\bibfnamefont {I.~L.~M.}\ \bibnamefont
  {Locht}}, \bibinfo {author} {\bibfnamefont {E.}~\bibnamefont {Lundin}},
  \bibinfo {author} {\bibfnamefont {O.}~\bibnamefont {Gr\aa{}n\"as}}, \bibinfo
  {author} {\bibfnamefont {O.}~\bibnamefont {Eriksson}},\ and\ \bibinfo
  {author} {\bibfnamefont {I.}~\bibnamefont {Di~Marco}},\ }\bibfield  {title}
  {\bibinfo {title} {Analytic continuation by averaging pad\'e approximants},\
  }\href {https://doi.org/10.1103/PhysRevB.93.075104} {\bibfield  {journal}
  {\bibinfo  {journal} {Phys. Rev. B}\ }\textbf {\bibinfo {volume} {93}},\
  \bibinfo {pages} {075104} (\bibinfo {year} {2016})}\BibitemShut {NoStop}%
\bibitem [{Note2()}]{Note2}%
  \BibitemOpen
  \bibinfo {note} {There is a certain similarity between our case and
  Morr-Pines scenario for the resonance peak in the cuprates~\cite
  {PhysRevLett.81.1086}. {We also note that the Higgs mode can, in principle,
  also decay into two quasiparticles of the phase mode $\theta _-$.~\cite
  {soto-garrido-wang-fradkin-cooper,fradkin-book} This is a higher-order
  process, not included in our analysis}}\BibitemShut {NoStop}%
\bibitem [{\citenamefont {Browne}\ and\ \citenamefont
  {Levin}(1983)}]{PhysRevB.28.4029}%
  \BibitemOpen
  \bibfield  {author} {\bibinfo {author} {\bibfnamefont {D.~A.}\ \bibnamefont
  {Browne}}\ and\ \bibinfo {author} {\bibfnamefont {K.}~\bibnamefont {Levin}},\
  }\bibfield  {title} {\bibinfo {title} {Collective modes in
  charge-density-wave superconductors},\ }\href
  {https://doi.org/10.1103/PhysRevB.28.4029} {\bibfield  {journal} {\bibinfo
  {journal} {Phys. Rev. B}\ }\textbf {\bibinfo {volume} {28}},\ \bibinfo
  {pages} {4029} (\bibinfo {year} {1983})}\BibitemShut {NoStop}%
\bibitem [{\citenamefont {Torchinsky}\ \emph {et~al.}(2013)\citenamefont
  {Torchinsky}, \citenamefont {Mahmood}, \citenamefont {Bollinger},
  \citenamefont {Bo{\v{z}}ovi{\'{c}}},\ and\ \citenamefont
  {Gedik}}]{Torchinsky2013}%
  \BibitemOpen
  \bibfield  {author} {\bibinfo {author} {\bibfnamefont {D.~H.}\ \bibnamefont
  {Torchinsky}}, \bibinfo {author} {\bibfnamefont {F.}~\bibnamefont {Mahmood}},
  \bibinfo {author} {\bibfnamefont {A.~T.}\ \bibnamefont {Bollinger}}, \bibinfo
  {author} {\bibfnamefont {I.}~\bibnamefont {Bo{\v{z}}ovi{\'{c}}}},\ and\
  \bibinfo {author} {\bibfnamefont {N.}~\bibnamefont {Gedik}},\ }\bibfield
  {title} {\bibinfo {title} {Fluctuating charge-density waves in a cuprate
  superconductor},\ }\href {https://doi.org/10.1038/nmat3571} {\bibfield
  {journal} {\bibinfo  {journal} {Nature Materials}\ }\textbf {\bibinfo
  {volume} {12}},\ \bibinfo {pages} {387} (\bibinfo {year} {2013})}\BibitemShut
  {NoStop}%
\bibitem [{\citenamefont {Tassini}\ \emph {et~al.}(2005)\citenamefont
  {Tassini}, \citenamefont {Venturini}, \citenamefont {Zhang}, \citenamefont
  {Hackl}, \citenamefont {Kikugawa},\ and\ \citenamefont
  {Fujita}}]{PhysRevLett.95.117002}%
  \BibitemOpen
  \bibfield  {author} {\bibinfo {author} {\bibfnamefont {L.}~\bibnamefont
  {Tassini}}, \bibinfo {author} {\bibfnamefont {F.}~\bibnamefont {Venturini}},
  \bibinfo {author} {\bibfnamefont {Q.-M.}\ \bibnamefont {Zhang}}, \bibinfo
  {author} {\bibfnamefont {R.}~\bibnamefont {Hackl}}, \bibinfo {author}
  {\bibfnamefont {N.}~\bibnamefont {Kikugawa}},\ and\ \bibinfo {author}
  {\bibfnamefont {T.}~\bibnamefont {Fujita}},\ }\bibfield  {title} {\bibinfo
  {title} {Dynamical properties of charged stripes in
  ${\mathrm{la}}_{2\ensuremath{-}x}{\mathrm{sr}}_{x}{\mathrm{cuo}}_{4}$},\
  }\href {https://doi.org/10.1103/PhysRevLett.95.117002} {\bibfield  {journal}
  {\bibinfo  {journal} {Phys. Rev. Lett.}\ }\textbf {\bibinfo {volume} {95}},\
  \bibinfo {pages} {117002} (\bibinfo {year} {2005})}\BibitemShut {NoStop}%
\bibitem [{\citenamefont {Sugai}\ \emph {et~al.}(2006)\citenamefont {Sugai},
  \citenamefont {Takayanagi},\ and\ \citenamefont
  {Hayamizu}}]{PhysRevLett.96.137003}%
  \BibitemOpen
  \bibfield  {author} {\bibinfo {author} {\bibfnamefont {S.}~\bibnamefont
  {Sugai}}, \bibinfo {author} {\bibfnamefont {Y.}~\bibnamefont {Takayanagi}},\
  and\ \bibinfo {author} {\bibfnamefont {N.}~\bibnamefont {Hayamizu}},\
  }\bibfield  {title} {\bibinfo {title} {Phason and amplitudon in the
  charge-density-wave phase of one-dimensional charge stripes in
  ${\mathrm{la}}_{2\ensuremath{-}x}{\mathrm{sr}}_{x}{\mathrm{cuo}}_{4}$},\
  }\href {https://doi.org/10.1103/PhysRevLett.96.137003} {\bibfield  {journal}
  {\bibinfo  {journal} {Phys. Rev. Lett.}\ }\textbf {\bibinfo {volume} {96}},\
  \bibinfo {pages} {137003} (\bibinfo {year} {2006})}\BibitemShut {NoStop}%
\bibitem [{\citenamefont {Lee}\ \emph {et~al.}(1973)\citenamefont {Lee},
  \citenamefont {Rice},\ and\ \citenamefont {Anderson}}]{PhysRevLett.31.462}%
  \BibitemOpen
  \bibfield  {author} {\bibinfo {author} {\bibfnamefont {P.~A.}\ \bibnamefont
  {Lee}}, \bibinfo {author} {\bibfnamefont {T.~M.}\ \bibnamefont {Rice}},\ and\
  \bibinfo {author} {\bibfnamefont {P.~W.}\ \bibnamefont {Anderson}},\
  }\bibfield  {title} {\bibinfo {title} {Fluctuation effects at a peierls
  transition},\ }\href {https://doi.org/10.1103/PhysRevLett.31.462} {\bibfield
  {journal} {\bibinfo  {journal} {Phys. Rev. Lett.}\ }\textbf {\bibinfo
  {volume} {31}},\ \bibinfo {pages} {462} (\bibinfo {year} {1973})}\BibitemShut
  {NoStop}%
\bibitem [{\citenamefont {Littlewood}\ and\ \citenamefont
  {Varma}(1982)}]{PhysRevB.26.4883}%
  \BibitemOpen
  \bibfield  {author} {\bibinfo {author} {\bibfnamefont {P.~B.}\ \bibnamefont
  {Littlewood}}\ and\ \bibinfo {author} {\bibfnamefont {C.~M.}\ \bibnamefont
  {Varma}},\ }\bibfield  {title} {\bibinfo {title} {Amplitude collective modes
  in superconductors and their coupling to charge-density waves},\ }\href
  {https://doi.org/10.1103/PhysRevB.26.4883} {\bibfield  {journal} {\bibinfo
  {journal} {Phys. Rev. B}\ }\textbf {\bibinfo {volume} {26}},\ \bibinfo
  {pages} {4883} (\bibinfo {year} {1982})}\BibitemShut {NoStop}%
\bibitem [{\citenamefont {Gr\"uner}(1988)}]{RevModPhys.60.1129}%
  \BibitemOpen
  \bibfield  {author} {\bibinfo {author} {\bibfnamefont {G.}~\bibnamefont
  {Gr\"uner}},\ }\bibfield  {title} {\bibinfo {title} {The dynamics of
  charge-density waves},\ }\href {https://doi.org/10.1103/RevModPhys.60.1129}
  {\bibfield  {journal} {\bibinfo  {journal} {Rev. Mod. Phys.}\ }\textbf
  {\bibinfo {volume} {60}},\ \bibinfo {pages} {1129} (\bibinfo {year}
  {1988})}\BibitemShut {NoStop}%
\bibitem [{\citenamefont {Cea}\ and\ \citenamefont
  {Benfatto}(2014)}]{PhysRevB.90.224515}%
  \BibitemOpen
  \bibfield  {author} {\bibinfo {author} {\bibfnamefont {T.}~\bibnamefont
  {Cea}}\ and\ \bibinfo {author} {\bibfnamefont {L.}~\bibnamefont {Benfatto}},\
  }\bibfield  {title} {\bibinfo {title} {Nature and raman signatures of the
  higgs amplitude mode in the coexisting superconducting and
  charge-density-wave state},\ }\href
  {https://doi.org/10.1103/PhysRevB.90.224515} {\bibfield  {journal} {\bibinfo
  {journal} {Phys. Rev. B}\ }\textbf {\bibinfo {volume} {90}},\ \bibinfo
  {pages} {224515} (\bibinfo {year} {2014})}\BibitemShut {NoStop}%
\bibitem [{\citenamefont {Grasset}\ \emph {et~al.}(2018)\citenamefont
  {Grasset}, \citenamefont {Cea}, \citenamefont {Gallais}, \citenamefont
  {Cazayous}, \citenamefont {Sacuto}, \citenamefont {Cario}, \citenamefont
  {Benfatto},\ and\ \citenamefont {M\'easson}}]{PhysRevB.97.094502}%
  \BibitemOpen
  \bibfield  {author} {\bibinfo {author} {\bibfnamefont {R.}~\bibnamefont
  {Grasset}}, \bibinfo {author} {\bibfnamefont {T.}~\bibnamefont {Cea}},
  \bibinfo {author} {\bibfnamefont {Y.}~\bibnamefont {Gallais}}, \bibinfo
  {author} {\bibfnamefont {M.}~\bibnamefont {Cazayous}}, \bibinfo {author}
  {\bibfnamefont {A.}~\bibnamefont {Sacuto}}, \bibinfo {author} {\bibfnamefont
  {L.}~\bibnamefont {Cario}}, \bibinfo {author} {\bibfnamefont
  {L.}~\bibnamefont {Benfatto}},\ and\ \bibinfo {author} {\bibfnamefont
  {M.-A.}\ \bibnamefont {M\'easson}},\ }\bibfield  {title} {\bibinfo {title}
  {Higgs-mode radiance and charge-density-wave order in
  $2h\ensuremath{-}{\mathrm{nbse}}_{2}$},\ }\href
  {https://doi.org/10.1103/PhysRevB.97.094502} {\bibfield  {journal} {\bibinfo
  {journal} {Phys. Rev. B}\ }\textbf {\bibinfo {volume} {97}},\ \bibinfo
  {pages} {094502} (\bibinfo {year} {2018})}\BibitemShut {NoStop}%
\bibitem [{\citenamefont {Cea}\ and\ \citenamefont
  {Benfatto}(2016)}]{PhysRevB.94.064512}%
  \BibitemOpen
  \bibfield  {author} {\bibinfo {author} {\bibfnamefont {T.}~\bibnamefont
  {Cea}}\ and\ \bibinfo {author} {\bibfnamefont {L.}~\bibnamefont {Benfatto}},\
  }\bibfield  {title} {\bibinfo {title} {Signature of the leggett mode in the
  ${A}_{1g}$ raman response: From ${\text{mgb}}_{2}$ to iron-based
  superconductors},\ }\href {https://doi.org/10.1103/PhysRevB.94.064512}
  {\bibfield  {journal} {\bibinfo  {journal} {Phys. Rev. B}\ }\textbf {\bibinfo
  {volume} {94}},\ \bibinfo {pages} {064512} (\bibinfo {year}
  {2016})}\BibitemShut {NoStop}%
\bibitem [{\citenamefont {Klein}\ and\ \citenamefont
  {Dierker}(1984)}]{PhysRevB.29.4976}%
  \BibitemOpen
  \bibfield  {author} {\bibinfo {author} {\bibfnamefont {M.~V.}\ \bibnamefont
  {Klein}}\ and\ \bibinfo {author} {\bibfnamefont {S.~B.}\ \bibnamefont
  {Dierker}},\ }\bibfield  {title} {\bibinfo {title} {Theory of raman
  scattering in superconductors},\ }\href
  {https://doi.org/10.1103/PhysRevB.29.4976} {\bibfield  {journal} {\bibinfo
  {journal} {Phys. Rev. B}\ }\textbf {\bibinfo {volume} {29}},\ \bibinfo
  {pages} {4976} (\bibinfo {year} {1984})}\BibitemShut {NoStop}%
\bibitem [{\citenamefont {Klein}(2010)}]{PhysRevB.82.014507}%
  \BibitemOpen
  \bibfield  {author} {\bibinfo {author} {\bibfnamefont {M.~V.}\ \bibnamefont
  {Klein}},\ }\bibfield  {title} {\bibinfo {title} {Theory of raman scattering
  from leggett's collective mode in a multiband superconductor: Application to
  ${\text{mgb}}_{2}$},\ }\href {https://doi.org/10.1103/PhysRevB.82.014507}
  {\bibfield  {journal} {\bibinfo  {journal} {Phys. Rev. B}\ }\textbf {\bibinfo
  {volume} {82}},\ \bibinfo {pages} {014507} (\bibinfo {year}
  {2010})}\BibitemShut {NoStop}%
\bibitem [{\citenamefont {Devereaux}\ and\ \citenamefont
  {Hackl}(2007)}]{RevModPhys.79.175}%
  \BibitemOpen
  \bibfield  {author} {\bibinfo {author} {\bibfnamefont {T.~P.}\ \bibnamefont
  {Devereaux}}\ and\ \bibinfo {author} {\bibfnamefont {R.}~\bibnamefont
  {Hackl}},\ }\bibfield  {title} {\bibinfo {title} {Inelastic light scattering
  from correlated electrons},\ }\href
  {https://doi.org/10.1103/RevModPhys.79.175} {\bibfield  {journal} {\bibinfo
  {journal} {Rev. Mod. Phys.}\ }\textbf {\bibinfo {volume} {79}},\ \bibinfo
  {pages} {175} (\bibinfo {year} {2007})}\BibitemShut {NoStop}%
\bibitem [{\citenamefont {Chubukov}\ and\ \citenamefont
  {Frenkel}(1995)}]{PhysRevB.52.9760}%
  \BibitemOpen
  \bibfield  {author} {\bibinfo {author} {\bibfnamefont {A.~V.}\ \bibnamefont
  {Chubukov}}\ and\ \bibinfo {author} {\bibfnamefont {D.~M.}\ \bibnamefont
  {Frenkel}},\ }\bibfield  {title} {\bibinfo {title} {Resonant two-magnon raman
  scattering in parent compounds of high-${\mathit{t}}_{\mathit{c}}$
  superconductors},\ }\href {https://doi.org/10.1103/PhysRevB.52.9760}
  {\bibfield  {journal} {\bibinfo  {journal} {Phys. Rev. B}\ }\textbf {\bibinfo
  {volume} {52}},\ \bibinfo {pages} {9760} (\bibinfo {year}
  {1995})}\BibitemShut {NoStop}%
\bibitem [{\citenamefont {Devereaux}\ and\ \citenamefont
  {Einzel}(1995)}]{PhysRevB.51.16336}%
  \BibitemOpen
  \bibfield  {author} {\bibinfo {author} {\bibfnamefont {T.~P.}\ \bibnamefont
  {Devereaux}}\ and\ \bibinfo {author} {\bibfnamefont {D.}~\bibnamefont
  {Einzel}},\ }\bibfield  {title} {\bibinfo {title} {Electronic raman
  scattering in superconductors as a probe of anisotropic electron pairing},\
  }\href {https://doi.org/10.1103/PhysRevB.51.16336} {\bibfield  {journal}
  {\bibinfo  {journal} {Phys. Rev. B}\ }\textbf {\bibinfo {volume} {51}},\
  \bibinfo {pages} {16336} (\bibinfo {year} {1995})}\BibitemShut {NoStop}%
\bibitem [{\citenamefont {Devereaux}\ \emph {et~al.}(1994)\citenamefont
  {Devereaux}, \citenamefont {Einzel}, \citenamefont {Stadlober}, \citenamefont
  {Hackl}, \citenamefont {Leach},\ and\ \citenamefont
  {Neumeier}}]{PhysRevLett.72.396}%
  \BibitemOpen
  \bibfield  {author} {\bibinfo {author} {\bibfnamefont {T.~P.}\ \bibnamefont
  {Devereaux}}, \bibinfo {author} {\bibfnamefont {D.}~\bibnamefont {Einzel}},
  \bibinfo {author} {\bibfnamefont {B.}~\bibnamefont {Stadlober}}, \bibinfo
  {author} {\bibfnamefont {R.}~\bibnamefont {Hackl}}, \bibinfo {author}
  {\bibfnamefont {D.~H.}\ \bibnamefont {Leach}},\ and\ \bibinfo {author}
  {\bibfnamefont {J.~J.}\ \bibnamefont {Neumeier}},\ }\bibfield  {title}
  {\bibinfo {title} {Electronic raman scattering in
  high-${\mathit{t}}_{\mathit{c}}$ superconductors: A probe of
  ${\mathit{d}}_{\mathit{x}}^{2}$-${\mathit{y}}^{2}$ pairing},\ }\href
  {https://doi.org/10.1103/PhysRevLett.72.396} {\bibfield  {journal} {\bibinfo
  {journal} {Phys. Rev. Lett.}\ }\textbf {\bibinfo {volume} {72}},\ \bibinfo
  {pages} {396} (\bibinfo {year} {1994})}\BibitemShut {NoStop}%
\bibitem [{\citenamefont {Mross}\ and\ \citenamefont
  {Senthil}(2012)}]{PhysRevB.86.115138}%
  \BibitemOpen
  \bibfield  {author} {\bibinfo {author} {\bibfnamefont {D.~F.}\ \bibnamefont
  {Mross}}\ and\ \bibinfo {author} {\bibfnamefont {T.}~\bibnamefont
  {Senthil}},\ }\bibfield  {title} {\bibinfo {title} {Stripe melting and
  quantum criticality in correlated metals},\ }\href
  {https://doi.org/10.1103/PhysRevB.86.115138} {\bibfield  {journal} {\bibinfo
  {journal} {Phys. Rev. B}\ }\textbf {\bibinfo {volume} {86}},\ \bibinfo
  {pages} {115138} (\bibinfo {year} {2012})}\BibitemShut {NoStop}%
\bibitem [{\citenamefont {Mross}\ and\ \citenamefont
  {Senthil}(2015)}]{PhysRevX.5.031008}%
  \BibitemOpen
  \bibfield  {author} {\bibinfo {author} {\bibfnamefont {D.~F.}\ \bibnamefont
  {Mross}}\ and\ \bibinfo {author} {\bibfnamefont {T.}~\bibnamefont
  {Senthil}},\ }\bibfield  {title} {\bibinfo {title} {Spin- and
  pair-density-wave glasses},\ }\href
  {https://doi.org/10.1103/PhysRevX.5.031008} {\bibfield  {journal} {\bibinfo
  {journal} {Phys. Rev. X}\ }\textbf {\bibinfo {volume} {5}},\ \bibinfo {pages}
  {031008} (\bibinfo {year} {2015})}\BibitemShut {NoStop}%
\bibitem [{Note3()}]{Note3}%
  \BibitemOpen
  \bibinfo {note} {The physics behind this is the same as that which underlies
  the disorder driven breaking of TRS in junctions in which the leading order
  Josephson coupling is frustrated.\cite
  {PhysRevB.61.5902,PhysRevB.108.L100505}}\BibitemShut {NoStop}%
\bibitem [{\citenamefont {Hirota}(2001)}]{HIROTA200161}%
  \BibitemOpen
  \bibfield  {author} {\bibinfo {author} {\bibfnamefont {K.}~\bibnamefont
  {Hirota}},\ }\bibfield  {title} {\bibinfo {title} {Neutron scattering studies
  of zn-doped ${\mathrm{la}}_{2-x}{\mathrm{sr}}_{x}{\mathbf{cuo}}_{4}$},\
  }\href {https://doi.org/https://doi.org/10.1016/S0921-4534(01)00195-2}
  {\bibfield  {journal} {\bibinfo  {journal} {Physica C: Superconductivity}\
  }\textbf {\bibinfo {volume} {357-360}},\ \bibinfo {pages} {61} (\bibinfo
  {year} {2001})}\BibitemShut {NoStop}%
\bibitem [{\citenamefont {Suchaneck}\ \emph {et~al.}(2010)\citenamefont
  {Suchaneck}, \citenamefont {Hinkov}, \citenamefont {Haug}, \citenamefont
  {Schulz}, \citenamefont {Bernhard}, \citenamefont {Ivanov}, \citenamefont
  {Hradil}, \citenamefont {Lin}, \citenamefont {Bourges}, \citenamefont
  {Keimer},\ and\ \citenamefont {Sidis}}]{PhysRevLett.105.037207}%
  \BibitemOpen
  \bibfield  {author} {\bibinfo {author} {\bibfnamefont {A.}~\bibnamefont
  {Suchaneck}}, \bibinfo {author} {\bibfnamefont {V.}~\bibnamefont {Hinkov}},
  \bibinfo {author} {\bibfnamefont {D.}~\bibnamefont {Haug}}, \bibinfo {author}
  {\bibfnamefont {L.}~\bibnamefont {Schulz}}, \bibinfo {author} {\bibfnamefont
  {C.}~\bibnamefont {Bernhard}}, \bibinfo {author} {\bibfnamefont
  {A.}~\bibnamefont {Ivanov}}, \bibinfo {author} {\bibfnamefont
  {K.}~\bibnamefont {Hradil}}, \bibinfo {author} {\bibfnamefont {C.~T.}\
  \bibnamefont {Lin}}, \bibinfo {author} {\bibfnamefont {P.}~\bibnamefont
  {Bourges}}, \bibinfo {author} {\bibfnamefont {B.}~\bibnamefont {Keimer}},\
  and\ \bibinfo {author} {\bibfnamefont {Y.}~\bibnamefont {Sidis}},\ }\bibfield
   {title} {\bibinfo {title} {Incommensurate magnetic order and dynamics
  induced by spinless impurities in
  ${\mathrm{yba}}_{2}{\mathrm{cu}}_{3}{\mathbf{o}}_{6.6}$},\ }\href
  {https://doi.org/10.1103/PhysRevLett.105.037207} {\bibfield  {journal}
  {\bibinfo  {journal} {Phys. Rev. Lett.}\ }\textbf {\bibinfo {volume} {105}},\
  \bibinfo {pages} {037207} (\bibinfo {year} {2010})}\BibitemShut {NoStop}%
\bibitem [{\citenamefont {Morr}\ and\ \citenamefont
  {Pines}(1998)}]{PhysRevLett.81.1086}%
  \BibitemOpen
  \bibfield  {author} {\bibinfo {author} {\bibfnamefont {D.~K.}\ \bibnamefont
  {Morr}}\ and\ \bibinfo {author} {\bibfnamefont {D.}~\bibnamefont {Pines}},\
  }\bibfield  {title} {\bibinfo {title} {The resonance peak in cuprate
  superconductors},\ }\href {https://doi.org/10.1103/PhysRevLett.81.1086}
  {\bibfield  {journal} {\bibinfo  {journal} {Phys. Rev. Lett.}\ }\textbf
  {\bibinfo {volume} {81}},\ \bibinfo {pages} {1086} (\bibinfo {year}
  {1998})}\BibitemShut {NoStop}%
\bibitem [{\citenamefont {Fradkin}(2021)}]{fradkin-book}%
  \BibitemOpen
  \bibfield  {author} {\bibinfo {author} {\bibfnamefont {E.}~\bibnamefont
  {Fradkin}},\ }\href@noop {} {\emph {\bibinfo {title} {Quantum field theory:
  an integrated approach}}}\ (\bibinfo  {publisher} {Princeton University
  Press},\ \bibinfo {year} {2021})\BibitemShut {NoStop}%
\bibitem [{\citenamefont {Zyuzin}\ and\ \citenamefont
  {Spivak}(2000)}]{PhysRevB.61.5902}%
  \BibitemOpen
  \bibfield  {author} {\bibinfo {author} {\bibfnamefont {A.}~\bibnamefont
  {Zyuzin}}\ and\ \bibinfo {author} {\bibfnamefont {B.}~\bibnamefont
  {Spivak}},\ }\bibfield  {title} {\bibinfo {title} {Theory of
  $\ensuremath{\pi}/2$ superconducting josephson junctions},\ }\href
  {https://doi.org/10.1103/PhysRevB.61.5902} {\bibfield  {journal} {\bibinfo
  {journal} {Phys. Rev. B}\ }\textbf {\bibinfo {volume} {61}},\ \bibinfo
  {pages} {5902} (\bibinfo {year} {2000})}\BibitemShut {NoStop}%
\bibitem [{\citenamefont {Yuan}\ \emph {et~al.}(2023)\citenamefont {Yuan},
  \citenamefont {Vituri}, \citenamefont {Berg}, \citenamefont {Spivak},\ and\
  \citenamefont {Kivelson}}]{PhysRevB.108.L100505}%
  \BibitemOpen
  \bibfield  {author} {\bibinfo {author} {\bibfnamefont {A.~C.}\ \bibnamefont
  {Yuan}}, \bibinfo {author} {\bibfnamefont {Y.}~\bibnamefont {Vituri}},
  \bibinfo {author} {\bibfnamefont {E.}~\bibnamefont {Berg}}, \bibinfo {author}
  {\bibfnamefont {B.}~\bibnamefont {Spivak}},\ and\ \bibinfo {author}
  {\bibfnamefont {S.~A.}\ \bibnamefont {Kivelson}},\ }\bibfield  {title}
  {\bibinfo {title} {Inhomogeneity-induced time-reversal symmetry breaking in
  cuprate twist junctions},\ }\href
  {https://doi.org/10.1103/PhysRevB.108.L100505} {\bibfield  {journal}
  {\bibinfo  {journal} {Phys. Rev. B}\ }\textbf {\bibinfo {volume} {108}},\
  \bibinfo {pages} {L100505} (\bibinfo {year} {2023})}\BibitemShut {NoStop}%
\end{thebibliography}%


\begin{thebibliography}{10}%
\makeatletter
\providecommand \@ifxundefined [1]{%
 \@ifx{#1\undefined}
}%
\providecommand \@ifnum [1]{%
 \ifnum #1\expandafter \@firstoftwo
 \else \expandafter \@secondoftwo
 \fi
}%
\providecommand \@ifx [1]{%
 \ifx #1\expandafter \@firstoftwo
 \else \expandafter \@secondoftwo
 \fi
}%
\providecommand \natexlab [1]{#1}%
\providecommand \enquote  [1]{``#1''}%
\providecommand \bibnamefont  [1]{#1}%
\providecommand \bibfnamefont [1]{#1}%
\providecommand \citenamefont [1]{#1}%
\providecommand \href@noop [0]{\@secondoftwo}%
\providecommand \href [0]{\begingroup \@sanitize@url \@href}%
\providecommand \@href[1]{\@@startlink{#1}\@@href}%
\providecommand \@@href[1]{\endgroup#1\@@endlink}%
\providecommand \@sanitize@url [0]{\catcode `\\12\catcode `\$12\catcode
  `\&12\catcode `\#12\catcode `\^12\catcode `\_12\catcode `\%12\relax}%
\providecommand \@@startlink[1]{}%
\providecommand \@@endlink[0]{}%
\providecommand \url  [0]{\begingroup\@sanitize@url \@url }%
\providecommand \@url [1]{\endgroup\@href {#1}{\urlprefix }}%
\providecommand \urlprefix  [0]{URL }%
\providecommand \Eprint [0]{\href }%
\providecommand \doibase [0]{https://doi.org/}%
\providecommand \selectlanguage [0]{\@gobble}%
\providecommand \bibinfo  [0]{\@secondoftwo}%
\providecommand \bibfield  [0]{\@secondoftwo}%
\providecommand \translation [1]{[#1]}%
\providecommand \BibitemOpen [0]{}%
\providecommand \bibitemStop [0]{}%
\providecommand \bibitemNoStop [0]{.\EOS\space}%
\providecommand \EOS [0]{\spacefactor3000\relax}%
\providecommand \BibitemShut  [1]{\csname bibitem#1\endcsname}%
\let\auto@bib@innerbib\@empty
\bibitem [{\citenamefont {Lin}\ and\ \citenamefont
  {Hu}(2012)}]{PhysRevLett.108.177005}%
  \BibitemOpen
  \bibfield  {author} {\bibinfo {author} {\bibfnamefont {S.-Z.}\ \bibnamefont
  {Lin}}\ and\ \bibinfo {author} {\bibfnamefont {X.}~\bibnamefont {Hu}},\
  }\bibfield  {title} {\bibinfo {title} {Massless leggett mode in three-band
  superconductors with time-reversal-symmetry breaking},\ }\href
  {https://doi.org/10.1103/PhysRevLett.108.177005} {\bibfield  {journal}
  {\bibinfo  {journal} {Phys. Rev. Lett.}\ }\textbf {\bibinfo {volume} {108}},\
  \bibinfo {pages} {177005} (\bibinfo {year} {2012})}\BibitemShut {NoStop}%
\bibitem [{\citenamefont {Soto-Garrido}\ \emph {et~al.}(2017)\citenamefont
  {Soto-Garrido}, \citenamefont {Wang}, \citenamefont {Fradkin},\ and\
  \citenamefont {Cooper}}]{soto-garrido-wang-fradkin-cooper}%
  \BibitemOpen
  \bibfield  {author} {\bibinfo {author} {\bibfnamefont {R.}~\bibnamefont
  {Soto-Garrido}}, \bibinfo {author} {\bibfnamefont {Y.}~\bibnamefont {Wang}},
  \bibinfo {author} {\bibfnamefont {E.}~\bibnamefont {Fradkin}},\ and\ \bibinfo
  {author} {\bibfnamefont {S.~L.}\ \bibnamefont {Cooper}},\ }\bibfield  {title}
  {\bibinfo {title} {Higgs modes in the pair density wave superconducting
  state},\ }\href {https://doi.org/10.1103/PhysRevB.95.214502} {\bibfield
  {journal} {\bibinfo  {journal} {Phys. Rev. B}\ }\textbf {\bibinfo {volume}
  {95}},\ \bibinfo {pages} {214502} (\bibinfo {year} {2017})}\BibitemShut
  {NoStop}%
\bibitem [{\citenamefont {Lee}\ \emph {et~al.}(1973)\citenamefont {Lee},
  \citenamefont {Rice},\ and\ \citenamefont {Anderson}}]{PhysRevLett.31.462}%
  \BibitemOpen
  \bibfield  {author} {\bibinfo {author} {\bibfnamefont {P.~A.}\ \bibnamefont
  {Lee}}, \bibinfo {author} {\bibfnamefont {T.~M.}\ \bibnamefont {Rice}},\ and\
  \bibinfo {author} {\bibfnamefont {P.~W.}\ \bibnamefont {Anderson}},\
  }\bibfield  {title} {\bibinfo {title} {Fluctuation effects at a peierls
  transition},\ }\href {https://doi.org/10.1103/PhysRevLett.31.462} {\bibfield
  {journal} {\bibinfo  {journal} {Phys. Rev. Lett.}\ }\textbf {\bibinfo
  {volume} {31}},\ \bibinfo {pages} {462} (\bibinfo {year} {1973})}\BibitemShut
  {NoStop}%
\bibitem [{\citenamefont {Littlewood}\ and\ \citenamefont
  {Varma}(1982)}]{PhysRevB.26.4883}%
  \BibitemOpen
  \bibfield  {author} {\bibinfo {author} {\bibfnamefont {P.~B.}\ \bibnamefont
  {Littlewood}}\ and\ \bibinfo {author} {\bibfnamefont {C.~M.}\ \bibnamefont
  {Varma}},\ }\bibfield  {title} {\bibinfo {title} {Amplitude collective modes
  in superconductors and their coupling to charge-density waves},\ }\href
  {https://doi.org/10.1103/PhysRevB.26.4883} {\bibfield  {journal} {\bibinfo
  {journal} {Phys. Rev. B}\ }\textbf {\bibinfo {volume} {26}},\ \bibinfo
  {pages} {4883} (\bibinfo {year} {1982})}\BibitemShut {NoStop}%
\bibitem [{\citenamefont {Gr\"uner}(1988)}]{RevModPhys.60.1129}%
  \BibitemOpen
  \bibfield  {author} {\bibinfo {author} {\bibfnamefont {G.}~\bibnamefont
  {Gr\"uner}},\ }\bibfield  {title} {\bibinfo {title} {The dynamics of
  charge-density waves},\ }\href {https://doi.org/10.1103/RevModPhys.60.1129}
  {\bibfield  {journal} {\bibinfo  {journal} {Rev. Mod. Phys.}\ }\textbf
  {\bibinfo {volume} {60}},\ \bibinfo {pages} {1129} (\bibinfo {year}
  {1988})}\BibitemShut {NoStop}%
\bibitem [{\citenamefont {Klein}\ and\ \citenamefont
  {Dierker}(1984)}]{PhysRevB.29.4976}%
  \BibitemOpen
  \bibfield  {author} {\bibinfo {author} {\bibfnamefont {M.~V.}\ \bibnamefont
  {Klein}}\ and\ \bibinfo {author} {\bibfnamefont {S.~B.}\ \bibnamefont
  {Dierker}},\ }\bibfield  {title} {\bibinfo {title} {Theory of raman
  scattering in superconductors},\ }\href
  {https://doi.org/10.1103/PhysRevB.29.4976} {\bibfield  {journal} {\bibinfo
  {journal} {Phys. Rev. B}\ }\textbf {\bibinfo {volume} {29}},\ \bibinfo
  {pages} {4976} (\bibinfo {year} {1984})}\BibitemShut {NoStop}%
\bibitem [{\citenamefont {Klein}(2010)}]{PhysRevB.82.014507}%
  \BibitemOpen
  \bibfield  {author} {\bibinfo {author} {\bibfnamefont {M.~V.}\ \bibnamefont
  {Klein}},\ }\bibfield  {title} {\bibinfo {title} {Theory of raman scattering
  from leggett's collective mode in a multiband superconductor: Application to
  ${\text{mgb}}_{2}$},\ }\href {https://doi.org/10.1103/PhysRevB.82.014507}
  {\bibfield  {journal} {\bibinfo  {journal} {Phys. Rev. B}\ }\textbf {\bibinfo
  {volume} {82}},\ \bibinfo {pages} {014507} (\bibinfo {year}
  {2010})}\BibitemShut {NoStop}%
\bibitem [{\citenamefont {Devereaux}\ and\ \citenamefont
  {Hackl}(2007)}]{RevModPhys.79.175}%
  \BibitemOpen
  \bibfield  {author} {\bibinfo {author} {\bibfnamefont {T.~P.}\ \bibnamefont
  {Devereaux}}\ and\ \bibinfo {author} {\bibfnamefont {R.}~\bibnamefont
  {Hackl}},\ }\bibfield  {title} {\bibinfo {title} {Inelastic light scattering
  from correlated electrons},\ }\href
  {https://doi.org/10.1103/RevModPhys.79.175} {\bibfield  {journal} {\bibinfo
  {journal} {Rev. Mod. Phys.}\ }\textbf {\bibinfo {volume} {79}},\ \bibinfo
  {pages} {175} (\bibinfo {year} {2007})}\BibitemShut {NoStop}%
\bibitem [{\citenamefont {Maiti}\ \emph {et~al.}(2017)\citenamefont {Maiti},
  \citenamefont {Chubukov},\ and\ \citenamefont
  {Hirschfeld}}]{PhysRevB.96.014503}%
  \BibitemOpen
  \bibfield  {author} {\bibinfo {author} {\bibfnamefont {S.}~\bibnamefont
  {Maiti}}, \bibinfo {author} {\bibfnamefont {A.~V.}\ \bibnamefont
  {Chubukov}},\ and\ \bibinfo {author} {\bibfnamefont {P.~J.}\ \bibnamefont
  {Hirschfeld}},\ }\bibfield  {title} {\bibinfo {title} {Conservation laws,
  vertex corrections, and screening in raman spectroscopy},\ }\href
  {https://doi.org/10.1103/PhysRevB.96.014503} {\bibfield  {journal} {\bibinfo
  {journal} {Phys. Rev. B}\ }\textbf {\bibinfo {volume} {96}},\ \bibinfo
  {pages} {014503} (\bibinfo {year} {2017})}\BibitemShut {NoStop}%
\bibitem [{\citenamefont {Yuan}\ \emph {et~al.}(2023)\citenamefont {Yuan},
  \citenamefont {Vituri}, \citenamefont {Berg}, \citenamefont {Spivak},\ and\
  \citenamefont {Kivelson}}]{PhysRevB.108.L100505}%
  \BibitemOpen
  \bibfield  {author} {\bibinfo {author} {\bibfnamefont {A.~C.}\ \bibnamefont
  {Yuan}}, \bibinfo {author} {\bibfnamefont {Y.}~\bibnamefont {Vituri}},
  \bibinfo {author} {\bibfnamefont {E.}~\bibnamefont {Berg}}, \bibinfo {author}
  {\bibfnamefont {B.}~\bibnamefont {Spivak}},\ and\ \bibinfo {author}
  {\bibfnamefont {S.~A.}\ \bibnamefont {Kivelson}},\ }\bibfield  {title}
  {\bibinfo {title} {Inhomogeneity-induced time-reversal symmetry breaking in
  cuprate twist junctions},\ }\href
  {https://doi.org/10.1103/PhysRevB.108.L100505} {\bibfield  {journal}
  {\bibinfo  {journal} {Phys. Rev. B}\ }\textbf {\bibinfo {volume} {108}},\
  \bibinfo {pages} {L100505} (\bibinfo {year} {2023})}\BibitemShut {NoStop}%
\end{thebibliography}%

\end{document}


\clearpage


\begin{center}
  \textbf{SUPPLEMENTARY MATERIAL FOR
  \\[0.2cm]
  \large Time-reversal symmetry breaking, collective modes, and Raman spectrum in  pair-density-wave states}
  \\[0.35cm]
  Yi-Ming Wu,$^{1}$ Andrey Chubukov,$^{2}$ Yuxuan Wang,$^{3}$ and Steven A. Kivelson$^{4}$
  \\[0.15cm]
  {\small \it $^{1}$Stanford Institute for Theoretical Physics, Stanford University, Stanford, California 94305, USA}
  \\[-1pt]
  {\small \it $^{2}$Department of Physics, University of Minnesota, Minneapolis, Minnesota 55455, USA}
  \\[-1pt]
  {\small \it $^{3}$Department of Physics, University of Florida, Gainesville, Florida 32611, USA}
   \\[-1pt]
  {\small \it $^{4}$Department of Physics, Stanford University, Stanford, California 94305, USA}

  \vspace{0.4cm}
  \parbox{0.85\textwidth}{In this Supplemental Material we 
  discuss the details about
  i) Mean field gap equations in various cases and TRSB in the PDW+SC case, 
  ii) fluctuations above the mean field solutions and various collective modes,
  iii) analysis about the energy hierarchy of amplitude modes in the PDW+SC case,
  iv) explicit calculations of the Raman susceptibility and v) discussion about disorder effect} 
\end{center}

\tableofcontents

\section{I. Mean field analysis}

\subsection{1. Bogoliubov quasiparticle band in the extended Brillouin zone} 
\label{sub:bogoliubov_quasiparticle_band_in_the_extended_brillouin_zone}
\begin{figure}
    \includegraphics[width=18cm]{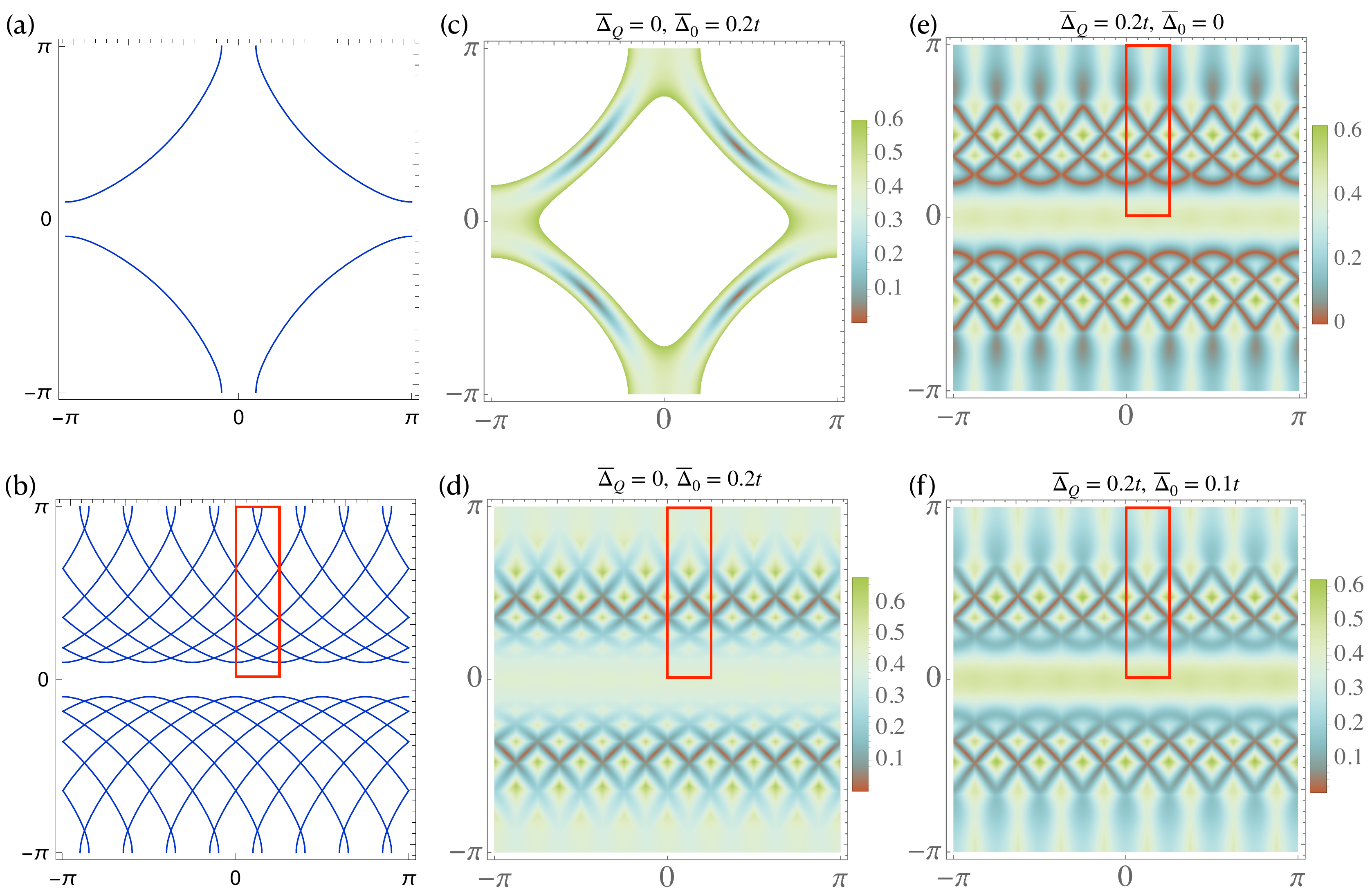}
    \caption{(a) Fermi surface of a square lattice $t-t'$ model at $1/8$ hole-doping. The same Fermi surface plotted in the folded Brillouin zone with period $8$ in the $x$-direction is shown in (b). (c) Bogoliubov dispersion $E(\bk)$ (in unites of $t$) for a $d$-wave SC, where $E(\bk)>0$ is the band above Fermi level, and $E(\bk)>0.6t$ parts are not shown. The gap closes at four nodal points. (d)$E(\bk)$ for a $d$-wave SC shown in the folded Brillouin zone. Here $E(\bk)$ is the first band just above the Fermi level, and hence shows repeated information of the low energy part of (c). (e) Plot of $E(\bk)$ for a unidirectional PDW with momentum $\bQ=(\frac{\pi}{4},0)$, where $E(\bk)$ is the first Bogoliubov band above Fermi level. (f) Plot of $E(\bk)$ for a unidirectional PDW with uniform SC, where $E(\bk)$ is the first Bogoliubov band above Fermi level. We have chosen $\varphi_{\bQ}-\varphi_{-\bQ}=0$ and $\varphi_0=(\varphi_{\bQ}+\varphi_{-\bQ})/2+\pi/2$, i.e. the TRSB configuration. In (b) and (d-e) the red rectangle indicates the regions shown in the main text. }\label{fig:PDW_BdG}
\end{figure}

We begin by introducing the non-interacting part of the fermions, which is a square lattice $t-t'$ model  described by the following Hamiltonian
\begin{equation}
    H_0=\sum_{\bk,\sigma}\xi(\bk)\psi^\dagger_{\sigma}(\bk)\psi_{\sigma}(\bk)
\end{equation}
where $\psi^\dagger_{\sigma}(\bk),\psi_\sigma(\bk)$ are fermion operators with spin $\sigma$ and
\begin{equation}
    \xi(\bk)=-2t\left[\cos k_x+\cos k_y\right]-4t'\cos k_x\cos k_y-\mu.
\end{equation}
Throughout this paper we choose $t'=-t/4$ and measure all energies scales in units of $t$.
Although it may not be important, we choose a particular chemical potential $\mu$ such that the total filling corresponds to 1/8 hole doping from half filling. The corresponding Fermi surface  is shown in Fig.\ref{fig:PDW_BdG}(a). Since we will study density wave orders with momentum $8\bQ=0$, we also present the Fermi surface in the reduced Brillouin zone (but in an extended way) in Fig.\ref{fig:PDW_BdG}(b).

Next we assume  that both the PDW and SC orders have the same pairing symmetry characterized by the same form factor $f_{\bk}$, such that at mean field level the order parameters couple to fermions as
\begin{equation}
     \Delta_{\overline \bq}f_{\bk}\psi_{\up}^\dagger(\bk+\frac{\overline\bq}{2})\psi_{\down}^\dagger(-\bk+\frac{\overline\bq}{2})+\text{h.c.}\label{eq:SCcoupling}
 \end{equation} 
 where we remind that $\overline{\bq}$ can be  $\bm{0},\pm\bQ$ and $\overline\Delta_{\overline{\bq}}$ is generally complex, i.e. it contains information about both mean field gap magnitude and phase. 
 Note a shift in $\bk$ must also affect $f_{\bk}$.
We will use the $d_{x^2-y^2}$ form factor
\begin{equation}
    f_{\bk}=\cos k_x - \cos k_y
\end{equation}
in our calculations when we discuss $d$-wave symmetry. To construct the BdG Hamiltonian, we must use the following  generalized fermion basis
\begin{equation}
    \Psi_{\sigma, \bk}^\dagger=[\psi_{\sigma}^\dagger(\bk), \psi_{\bQ,\sigma}^\dagger(\bk),\psi_{2\bQ,\sigma}^\dagger(\bk),...,\psi_{7\bQ,\sigma}^\dagger(\bk)],\label{eq:basis1}
\end{equation}
where we have used $\sigma$ to denote spin and $\psi_{n\bQ}(\bk)=\psi(n\bQ+\bk)$ for brevity. The BdG Hamiltonian formally reads
\begin{equation}
  H_\text{BdG}=\sum_{\bk\in\text{fBZ},\sigma} (\Psi_{\up,k}^\dagger,\Psi^\text{T}_{\down,-k})\begin{pmatrix}
    \hat\xi_p(\bk) & \hat\Delta\\
    \hat\Delta^\dagger & \hat\xi_h(\bk) 
  \end{pmatrix}\begin{pmatrix}
        \Psi_{\up,k'}\\
        {\Psi^\dagger}^\text{T}_{\down,-k'}
    \end{pmatrix}\label{eq:BdG}
\end{equation}
where $\hat\xi_p(\bk)$ and $\hat\xi_h(\bk)$ are (diagonal matrix) particle and hole dispersion respectively, and they satisfy $-\hat\xi_h(-\bk)=\hat\xi_p(\bk)$. Note that the summation is now restricted to the folded Brillouin zone (fBZ). The mean field gap $\hat\Delta$ takes the following matrix form
\begin{equation}
  \hat\Delta=\begin{pmatrix}
         \overline\Delta_{0}f_{\bk} &  \overline\Delta_{\bQ}f_{\bk-\frac{\bQ}{2}} & 0 & 0 & 0 & 0 & 0 &  \overline\Delta_{-\bQ}f_{\bk+\frac{\bQ}{2}}\\
         \overline\Delta_{\bQ}f_{\bk+\frac{\bQ}{2}} & 0 & 0 & 0 & 0 & 0 & \overline\Delta_{-\bQ}f_{\bk+\frac{3\bQ}{2}} &  \overline\Delta_{0} f_{\bk+\bQ}\\
        0 & 0 & 0 & 0 & 0 & \overline\Delta_{-\bQ}f_{\bk+\frac{5\bQ}{2}} &  \overline\Delta_{0}f_{\bk+2\bQ} &  \overline\Delta_{\bQ}f_{\bk+\frac{3\bQ}{2}}\\
        0 & 0 & 0 & 0 & \overline\Delta_{-\bQ}f_{\bk+\frac{7\bQ}{2}} &  \overline\Delta_{0}f_{\bk+3\bQ} &  \overline\Delta_{\bQ}f_{\bk+\frac{5\bQ}{2}} & 0\\
        0 & 0 & 0 & \overline\Delta_{-\bQ}f_{\bk+\frac{9\bQ}{2}} &  \overline\Delta_{0}f_{\bk+4\bQ} &  \overline\Delta_{\bQ}f_{\bk+\frac{7\bQ}{2}} & 0 & 0\\
        0 & 0 & \overline\Delta_{-\bQ}f_{\bk+\frac{11\bQ}{2}} &  \overline\Delta_{0}f_{\bk+5\bQ} &  \overline\Delta_{\bQ}f_{\bk+\frac{9\bQ}{2}} & 0 & 0 & 0\\
        0 & \overline\Delta_{-\bQ}f_{\bk+\frac{13\bQ}{2}} &  \overline\Delta_{0}f_{\bk+6\bQ} &  \overline\Delta_{\bQ}f_{\bk+\frac{11\bQ}{2}} & 0 & 0 & 0 & 0\\
        \overline\Delta_{-\bQ}f_{\bk+\frac{15\bQ}{2}} &  \overline\Delta_{0}f_{\bk+7\bQ} &  \overline\Delta_{\bQ}f_{\bk+\frac{13\bQ}{2}} & 0 & 0 & 0 & 0 & 0\\
    \end{pmatrix}.
\end{equation}

Diagonalizing Eq.\eqref{eq:BdG} results in 16 Bogoliubov bands--8 above Fermi level and 8 below Fermi level, and the whole bands are particle-hole symmetric, just as in the usual uniform superconductivity case. For the resulting 16 bands, we choose to present the band
the band just above the Fermi level which we denote by $E(\bk)$. In the main text we present $E(\bk)$ in the folded Brillouin zone. Here we present $E(\bk)$ in the extended BZ (i.e. the original BZ), from which the translation symmetry breaking becomes manifest. 

 In the simplest case when $\overline\Delta_{\bQ}=0$, the system is just a $d$-wave superconductor, for which the Bogoliubov spectrum can be plotted in the trivial basis [see Fig.\ref{fig:PDW_BdG}(c)] or in the folded basis [see Fig.\ref{fig:PDW_BdG}(d)]. The latter is nothing but repeated plot for the low energy part of the former. For a pure unidirectional $d$-wave PDW ($\overline\Delta_{\bQ}\neq0, \overline\Delta_0=0$), the Bogoniubov spectrum shows some residual Fermi surfaces that are not gapped by the PDW order parameter, as shown in Fig.\ref{fig:PDW_BdG}(e). In the general case when both $\overline\Delta_{\bQ}$ and $\overline\Delta_0$ are both nonzero, the system is generally gapped with the four $d$-wave notes from the uniform SC component, as shown in Fig.\ref{fig:PDW_BdG}(f). For comparison with the plots in the main text, we have highlighted using red rectangle the areas that are presented in the main text.

\subsection{2. Mean field gap equations} 
\label{sub:mean_field_gap_equations}

To obtain the mean field gap equations for the order parameters, we integrate out fermions, and restrict to the non-fluctuating part of the resulting effective action. The total action for our departure, which includes both the PDW and SC order parameters, fermions in the basis of Eq.\eqref{eq:basis1}, and the coupling between these two sectors, is given by 
\begin{equation}
    \begin{aligned}
        S=&\frac{1}{g_1}\sum_q\left(|\Delta_{\bQ}(q)|^2+|\Delta_{-\bQ}(q)|^2\right)+\frac{1}{g_2}\sum_q|\Delta_0(q)|^2\label{eq:action1}\\
   & +\frac{1}{8}\sum_{k,k'} (\Psi_{\up,k}^\dagger,\Psi^\text{T}_{\down,-k})\hat{\mathcal{K}}_{k,k'}\begin{pmatrix}
        \Psi_{\up,k'}\\
        {\Psi^\dagger}^\text{T}_{\down,-k'}
    \end{pmatrix}
    \end{aligned}
\end{equation}
and 
\begin{equation}
\hat{\mathcal{K}}_{k,k'}=
    \begin{pmatrix}
        \hat G_p^{-1}(k)\delta_{k,k'} & \hat\Phi(k,k')\\
        \hat\Phi^\dagger(k',k) & \hat G_h^{-1}(k)\delta_{k,k'}
    \end{pmatrix}.\label{eq:mathcalK}
\end{equation}
Note that here we have included fluctuations by making $\Delta_{\overline{\bq}}$ $q$-dependent. We adopt imaginary time formalism so $q=(\omega_n,\bq)$ is the combination of both Matsubara frequency and momentum.
Therefore $\sum_k\equiv \frac{T}{V}\sum_{k}=T\sum_{n}\int_\text{BZ} \frac{d^2\bk}{(2\pi)^2}$. The momentum summation is over the original (unfolded) first Brillouin zone.
Here we have introduced two interactions, $g_1$ and $g_2$, which are responsible for the onsite of PDW and uniform SC respectively. 
The particle Green's function $\hat G_p$, the hole Green's function $\hat G_h$  introduced in Eq.\eqref{eq:mathcalK} are given by
\begin{equation}
    \hat{G}_p(k)=\begin{pmatrix}
        G_{p,0}(k) & 0 & 0 & 0 & 0 & 0 & 0 & 0\\
        0 & G_{p,1}(k) & 0 & 0 & 0 & 0 & 0 & 0\\
        0 & 0 & G_{p,2}(k) & 0 & 0 & 0 & 0 & 0\\
        0 & 0 & 0 & G_{p,3}(k) & 0 & 0 & 0 & 0\\
        0 & 0 & 0 & 0 & G_{p,4}(k) & 0 & 0 & 0\\
        0 & 0 & 0 & 0 & 0 & G_{p,5}(k) & 0 & 0\\
        0 & 0 & 0 & 0 & 0 & 0 & G_{p,6}(k) & 0\\
        0 & 0 & 0 & 0 & 0 & 0 & 0 & G_{p,7}(k)\\
    \end{pmatrix},\label{eq:matrixGp}
\end{equation}
\begin{equation}
    \hat{G}_h(k)=\begin{pmatrix}
        G_{h,0}(k) & 0 & 0 & 0 & 0 & 0 & 0 & 0\\
        0 & G_{h,1}(k) & 0 & 0 & 0 & 0 & 0 & 0\\
        0 & 0 & G_{h,2}(k) & 0 & 0 & 0 & 0 & 0\\
        0 & 0 & 0 & G_{h,3}(k) & 0 & 0 & 0 & 0\\
        0 & 0 & 0 & 0 & G_{h,4}(k) & 0 & 0 & 0\\
        0 & 0 & 0 & 0 & 0 & G_{h,5}(k) & 0 & 0\\
        0 & 0 & 0 & 0 & 0 & 0 & G_{h,6}(k) & 0\\
        0 & 0 & 0 & 0 & 0 & 0 & 0 & G_{h,7}(k)\\
    \end{pmatrix}\label{eq:matrixGh}
\end{equation}
where the elementary Green's functions are given by
\begin{equation}
    G_{p,m}(k)=\frac{1}{i\epsilon_n-\xi_{\bk+m\bQ}}, ~~G_{h,m}(k)=\frac{1}{i\epsilon_n+\xi_{-\bk+m\bQ}}.\label{eq:phGreen}
\end{equation}
For the pairing vertex matrix, 
it is more convenient to decompose $\hat\Phi(k,k')$ as $\hat\Phi(k,k')=\hat\Phi_{\bQ}(k,k')+\hat\Phi_{-\bQ}(k,k')+\hat\Phi_{0}(k,k')$, where each of the components is 
\begin{equation}
    \hat\Phi_{\bQ}(k,k')=\Delta_{\bQ}(q)\begin{pmatrix}
        0 &  f_{\bk-\frac{\bQ}{2}} & 0 & 0 & 0 & 0 & 0 &  0\\
         f_{\bk+\frac{\bQ}{2}} & 0 & 0 & 0 & 0 & 0 & 0 &  0\\
        0 & 0 & 0 & 0 & 0 & 0 &  0 &  f_{\bk+\frac{3\bQ}{2}}\\
        0 & 0 & 0 & 0 & 0 &  0 &  f_{\bk+\frac{5\bQ}{2}} & 0\\
        0 & 0 & 0 & 0 &  0 &  f_{\bk+\frac{7\bQ}{2}} & 0 & 0\\
        0 & 0 & 0 &  0 &  f_{\bk+\frac{9\bQ}{2}} & 0 & 0 & 0\\
        0 & 0 &  0 &  f_{\bk+\frac{11\bQ}{2}} & 0 & 0 & 0 & 0\\
        0 &  0 &  f_{\bk+\frac{13\bQ}{2}} & 0 & 0 & 0 & 0 & 0\\
    \end{pmatrix},\label{eq:PhiQ}
\end{equation}
\begin{equation}
    \hat\Phi_{-\bQ}(k,k')=\Delta_{-\bQ}(q)\begin{pmatrix}
        0 & 0 & 0 & 0 & 0 & 0 & 0 &   f_{\bk+\frac{\bQ}{2}}\\
        0 & 0 & 0 & 0 & 0 & 0 &  f_{\bk+\frac{3\bQ}{2}} &  0\\
        0 & 0 & 0 & 0 & 0 &  f_{\bk+\frac{5\bQ}{2}} &  0 & 0\\
        0 & 0 & 0 & 0 &  f_{\bk+\frac{7\bQ}{2}} &  0 & 0 & 0\\
        0 & 0 & 0 &  f_{\bk+\frac{9\bQ}{2}} &  0 & 0 & 0 & 0\\
        0 & 0 &  f_{\bk+\frac{11\bQ}{2}} &  0 & 0 & 0 & 0 & 0\\
        0 &  f_{\bk+\frac{13\bQ}{2}} &  0 & 0 & 0 & 0 & 0 & 0\\
         f_{\bk+\frac{15\bQ}{2}} &  0 & 0 & 0 & 0 & 0 & 0 & 0\\
    \end{pmatrix},\label{eq:PhimQ}
\end{equation}
\begin{equation}
    \hat\Phi_{0}(k,k')=\Delta_{0}(q)\begin{pmatrix}
         f_{\bk} & 0 & 0 & 0 & 0 & 0 & 0 &  0\\
        0 & 0 & 0 & 0 & 0 & 0 & 0 &   f_{\bk+\bQ}\\
        0 & 0 & 0 & 0 & 0 & 0 &   f_{\bk+2\bQ} & 0\\
        0 & 0 & 0 & 0 & 0 &   f_{\bk+3\bQ} & 0 & 0\\
        0 & 0 & 0 & 0 &   f_{\bk+4\bQ} & 0 & 0 & 0\\
        0 & 0 & 0 &   f_{\bk+5\bQ} & 0 & 0 & 0 & 0\\
        0 & 0 &   f_{\bk+6\bQ} & 0 & 0 & 0 & 0 & 0\\
        0 &   f_{\bk+7\bQ} & 0 & 0 & 0 & 0 & 0 & 0\\
    \end{pmatrix}.\label{eq:Phi00}
\end{equation}
Here we have used $q=k-k'$

Integrating out the fermions in Eq.\eqref{eq:action1} results in an effective action for the pairing fields only which we write as $S=S_\text{MF}+S_\text{FL}$.
The former does not contain fluctuations and thus comes from the diagonal term $\delta_{k,k'}\hat\Phi(k,k)$, which we name as $\overline\Phi$ hereafter. Explicitly, we can also decompose $\overline\Phi$ as $\overline\Phi=\overline\Phi_0+\overline\Phi_{\bQ}+\overline\Phi_{-\bQ}$ and the three components are given by taking $q=0$ in Eqs.\eqref{eq:PhiQ}, \eqref{eq:PhimQ} and \eqref{eq:Phi00}.
If we define $q=k'-k$, $S_\text{MF}$ corresponds to the $q=0$ part. Consequently, the ingredients in the matrix $\overline\Phi$ are the mean-field gaps $\overline\Delta_0$ and $\overline\Delta_{\pm\bQ}$.
In contrast, the fluctuating part comes from $q\neq0$ contribution and we denote it by $\tilde\Phi(k,k')$, which includes $h_{0}(q)$ and  $h_{\pm\bQ}(q)$. Explicitly, we have
\begin{equation}
    \begin{aligned}
        S_\text{MF}&=\frac{1}{g_1}\left(|\overline\Delta_{\bQ}|^2+|\overline\Delta_{-\bQ}|^2\right)+\frac{1}{g_2}|\overline\Delta_0|^2\\
        &~~~-\frac{1}{8}\text{Tr}\ln\left(1+\mathcal{G}_0\overline\Sigma\right),\\
        S_\text{FL}&=\frac{1}{g_1}\sum_q\left(|\overline\Delta_{\bQ}h_{\bQ}(q)|^2+|\overline\Delta_{- \bQ}h_{-\bQ}(q)|^2\right)\\
        &~~~+\frac{1}{g_2}\sum_q |\overline\Delta_0h_0(q)|^2+\frac{1}{8}\text{Tr}\sum_{n=1}^\infty\frac{(-\overline{\mathcal{G}}\tilde\Sigma)^n}{n}.
    \end{aligned}\label{eq:action2}
\end{equation}
The free Green's function $\mathcal{G}_0$, the mean-field anomalous self energy $\overline\Sigma$, the full Green's function $\mathcal{G}$ and the fluctuating self energy $\tilde\Sigma$ are given by  
\begin{equation}
    [{\mathcal{G}}_0]_{k,k'}=\delta_{k,k'}\begin{pmatrix}
        \hat G_p(k) & 0\\
        0 & \hat G_h(k)
    \end{pmatrix}, 
    [\overline{\Sigma}]_{k,k'} =\delta_{k,k'}
    \begin{pmatrix}
        0 & \overline \Phi\\
        {\overline \Phi}^\dagger & 0\\
    \end{pmatrix}
\end{equation}
and
 \begin{equation}
     \overline{\mathcal{G}}=\left(\mathcal{G}_0^{-1}+\overline{\Sigma}\right)^{-1}, ~ [\tilde\Sigma]_{k,k'}=\begin{pmatrix}
         0 & \tilde\Phi(k,k')\\
         \tilde{\Phi}^\dagger(k',k) & 0
     \end{pmatrix}
     .\label{eq:Gh}
 \end{equation}
 respectively. 

 The mean field gap equations are the saddle point equations for $S_\text{MF}$, namely
 \begin{equation}
    \begin{aligned}
         \frac{\delta S_\text{MF}}{\delta {\overline\Delta}^*_{\pm\bQ}}&=\frac{\overline\Delta_{\pm\bQ}}{g_1}-\frac{1}{8}\text{Tr}\left\{{\mathcal{G}}\frac{\delta \overline{\Sigma}}{\delta {\overline\Delta}^*_{\pm\bQ}}\right\}=0,\\
          \frac{\delta S_\text{MF}}{\delta {\overline\Delta}^*_{0}}&=\frac{\overline\Delta_{0}}{g_2}-\frac{1}{8}\text{Tr}\left\{{\mathcal{G}}\frac{\delta \overline{\Sigma}}{\delta {\overline\Delta}^*_{0}}\right\}=0.
    \end{aligned}\label{eq:gap2}
\end{equation}
The trace operation combines both taking matrix trace and summing all frequency-momentum. 
For a given set of $g_1$ and $g_2$, the gap functions can be found from simple numerical iteration. But it is more convenient in our purpose to fix the gap magnitudes $|\overline\Delta_{\overline\bq}|$ and solve for $g_{1,2}$. Therefore, we use Eq.\eqref{eq:gap2} to determine $g_{1,2}$ and $\varphi_{\overline{\bq}}$ for a given set of $|\overline\Delta_{\overline\bq}|$. 
Note that since both $\overline\Delta_{\bQ}$ and $\overline\Delta_{-\bQ}$ are induced by the same $g_1$ interaction, we will have to choose $|\overline\Delta_{\bQ}|=|\overline\Delta_{-\bQ}|$. 

\begin{figure}
    \includegraphics[width=8.5cm]{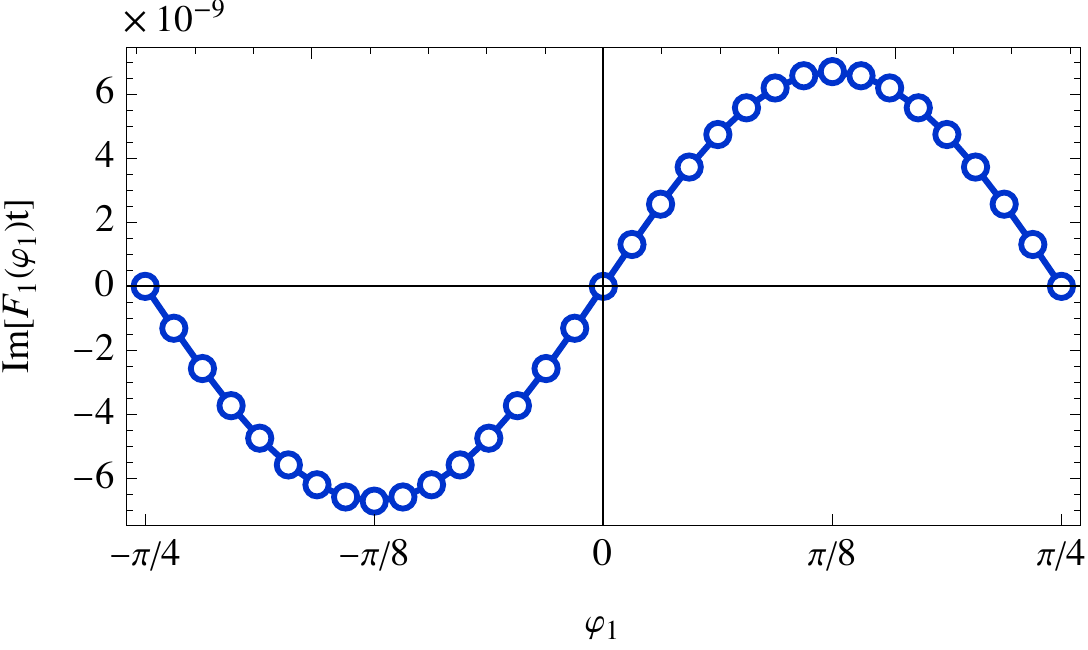}
    \caption{ $\text{Im}F_1(\varphi_1)$ defined in Eq.\eqref{eq:MFgap_F}, for $|\overline\Delta_{\bQ}|=0.15$. Here we have defined $\varphi_1=\varphi_{\bQ}-\varphi_{-\bQ}$. In the absence of $\overline\Delta_0$, the mean-field gap equation has solutions only when Im$F$ vanishes. }\label{fig:phi1}
\end{figure}

In fact, the mean field gap equations do not always admit a set of solution for arbitrary $\varphi_{\overline{\bq}}$. To see this, we consider the simplest case when $\overline\Delta_0=0$. In this case, only $\varphi_{\bQ}-\varphi_{-\bQ}$ enters the mean field equations. The gap equation for $g_1$ can be rewritten as
\begin{equation}
    F_1(\varphi_{\bQ}-\varphi_{-\bQ}):=\frac{1}{8{\overline\Delta}_{\bQ}}\text{Tr}\left.\left\{\overline{\mathcal{G}}\frac{\delta \overline{\Sigma}}{\delta {\overline\Delta}^*_{\bQ}}\right\}\right|_{|\overline\Delta_{\overline\bq}|\text{ fixed}}.\label{eq:MFgap_F}
\end{equation}
Then the nonlinear gap equation for $\Delta_{\bQ}$ becomes simply $g_1F_1(\varphi_{\bQ}-\varphi_{-\bQ})=1$.
Since $g_1$ is real, this equation has solution only when $\text{Im}[F(\varphi_{\bQ}-\varphi_{-\bQ})]=0$. In Fig.\ref{fig:phi1} we show the plot of $\text{Im}[F(\varphi_{\bQ}-\varphi_{-\bQ})]$ as a function of $\varphi_{\bQ}-\varphi_{-\bQ}$, from which we see that $\text{Im}F$ depends on $\varphi_{\bQ}-\varphi_{-\bQ}$ through $\sin[4(\varphi_{\bQ}-\varphi_{-\bQ})]$, so the mean field gap equation has solution only when $\varphi_{\bQ}-\varphi_{-\bQ}=0,\pm\pi/4, ...$. We will see below that $\varphi_{\bQ}-\varphi_{-\bQ}=0$ mod $n\pi/2$ corresponds to a stable solution. This can be examined since with the commensurate ordering wavevector  $8\bQ=0$, there is an allowed term in the free energy $\upsilon[(\overline\Delta_{\bQ}\overline\Delta_{-\bQ})^4e^{i4(\varphi_{\bQ}-\varphi_{-\bQ})}+h.c.]$ and if $\upsilon<0$ the systems favors $\varphi_{\bQ}-\varphi_{-\bQ}=0~\text{mod}~n\pi/2$; while if $\upsilon>0$ we have $\varphi_{\bQ}-\varphi_{-\bQ}=\pi/4~\text{mod}~n\pi/2$ instead.

When a nonzero $\overline\Delta_0$ is also present, one can follow the same strategy to study for what values of $\varphi_{\overline{\bq}}$ Eq.\eqref{eq:gap2} has solutions.
But perhaps a more transparent way is to obtain  the effective Ginzburg-Landau free energy density by expanding $S_\text{ML}$ to quartic order
\begin{equation}
    \begin{aligned}
        \mathcal{F}_\text{GL}=&\alpha_1 (|\overline\Delta_{\bQ}|^2+|\overline\Delta_{-\bQ}|^2)+\alpha_2|\overline\Delta_{0}|^2\\
        &+\beta_1(|\overline\Delta_{\bQ}|^4+|\overline\Delta_{-\bQ}|^4)+\beta_2|\overline\Delta_{0}|^4\\
        &+\beta_3|\overline\Delta_{\bQ}|^2|\overline\Delta_{-\bQ}|^2
        +\beta_4|\overline\Delta_0|^2\left(|\overline\Delta_{\bQ}|^2+|\overline\Delta_{-\bQ}|^2\right)\\
        &+\beta_5\left(\overline\Delta_0^2\overline\Delta_{-\bQ}\overline\Delta_{\bQ}+h.c.\right)+ ... 
    \end{aligned}\label{eq:FGL}
\end{equation}
If $\beta_5>0$ the system favors 
\begin{equation}
    \begin{aligned}
        \varphi_{\bf 0}-\frac{\varphi_{\bQ}+\varphi_{-\bQ}}2 =
        \frac \pi 2+n\pi,
        ~~~\varphi_{\bQ}-\varphi_{-\bQ}=\frac{m\pi}{2}. 
    \end{aligned}\label{eq:phi1phi2}
\end{equation}
Unlike the PDW case, a nonzero $\varphi_{\bf 0}-\frac{\varphi_{\bQ}+\varphi_{-\bQ}}2$ implies a true TRSB in the system. As we will see below in our analysis of the collective modes in this system, Eq.\eqref{eq:phi1phi2} is also consistent with the stability of the phase modes, i.e. having choices other than Eq.\eqref{eq:phi1phi2} leads to unstable phase modes. 


\subsection{3. Evidence of TRSB: explicit calculation of $\beta_5$} 
\label{sub:evidence_of_trsb_explicit_calculation_of_}

\begin{figure*}
    \includegraphics[width=13cm]{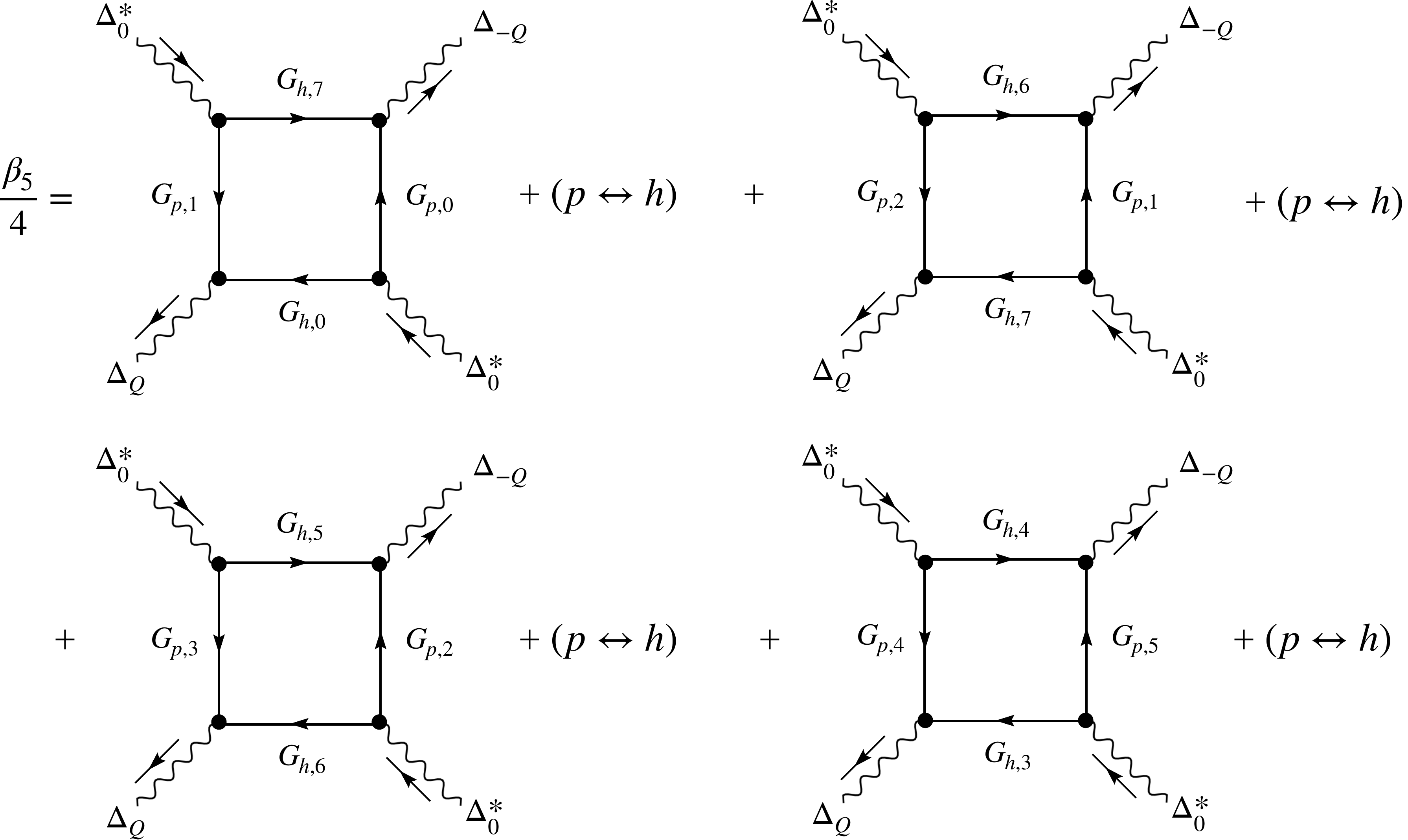}
    \caption{Diagrammatic representation of the contributions in $\beta_5$ term. Here the wavy lines represent the superconducting order parameters, which carry finite momentum $\pm\bQ$ for the PDW case, and zero momentum for the uniform SC case. The momentum is conserved (mod $8\bQ$) at each vertex. The solid lines are for the particle and hole Green's functions $G_{p,n}$ and $G_{h,n}$, defined in Eq.\eqref{eq:phGreen}. The black dot at each vertex represents the form factor $f_{\bk}$ for the $d$-wave pairing symmetry.}\label{fig:beta5}
\end{figure*}

By explicitly expanding $S_\text{MF}$ to the quartic order, we found that the coefficient $\beta_5$ introduced in Eq.\eqref{eq:FGL} is given by
\begin{equation}
    \begin{aligned}
        \beta_5=4T\sum_{n}\int\frac{d^2\bk}{(2\pi)^2}&\left[f_0(\bk)f_1^2(\bk)f_2(\bk)G_{p,0}(k)G_{p,1}(k)G_{h,7}(k)(k)G_{h,0}(k)\right.\\
        +&~f_2(\bk)f_3^2(\bk)f_4(\bk)G_{p,1}(k)G_{p,2}(k)G_{h,6}(k)(k)G_{h,7}(k)\\
        +&~f_4(\bk)f_5^2(\bk)f_6(\bk)G_{p,2}(k)G_{p,3}(k)G_{h,5}(k)(k)G_{h,6}(k)\\
        +&~f_6(\bk)f_7^2(\bk)f_8(\bk)G_{p,3}(k)G_{p,4}(k)G_{h,4}(k)(k)G_{h,5}(k)\\
        +&~f_8(\bk)f_9^2(\bk)f_{10}(\bk)G_{p,4}(k)G_{p,5}(k)G_{h,3}(k)(k)G_{h,4}(k)\\
        +&~f_{10}(\bk)f_{11}^2(\bk)f_{12}(\bk)G_{p,5}(k)G_{p,6}(k)G_{h,2}(k)(k)G_{h,3}(k)\\
        +&~f_{12}(\bk)f_{13}^2(\bk)f_{14}(\bk)G_{p,6}(k)G_{p,7}(k)G_{h,1}(k)(k)G_{h,2}(k)\\
        +&\left.f_{14}(\bk)f_{15}^2(\bk)f_0(\bk)G_{p,7}(k)G_{p,0}(k)G_{h,0}(k)(k)G_{h,1}(k)\right].
    \end{aligned}
\end{equation}
Here we have used the abbreviation that $f_n(\bk)=f_{k+\frac{n}{2}\bQ}$.
We note that  half of the contributions can be related to the other half by interchanging the particle and hole Green's functions $G_{p,n}$ and $G_{h,n}$, which are defined in Eq.\eqref{eq:phGreen}. In Fig.\ref{fig:beta5} show the contributions to $\beta_5$ diagrammatically. We have found from numerical evaluations that all these 8 terms are equal. 

\begin{figure*}
    \includegraphics[width=8cm]{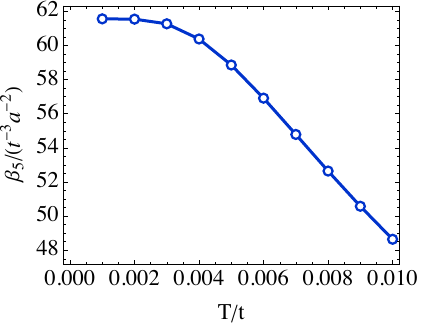}
    \caption{Numerical evaluation of $\beta_5$ as a function of $T$. Here $\beta_5$ is in units of $t^{-3}a^{-2}$ where $a$ is the lattice constant. }\label{fig:Nbeta5}
\end{figure*}

In Fig.\ref{fig:Nbeta5} we present the numerical calculation of $\beta_5$ as a function of $T$, at the particular $1/8$ hole doping. In all temperature range considered, $\beta_5$ is positive, implying the phase difference between $\overline\Delta_0$ and $\overline\Delta_{\pm\bQ}$ has to be nonzero to minimize the free energy. Therefore, the system energetically prefers a TRSB ground state. 

\section{II. Fluctuation effects} 
\label{sec:ii_fluctuation_effects}

\subsection{1. Pure PDW} 
\label{sub:pure_pdw}
We now discuss collective modes from the fluctuating part $S_\text{FL}$. Let's first consider the case when only the unidirectional PDW $\Delta_{\pm\bQ}$ is present. Their mean field gap values are solved from Eq.\eqref{eq:gap2} by setting $\overline\Delta_0=0$.
Assuming the fluctuating fields $h_{\pm\bQ}(q)$ are small, we truncate the $S_\text{FL}$ at Gaussian level. It is convenient to work in the following basis for the fluctuating fields
\begin{equation}
     \zeta(q)=[\mathcal{A}_{\bQ}(q), \mathcal{A}_{-\bQ}(q), \theta_{\bQ}(q), \theta_{-\bQ}(q)]^T\label{eq:flucbasis1}
 \end{equation}
 where we remind that $\mathcal{A}_{\pm\bQ}$ are for amplitudes and $\theta_{\pm\bQ}$ are for phases.
In terms of $\zeta(q)$, $\tilde\Sigma$ in Eq.\eqref{eq:Gh} can be rewritten as 
\begin{equation}
    [\tilde\Sigma]_{k,k'}=\left(\hat{D}_{1},\hat{D}_{2},\hat{D}_{3},\hat{D}_{4}\right)\cdot \zeta(q).\label{eq:zetabasis}
\end{equation}
where the matrix components $\hat D_i$ are are
\begin{equation}
    \begin{aligned}
        &\hat{D}_{1}=\begin{pmatrix}
         0 & \overline{\Phi}_{\bQ}\\
         \overline{\Phi}_{\bQ}^\dagger & 0\\
     \end{pmatrix},~\hat{D}_{2}=\begin{pmatrix}
         0 & \overline{\Phi}_{-\bQ}\\
         \overline{\Phi}_{-\bQ}^\dagger & 0\\
     \end{pmatrix},~ \hat{D}_{3}=\begin{pmatrix}
         0 & i\overline{\Phi}_{\bQ}\\
         -i\overline{\Phi}_{\bQ}^\dagger & 0\\
     \end{pmatrix},~\hat{D}_{4}=\begin{pmatrix}
         0 & i\overline{\Phi}_{-\bQ}\\
         -i\overline{\Phi}_{-\bQ}^\dagger & 0\\
     \end{pmatrix},
    \end{aligned}\label{eq:Dmatrix}
\end{equation}
and $\overline\Phi_{\bQ}$ and $\overline\Phi_{-\bQ}$ are the $q=0$ components of $\hat\Phi_{\bQ}$ and $\hat\Phi_{-\bQ}$ given in Eqs.\eqref{eq:PhiQ} and \eqref{eq:PhimQ} respectively.

The fluctuating part, to quadratic order, is 
\begin{equation}
    \begin{aligned}
        S^{(2)}_\text{FL}
        &=\sum_q \zeta^T(q) \hat\Gamma^{-1}(q) \zeta(-q).
    \end{aligned}
\end{equation}
Here
\begin{equation}
     \hat\Gamma^{-1}(q)=\hat U^{-1}+\hat\Pi(q)\label{eq:invGamma}
 \end{equation} 
is the inverse RPA type interaction dressed by collective modes and
  the inverse interaction matrix $\hat U^{-1}$
 \begin{equation}
     \hat U^{-1}=\frac{1}{g_1}\begin{pmatrix}
         |\overline\Delta_{\bQ}|^2 & 0 & 0 & 0\\
         0 & |\overline\Delta_{-\bQ}|^2 & 0 & 0\\
         0 & 0 & |\overline\Delta_{\bQ}|^2 & 0\\
         0 & 0 & 0 & |\overline\Delta_{-\bQ}|^2\\
     \end{pmatrix}.
 \end{equation}
and the four-by-four matrix $\hat\Pi(q)$ is given by
 \begin{equation}
     \Pi_{i,j}(q)=\frac{1}{16}\sum_{k}\text{Tr}\left[\overline{\mathcal{G}}_{k+\frac{q}{2}}\hat D_i\overline{\mathcal{G}}_{k-\frac{q}{2}}\hat D_j\right]\label{eq:M1}
 \end{equation}
 where $i,j=1,2,3,4$ and the momentum summation is for the whole original BZ. From this definition it is easy to see $\Pi_{i,j}(q)=\Pi_{j,i}(-q)$.
 In practice, evaluations of the $\hat\Pi(q)$ matrix and hence $\hat\Gamma^{-1}$ requires solving the mean field gap equations \eqref{eq:gap2} first, which admits solutions only for a discrete set of $\varphi_1$ as we discussed above. Thus, the matrix $\hat\Gamma^{-1}$ depends on $\varphi_1$ implicitly.
Depending on whether or not the matrix element denotes the amplitude-phase coupling, $\Pi_{ij}(q)$ have distinct behaviors: 
\begin{enumerate}
    \item If $\Pi_{i,j}$ couples amplitude and phase modes (e.g. $\Pi_{1,3}$), it obeys $\Pi_{i,j}(q)=-\Pi_{j,i}(q)$ and $\Pi_{i,j}(q)=-\Pi_{i,j}(-q)$.
    \item If $\Pi_{i,j}$ is the coupling between amplitude modes, or between phase modes (e.g. $\Pi_{1,2}$ or $\Pi_{3,4}$), it obeys $\Pi_{i,j}(q)=\Pi_{j,i}(q)$ and $\Pi_{i,j}(q)=\Pi_{i,j}(-q)$.
\end{enumerate}
Moreover, all $\Pi_{i,j}(q)$ are real. Thus, due to amplitude-phase coupling,  the matrix $\hat\Gamma^{-1}(q)$ is in general a mixture of symmetric and anti-symmetric real matrix whose eigenvalues are in general complex-valued. Perhaps a better way to understand the fact that amplitude-phase coupling $\Pi_{i,j}(q)$ being real and odd in $q$ is to move back to the space-time coordinates. We remind that both $\mathcal{A}_{\overline\bq}(R)$ and $\theta_{\overline\bq}(R)$ are defined as real functions, therefore, the term $\tilde \Pi_{{\overline\bq},{\overline\bq}'}(R_1,R_2)\mathcal{A}_{{\overline\bq}}(R_1)\theta_{{\overline\bq}'}(R_2)$ is consistent with a general amplitude-phase coupling $\sim i\mathcal{A}_{{\overline\bq}}(R_1)\theta_{{\overline\bq}'}(R_2)$ only when $\tilde \Pi_{{\overline\bq},{\overline\bq}'}(R_1,R_2)$ is purely imaginary. This is fulfilled in the present case as the Fourier transform of the real-valued $\Pi_{i,j}(q)$, when odd in $q$, is guaranteed to be purely imaginary. 

The signals of various long wavelength ($\bq\to0$) amplitude and phase modes are reflected by the proximity of the eigenvalues of $\hat\Gamma(\omega,\bq=0)$ (in real frequency axis) to nearby poles on the complex frequency plane. For a simple $s$-wave uniform SC order $\overline\Delta_{s,0}$ at zero temperature $T=0$, there are two poles along the real frequency axis for $\omega\in[0,\infty)$: One is located at $\omega=0$ and corresponds to the gapless BAG mode; the other is located at $\omega=2\overline\Delta_{s,0}$ which is from the massive Higgs mode. At finite temperature, the Higgs pole moves down to the lower half plane. The proximity to these poles gives rise to sharp peak features in the spectrum of related response functions. 
Here in our PDW system we study different eigenvalues of $\hat\Gamma(\omega)$ at finite but small temperatures. 


Here we focus on small $q$ limit. In this case, as long as the mean field gaps $\overline\Delta_{\overline\bq}\ll E_F\sim t$, the amplitude-phase couplings are small. This is because 
 the matrix entries $M_{ij}$ between amplitude and phase fluctuations are odd functions in $q$, as we mentioned above, and they vanish in the limit when $q\to0$. Namely, for $q=0$, can write $\hat\Gamma^{-1}$ as
\begin{equation}
    \hat \Gamma^{-1}(0)=\begin{pmatrix}
        \hat\Gamma^{-1}_A(0) & 0\\
        0 & \hat\Gamma^{-1}_p(0)
    \end{pmatrix}.\label{eq:A-p}
\end{equation}
For small $q$, we work in the regime where the off-diagonal terms in Eq.\eqref{eq:A-p} are still negligible, 
which allows us to treat the amplitude and phase modes separately.

We first discuss the phase mode in this decoupling limit, for which we find that the phase part must take the following form,
\begin{equation}
   \hat\Gamma^{-1}_{p}(0)= \begin{pmatrix}
        A & -A\\
        -A & A
    \end{pmatrix}.\label{eq:PDWphase}
\end{equation}
where $A=M_{3,4}(0)$ depending implicitly on the relative phase $\varphi_1$.

\begin{itemize}
    \item {\bf Proof.} First of all, we notice that since $\Gamma^{-1}_{3,3}$ and $\Gamma^{-1}_{4,4}$ describes the phase fluctuations which are gapless if they are decoupled. However, they do not vanish in the limit when $q\to0$ in our case. To see this we examine $\Gamma^{-1}_{3,3}(q\to0)$:
\begin{equation}
    \Gamma^{-1}_{3,3}(0)=\frac{1}{16}\sum_k \text{Tr}[\overline{\mathcal{G}}_k \hat D_3\overline{\mathcal{G}}_k \hat D_3]+\frac{|\overline\Delta_{\bQ}|^2}{g_1}.
\end{equation}
To investigate this value, we use the first line of Eq.\eqref{eq:gap2}, which is equivalent to 
\begin{equation}
    \begin{aligned}
        &\frac{\overline \Delta_{\bQ}}{g_1}-\frac{i}{8}\sum_k\text{Tr}\left[\overline{\mathcal{G}}_k\frac{\delta \hat D_3}{\delta\overline\Delta_{\bQ}^*}\right]=\frac{\overline \Delta_{\bQ}}{g_1}-\frac{i}{8}\frac{\delta}{\delta\overline\Delta_{\bQ}^*}\sum_k\text{Tr}\left[\overline{\mathcal{G}}_k{ \hat D_3}\right]+\frac{i}{8}\sum_k\text{Tr}\left[\frac{\delta \overline{\mathcal{G}}_k}{\delta\overline\Delta_{\bQ}^*}\hat D_3\right]=0.
    \end{aligned}\label{eq:M33}
\end{equation}
Using $|\overline \Delta_{\bQ}|=|\overline \Delta_{-\bQ}|$ it is easy to show
\begin{equation}
    \begin{aligned}
        \text{Tr}\left[\overline{\mathcal{G}}_k{ \hat D_3}\right]&=0, \\
        \frac{\delta[\overline{\mathcal{G}}_k^{-1}\overline{\mathcal{G}}_k\overline{\mathcal{G}}_k^{-1}]}{\delta\overline\Delta_{\bQ}^*}&=i\frac{\delta \hat D_3}{\delta\overline\Delta_{\bQ}^*}+i\frac{\delta \hat D_4}{\delta\overline\Delta_{\bQ}^*} 
    \end{aligned}
\end{equation}
The second line above indicates
\begin{equation}
    \frac{\delta\overline{\mathcal{G}}_k}{\delta\overline\Delta_{\bQ}^*}=-i\overline{\mathcal{G}}_k\frac{\delta \hat D_3}{\delta\overline\Delta_{\bQ}^*} \overline{\mathcal{G}}_k-i\overline{\mathcal{G}}_k\frac{\delta \hat D_4}{\delta\overline\Delta_{\bQ}^*} \overline{\mathcal{G}}_k
\end{equation}
and therefore from Eq.\eqref{eq:M33}
\begin{equation}
    \frac{|\overline \Delta_{\bQ}|^2}{g_1}+\frac{1}{8}\sum_k\text{Tr}\left[\overline{\mathcal{G}}_k\overline\Delta_{\bQ}^*\frac{\delta (\hat D_3+\hat D_4)}{\delta\overline\Delta_{\bQ}^*} \overline{\mathcal{G}}_k\hat D_3\right]=0.\label{eq:D3D4}
\end{equation}
Because 
\begin{equation}
    \overline\Delta_{\bQ}^*\frac{\delta \hat D_3}{\delta\overline\Delta_{\bQ}^*}=\frac{1}{2}\left(\hat D_3-i\hat D_1\right), ~\overline\Delta_{\bQ}^*\frac{\delta \hat D_4}{\delta\overline\Delta_{\bQ}^*}=\frac{1}{2}\left(\hat D_4-i\hat D_2\right)
\end{equation}
and 
\begin{equation}
    \text{Tr}\left[\overline{\mathcal{G}}_k\hat D_1 \overline{\mathcal{G}}_k\hat D_3\right]=\text{Tr}\left[\overline{\mathcal{G}}_k\hat D_2 \overline{\mathcal{G}}_k\hat D_3\right]=0 ~ \text{at each } k.\label{eq:TrD1D3}
\end{equation}
If $|\overline \Delta_{\bQ}|=|\overline \Delta_{-\bQ}|$, we obtain from Eq.\eqref{eq:D3D4} that
\begin{equation}
     \begin{aligned}
         \frac{|\overline \Delta_{\bQ}|^2}{g_1}&+\frac{1}{16}\sum_k\text{Tr}\left[\overline{\mathcal{G}}_k\hat D_3 \overline{\mathcal{G}}_k\hat D_3\right] +\frac{1}{16}\sum_k\text{Tr}\left[\overline{\mathcal{G}}_k\hat D_4 \overline{\mathcal{G}}_k\hat D_3\right]=0.
     \end{aligned}
\end{equation}
This relation immediately implies 
\begin{equation}
     \Gamma^{-1}_{3,3}(0)+\Gamma^{-1}_{3,4}(0)=0
 \end{equation} 
 and hence Eq.\eqref{eq:PDWphase}.
\end{itemize}

The same form of matrix also appears in the discussion of phase fluctuations in a two component uniform SC system where the Leggett emerges due to the off-diagonal coupling. 
Eq.\eqref{eq:PDWphase} has obvious eigenvalues, $0$ and $2A$, which correspond to the following two eigen modes:
\begin{equation}
    \begin{aligned}
       \theta_\pm(q)=\frac{1}{\sqrt{2}}\left[\theta_{\bQ}(q)\pm\theta_{-\bQ}(q)\right].\\
    \end{aligned}\label{eq:phase12}
\end{equation}
The first phase mode $\theta_+$ has eigenvalue $0$ in the $q\to0$ limit and thus corresponds to a gapless BAG mode, while the second mode, if $A>0$, corresponds to a gapped phase mode for the composite commensurate CDW order (reminiscent of Leggett mode in a two component SC case). If $A<0$, the second eigenvalue becomes negative, and the resultant action describes an unstable saddle point where fluctuations tend to destroy the system. 
Although the mean field Eq.\eqref{eq:gap2} has solutions for $\varphi_1=0,\pm\pi/4, ...$, we have confirmed that $A|_{\varphi_1=\pm\frac{\pi}{4}}<0$ while $A|_{\varphi_1=0}>0$. Therefore, from analyzing the stability of phase fluctuations, we arrive at the same conclusion as we discussed above [see Eq.\eqref{eq:phi1phi2}]: the system is stable only when $\varphi_1\equiv\varphi_{\bQ}-\varphi_{-\bQ}=0$ and other equivalent values. 

We now discuss the amplitude modes, neglecting the amplitude-phase coupling. By diagonalizing $\hat\Gamma_A^{-1}(q)$, we find the first eigen mode $\mathcal{A}_+$ corresponds to the two fluctuations moving in the same direction and the second one $\mathcal{A}_-$ is when they move in opposite directions, i.e. 
\begin{equation}
    \begin{aligned}
    \mathcal{A}_\pm(q)=\frac{1}{\sqrt{2}}\left[\mathcal{A}_{\bQ}(q)\pm\mathcal{A}_{-\bQ}(q)\right],
    \end{aligned}\label{eq:amplitude12}
\end{equation}
which are formally the same as the phase modes on Eq.\eqref{eq:phase12}.

\subsection{2. PDW+SC} 
\label{sub:2_pdw_sc}

When the uniform SC component is also present in addition to the PDW state, more collective modes from the SC fluctuations need to be included. The following discussion is parallel to what we have outlined above. 
We define a new basis that incorporate the $\mathcal{A}_0$ and $\theta_0$ from fluctuating SC, 
\begin{equation}
    \zeta'(q)=[\mathcal{A}_{\bQ}(q), \mathcal{A}_{-\bQ}(q),  \mathcal{A}_{0}(q),\theta_{\bQ}(q), \theta_{-\bQ}(q), \theta_0(q)]^T\label{eq:flucbasis2}
\end{equation}
and express $\tilde\Sigma$ as
\begin{equation}
    [\tilde\Sigma]_{k,k'}=\left(\hat{E}_{1},\hat{E}_{2},\hat{E}_{3},\hat{E}_{4},\hat{E}_{5},\hat{E}_{6}\right)\cdot{\zeta'}(q).\label{eq:fluct2}
\end{equation}
The explicit forms of these $\hat E$-matrices are given by
\begin{equation}
    \begin{aligned}
        &\hat{E}_{1}=\hat{D}_1,~\hat{E}_{2}=\hat{D}_2,~
     \hat{E}_{3}=\begin{pmatrix}
         0 & \overline{\Phi}_{0}\\
         \overline{\Phi}_{0}^\dagger & 0\\
     \end{pmatrix},\hat{E}_{4}=\hat{D}_3,~
     \hat{E}_{5}=\hat{D}_4, \hat{E}_{6}=\begin{pmatrix}
         0 & i\overline{\Phi}_{0}\\
         -i\overline{\Phi}_{0}^\dagger & 0\\
     \end{pmatrix}
    \end{aligned}\label{eq:Ematrix}
\end{equation}
where $\overline\Phi_0$ is the $q=0$ component of $\hat\Phi_0$ in Eq.\eqref{eq:Phi00}.


With the new basis, the fluctuating action truncated at the Gaussian level can be compactly written as
\begin{equation}
    S^{(2)}_\text{FL}=\sum_q {\zeta'}^T(q) \hat\Gamma^{-1}(q) \zeta'(-q)
\end{equation}
where the kernel $\hat\Gamma^{-1}(q)$, although being a $6\times6$ matrix now,  is still formally given by Eq.\eqref{eq:invGamma}.
In the present case the inverse bare interaction matrix becomes 
\begin{equation}
     \hat U^{-1}=\begin{pmatrix}
         \frac{|\overline\Delta_{\bQ}|^2}{g_1} & 0 & 0 & 0 & 0 & 0\\
         0 & \frac{|\overline\Delta_{-\bQ}|^2}{g_1} & 0 & 0 & 0 & 0\\
         0 & 0 & \frac{|\overline\Delta_{0}|^2}{g_2} & 0 & 0 & 0\\
         0 & 0 & 0 & \frac{|\overline\Delta_{\bQ}|^2}{g_1} & 0 & 0\\
         0 & 0 & 0 & 0 & \frac{|\overline\Delta_{-\bQ}|^2}{g_1} & 0\\
         0 & 0 & 0 & 0 & 0 & \frac{|\overline\Delta_{0}|^2}{g_2}\\
     \end{pmatrix},
 \end{equation}
and the $\hat\Pi$ matrix is now given by
\begin{equation}
     \Pi_{i,j}(q)=\frac{1}{16}\sum_{k}\text{Tr}\left[\overline{\mathcal{G}}_{k+\frac{q}{2}}\hat E_i\overline{\mathcal{G}}_{k-\frac{q}{2}}\hat E_j\right]\label{eq:M2}
 \end{equation}
 where $i,j=1,2,3,4,5,6$ and the summation of momentum is in the original BZ. Albeit taking a different form, the matrix elements of $\Pi(q)$ also obeys the properties that we discussed above. First of all, they are all real.  Moreover, when $\Pi_{i,j}$ denotes the amplitude-phase coupling, it is odd in $q$, and odd under $i\leftrightarrow j$; other wise it is even in $q$ and even under $i\leftrightarrow j$. In all cases it must obey $ \Pi_{i,j}(q)=\Pi_{j,i}(-q)$ as seen from its definition. 

 Again this implies that the amplitude-phase coupling vanishes in $q\to0$ limit and $\hat \Gamma ^{-1}(0)$ formally become that in Eq.\eqref{eq:A-p}, and the phase fluctuation part must take the form
 \begin{equation}
   \hat\Gamma^{-1}_{p}(0)= \begin{pmatrix}
        B & -B+C/2 & -C/2\\
        -B+/2 & B & -C/2\\
        -C/2 & -C/2 & C
    \end{pmatrix}.\label{eq:SCPDWphase}
\end{equation}
where $B=\Gamma^{-1}_{4,4}(0)$ and $C=\Gamma^{-1}_{6,6}(0)$ and both depend on the two relative phases $\varphi_{\bm{0}}$ and $\varphi_{\pm\bQ}$ implicitly. 
\begin{itemize}
    \item{\bf Proof. } We use the first saddle point equation as an example, and play the integrating-by-part trick to obtain
\begin{equation}
    \frac{\overline \Delta_{\bQ}}{g_1}-\frac{i}{8}\frac{\delta}{\delta\overline\Delta_{\bQ}^*}\sum_k\text{Tr}\left[\overline{\mathcal{G}}_k{ \hat E_4}\right]+\frac{i}{8}\sum_k\text{Tr}\left[\frac{\delta \overline{\mathcal{G}}_k}{\delta\overline\Delta_{\bQ}^*}\hat E_4\right]=0.\label{eq:M44}
\end{equation}
It is easy to show that $\text{Tr}\left[\overline{\mathcal{G}}_k{ \hat E_4}\right]=0$ at each $k$ so the middle term does not contribute. We will further stick to the case when $|\overline \Delta_{\bQ}|=|\overline \Delta_{-\bQ}|$. Moreover, because the uniform gap $\overline \Delta_0$ is coupled to $\overline\Delta_{\pm\bQ}$ through the gap equation, we can think  $\overline \Delta_0$ ($\overline \Delta_0^*$) is an implicit function of $\overline\Delta_{\bQ}$ ($\overline\Delta_{\bQ}^*$) such that $\delta \overline \Delta_0^*/\delta \overline\Delta_{\bQ}^*$ is nonzero. As a result, it is easy to obtain 
\begin{equation}
    \frac{\delta\overline{\mathcal{G}}_k}{\delta\overline\Delta_{\bQ}^*}=-i\overline{\mathcal{G}}_k\frac{\delta \hat E_4}{\delta\overline\Delta_{\bQ}^*} \overline{\mathcal{G}}_k-i\overline{\mathcal{G}}_k\frac{\delta \hat E_5}{\delta\overline\Delta_{\bQ}^*} \overline{\mathcal{G}}_k-i\overline{\mathcal{G}}_k\frac{\delta \hat E_6}{\delta\overline\Delta_{\bQ}^*} \overline{\mathcal{G}}_k.
\end{equation}
Substituting this relation back to the gap equation one gets, 
\begin{equation}
    \frac{|\overline \Delta_{\bQ}|^2}{g_1}+\frac{1}{8}\sum_k\text{Tr}\left[\overline{\mathcal{G}}_k\overline\Delta_{\bQ}^*\frac{\delta (\hat E_4+\hat E_5+\hat E_6)}{\delta\overline\Delta_{\bQ}^*} \overline{\mathcal{G}}_k\hat E_4\right]=0.\label{eq:D456D4}
\end{equation}
As before we have 
\begin{equation}
    \overline\Delta_{\bQ}^*\frac{\delta \hat E_4}{\delta\overline\Delta_{\bQ}^*}=\frac{1}{2}\left(\hat E_4-i\hat E_1\right), ~\overline\Delta_{\bQ}^*\frac{\delta \hat E_5}{\delta\overline\Delta_{\bQ}^*}=\frac{1}{2}\left(\hat E_5-i\hat E_2\right)
\end{equation}
and if $\overline \Delta_0$ ($\overline \Delta_0^*$) scales linearly with $\overline\Delta_{\bQ}$ ($\overline\Delta_{\bQ}^*$) we will have 
\begin{equation}
    \overline\Delta_{\bQ}^*\frac{\delta \hat E_6}{\delta\overline\Delta_{\bQ}^*}=\frac{1}{2}\left(\hat E_6-i\hat E_3\right).\label{eq:D63}
\end{equation}
Using the fact that 
\begin{equation}
    \text{Tr}\left[\overline{\mathcal{G}}_k\hat E_1 \overline{\mathcal{G}}_k\hat E_4\right]=\text{Tr}\left[\overline{\mathcal{G}}_k\hat E_2 \overline{\mathcal{G}}_k\hat E_4\right]=\text{Tr}\left[\overline{\mathcal{G}}_k\hat E_3 \overline{\mathcal{G}}_k\hat E_4\right]=0
\end{equation}
it is then easy to see from Eq.\eqref{eq:D456D4} that
\begin{equation}
     \begin{aligned}
         &\frac{|\overline \Delta_{\bQ}|^2}{g_1}+\frac{1}{16}\sum_k\text{Tr}\left[\overline{\mathcal{G}}_k\hat E_4 \overline{\mathcal{G}}_k\hat E_4\right]+\\
         &\frac{1}{16}\sum_k\text{Tr}\left[\overline{\mathcal{G}}_k\hat E_4 \overline{\mathcal{G}}_k\hat E_5\right]+\frac{1}{16}\sum_k\text{Tr}\left[\overline{\mathcal{G}}_k\hat E_4 \overline{\mathcal{G}}_k\hat E_6\right]=0,
     \end{aligned}
\end{equation}
which means
\begin{equation}
     \Gamma^{-1}_{4,4}(0)+\Gamma^{-1}_{4,5}(0)+\Gamma^{-1}_{4,6}(0)=0.
 \end{equation}

The only thing that needs to be justified in the above proof is whether  $\overline \Delta_0$ ($\overline \Delta_0^*$) scales linearly with $\overline\Delta_{\bQ}$ ($\overline\Delta_{\bQ}^*$). From the gap equations we have
\begin{equation}
    \frac{\overline\Delta_0}{\overline \Delta_{\bQ}}\frac{g_1}{g_2}=\frac{\sum_k \text{Tr}\left\{[G_{0}^{-1}(k)+\overline\Sigma]^{-1}\frac{\delta \hat{E}_6}{\delta \overline{\Delta}^*_{0}}\right\}}{\sum_k \text{Tr}\left\{[G_{0}^{-1}(k)+\overline\Sigma]^{-1}\frac{\delta \hat{E}_4}{\delta \overline{\Delta}^*_{0}}\right\}}=\mathcal{S}\left(\frac{\overline\Delta_0}{\overline \Delta_{\bQ}}\right).
\end{equation}
with $\mathcal{S}(...)$ being some function deduced from the ratio. Therefore, $\overline\Delta_0$ scales linearly with $\overline\Delta_{\bQ}$, and Eq.\eqref{eq:D63} is indeed true. 
Following similar routes, it is also straightforward to see 
\begin{equation}
     \begin{aligned}
         \Gamma^{-1}_{5,4}(0)+\Gamma^{-1}_{5,5}(0)+\Gamma^{-1}_{5,6}(0)=0,\\
         \Gamma^{-1}_{6,4}(0)+\Gamma^{-1}_{6,5}(0)+\Gamma^{-1}_{6,6}(0)=0,
     \end{aligned}
 \end{equation}
 and consequently 
 \begin{equation}
     \Gamma^{-1}_{6,4}(0)=\Gamma^{-1}_{6,5}(0).
 \end{equation}
 The last two equations complete the proof of Eq.\eqref{eq:SCPDWphase}.
\end{itemize}

\begin{figure}
    \includegraphics[width=16cm]{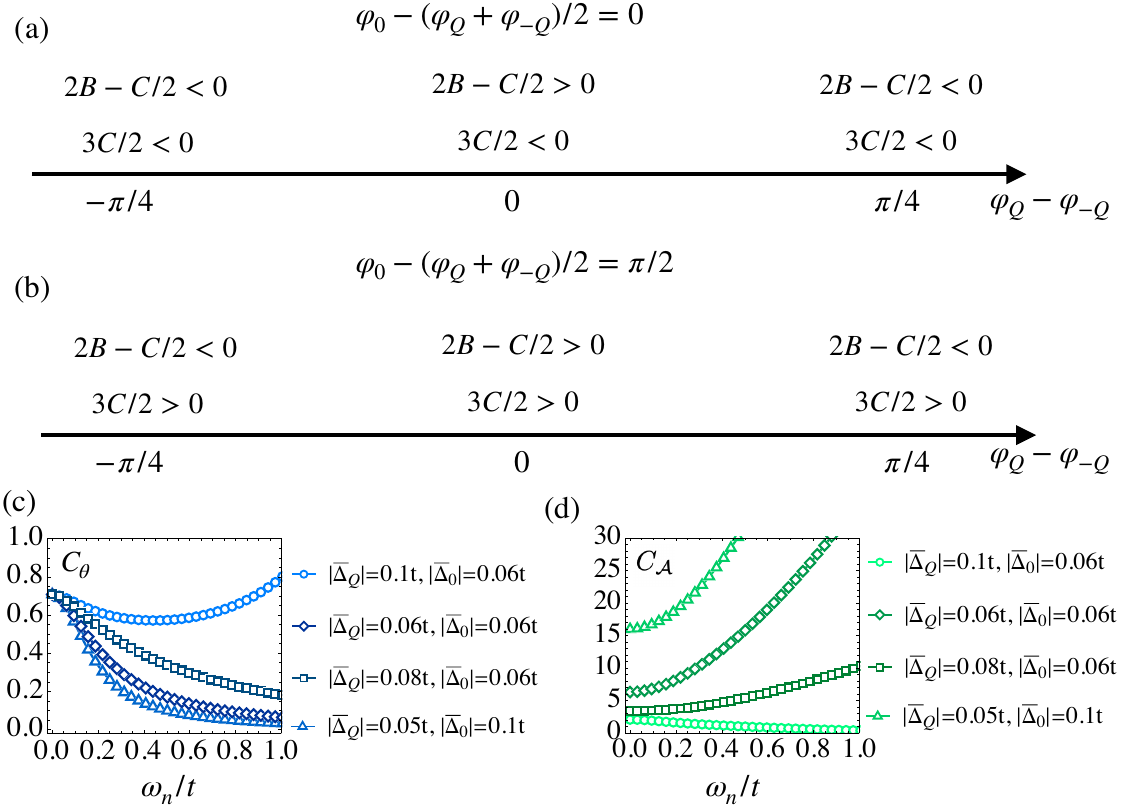}
    \caption{(a) The sign of the two nonzero eigenvalues of $\hat\Gamma_p^{-1}(0)$, $3C/2$ and $2B-C/2$,  as a function of $\varphi_1$ for $\varphi_2-\varphi_1/2=0$. Note that due to the constraint from the mean field gap equation, $\varphi_1$ can only take the distinct values $0,\pm\pi/4, ...$.
    (b) Same as (a) but for $\varphi_2-2\varphi_1=\pi$.
     (c) $C_\theta$, defined in Eq.\eqref{eq:phase123}, as a function of Matsubara frequency $\varepsilon_n$ at $\bq\to0$ and $T=5\times10^{-3}$. (d) $C_\lambda$, defined in Eq.\eqref{eq:amplitude123}, as a function of Matsubara frequency $\varepsilon_n$ at $\bq\to0$ and $T=5\times10^{-3}$.}\label{fig:eigBC}
\end{figure}

We note that the same matrix also appears in the discussion of three component uniform SC system in Ref\cite{PhysRevLett.108.177005}.
The eigenvalues of this matrix are obviously $0$, $2B-C/2$ and $\frac{3}{2}C$ and the corresponding eigen modes are
\begin{equation}
    \begin{aligned}
       & \theta_{+,\pm}(0)\propto\frac{\theta_{\bQ}(0)+\theta_{-\bQ}(0)}{\sqrt{2}}\pm\left[1/\sqrt{2}\right]^{\pm1}\theta_{0}(0),\\
       & \theta_-(0)\propto\theta_{\bQ}(0)-\theta_{-\bQ}(0).
    \end{aligned}\label{eq:phase1230}
\end{equation}
Among these $\theta_{+,+}$ corresponds to the gapless BAG mode. $\theta_-$ is the same as that in Eq.\eqref{eq:phase12}, which is actually the phase mode of the composite CDW order. The last one $\theta_{+,-}$ is the phase mode when the uniform SC phase and the PDW phase move in opposite direction, and thus represent a true Leggett mode. 
Like we have discussed above, the stability of the system requires that all the eigenvalues to be positive, and the maximally positive eigenvalues correspond to the most stable case. We therefore look for situations when both $2B-C/2$ and $3C/2$ are maximized by $\varphi_{\bQ}-\varphi_{-\bQ}$ and $\varphi_{\bm{0}}$.  In Fig.\ref{fig:eigBC}(a) and (b) we plot the two eigenvalues as a function of $\varphi_{\bQ}-\varphi_{-\bQ}$ with two different constraints: $\varphi_{\bm{0}}=(\varphi_{\bQ}-\varphi_{-\bQ})/2$ and $\varphi_{\bm{0}}=(\varphi_{\bQ}-\varphi_{-\bQ})/2+\pi/2$. In both cases, these eigenvalues are periodic in $\varphi_{\bQ}-\varphi_{-\bQ}$ with a period $\pi/2$ and reach extreme values at $\varphi_{\bQ}-\varphi_{-\bQ}=0,\pm\pi/4,...$. But for $\varphi_{\bQ}-\varphi_{-\bQ}=\pm\pi/4$, $3C/2$ is always negative. Thus the stable system favors $\varphi_{\bm{0}}-(\varphi_{\bQ}-\varphi_{-\bQ})/2=\pi/2$ with $\varphi_{\bQ}-\varphi_{-\bQ}=0$. This results is consistent with Eq.\eqref{eq:phi1phi2} and serve as another proof of favorable time-reversal symmetry breaking in this system. 

At finite $q$, the linear combination coefficients in $\theta_{+,\pm}$ acquire a weak $q$-dependence, while $\theta_-$ remains the same. Therefore, Eq.\eqref{eq:phase1230} is further modified into
\begin{equation}
    \begin{aligned}
        & \theta_{+,\pm}(q)\propto\frac{\theta_{\bQ}(q)+\theta_{-\bQ}(q)}{\sqrt{2}}\pm\left[C_\theta(q)\right]^{\pm1}\theta_{0}(q),\\
       & \theta_-(q)\propto\theta_{\bQ}(q)-\theta_{-\bQ}(q),
    \end{aligned}\label{eq:phase123}
\end{equation}
Where $C_\theta(q)>0$ is a number that depends on $q$, as shown in the main text. In the limit when $q\to0$, $C_\theta(q)$  must approach $1/\sqrt{2}$, as we see from Eq.\eqref{eq:phase1230} analytically. Numerically we have evaluated $C_{\theta}$ at $\bq=0$ and as a function of Matsubara frequency $\omega_n$, as shown in Fig.\ref{fig:eigBC} (c). Clearly we see regardless of different values of the mean-field gaps, at $\omega_n\to0$ the all the curves converge to $1/\sqrt{2}\approx0.707$. 

\begin{figure*}
    \includegraphics[width=16cm]{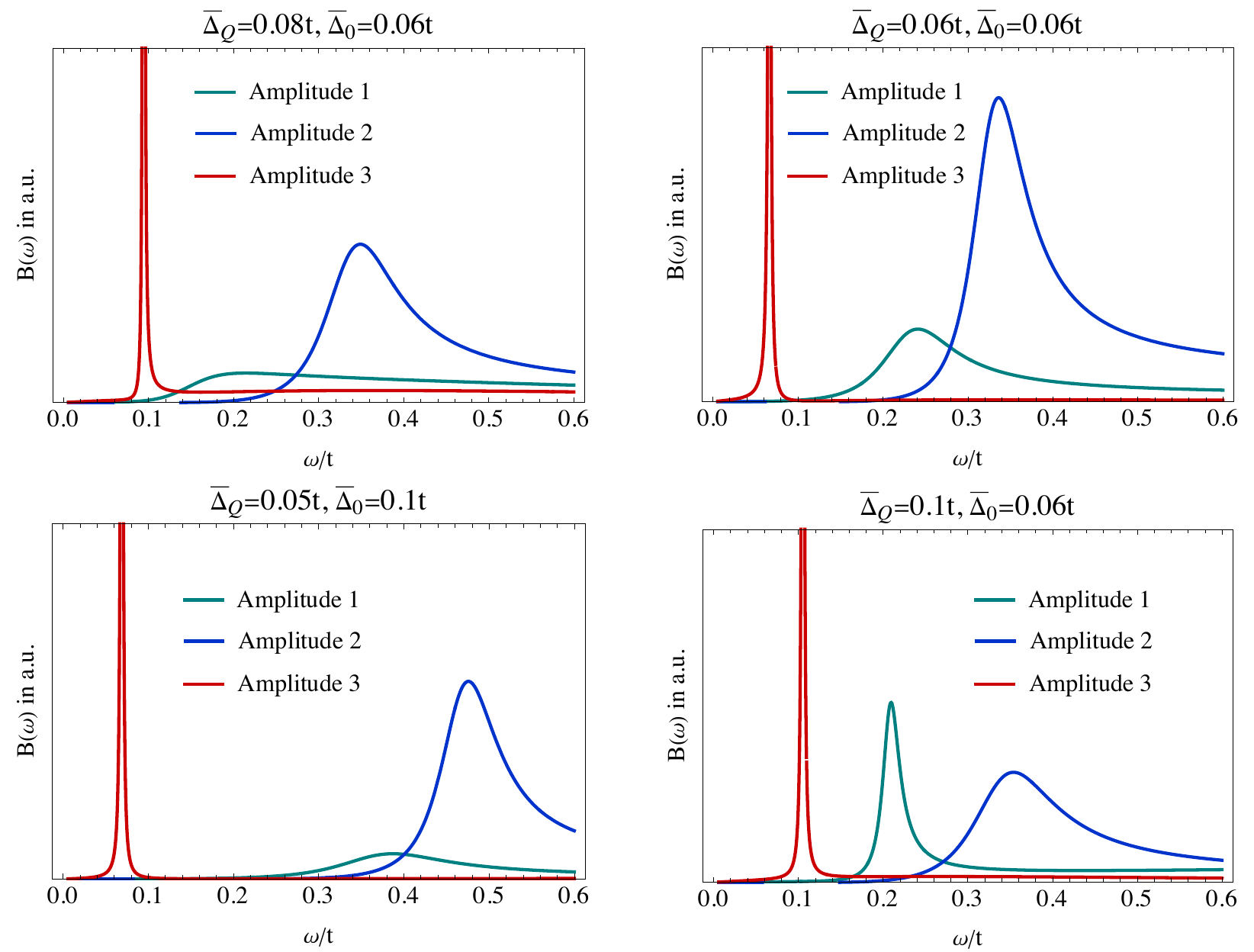}
    \caption{Spectral functions for the amplitude modes when both unidirectional PDW ($d$-wave) and uniform SC ($d$-wave) are present. See Eq.\eqref{eq:amplitude123} for details. In all plots, we take $T=5\times10^{-3}t$. }\label{fig:SCPDWmodes}
\end{figure*}

For the amplitudes fluctuations, the corresponding eigen modes can also be found by diagonalizing $\hat\Gamma_{A}^{-1}(q)$ at small $q$. From the eigenvectors we are able to identify the following three modes:
\begin{equation}
    \begin{aligned}
       &\mathcal{A}_{+,\pm}(q)\propto\frac{\mathcal{A}_{\bQ}(q)+\mathcal{A}_{-\bQ}(q)}{\sqrt{2}}\pm\left[C_{\mathcal{A}}(q)\right]^{\pm1}\mathcal{A}_{0}(q),\\
       & \mathcal{A}_-(q)\propto\mathcal{A}_{\bQ}(q)-\mathcal{A}_{-\bQ}(q).
    \end{aligned}\label{eq:amplitude123}
\end{equation}
Where $C_{\mathcal{A}}(q)$ is a numbers that depends on $q$. Note the similarity between these amplitude modes and the phase modes in Eq.\eqref{eq:phase123}, and also note that $\mathcal{A}_-$ is the same as that of the pure PDW case discussed in Eq.\eqref{eq:amplitude12}. For $C_{\mathcal{A}}(q)$ there is no constraints for the $q\to0$ limit, and therefore these coefficients are generally different for different gap magnitudes $(|\overline\Delta_{\bQ}|,|\overline\Delta_0|)$ in this limit. In Fig.\ref{fig:eigBC}(d) we present the numerical results for $C_{\mathcal{A}}$ at $\bq=0$ and finite $\omega_n$. Depending on the relative magnitudes of $|\Delta_{\bQ}|$ and $|\Delta_0|$, this quantities can be either much larger than 1, or much smaller than $1$.

In Fig.\ref{fig:SCPDWmodes} we show the Higgs susceptibilities along the real frequency axis for the three amplitudes in Eq.\eqref{eq:amplitude123}. We choose four representative mean-field gaps.  As in the pure PDW case, inclusion of the couplings among these order parameters results in splittings of these peaks. In particular, we find that the most pronounced peak feature corresponds to $\mathcal{A}_{+,-}$ in Eq.\eqref{eq:amplitude123}, and the peak position seems always located  below $\min\{2|\overline\Delta_{\bQ}|,2|\overline\Delta_0|\}$. The second pronounced peak corresponds to $\mathcal{A}_-$ in Eq.\eqref{eq:amplitude123}, which is located at some frequencies higher than $\max\{2|\overline\Delta_{\bQ}|,2|\overline\Delta_0|\}$. Note that, interestingly, this mode corresponds to the breathing mode between $\lambda_{\bQ}$ and $\lambda_{-\bQ}$, and had we turned off the uniform SC component, its peak positions would fall below $2\overline\Delta_{\bQ}$, as we have seen from the plots in the main text. The least weighted features are the one corresponding to $\mathcal{A}_{+,+}$ in Eq.\eqref{eq:amplitude123}. This mode is when all the three components, $\lambda_{\pm\bQ}$ and $\lambda_0$ moves in the same direction. This one being the smallest is consistent with the pure PDW case discussed above.

\begin{figure}
    \includegraphics[width=18cm]{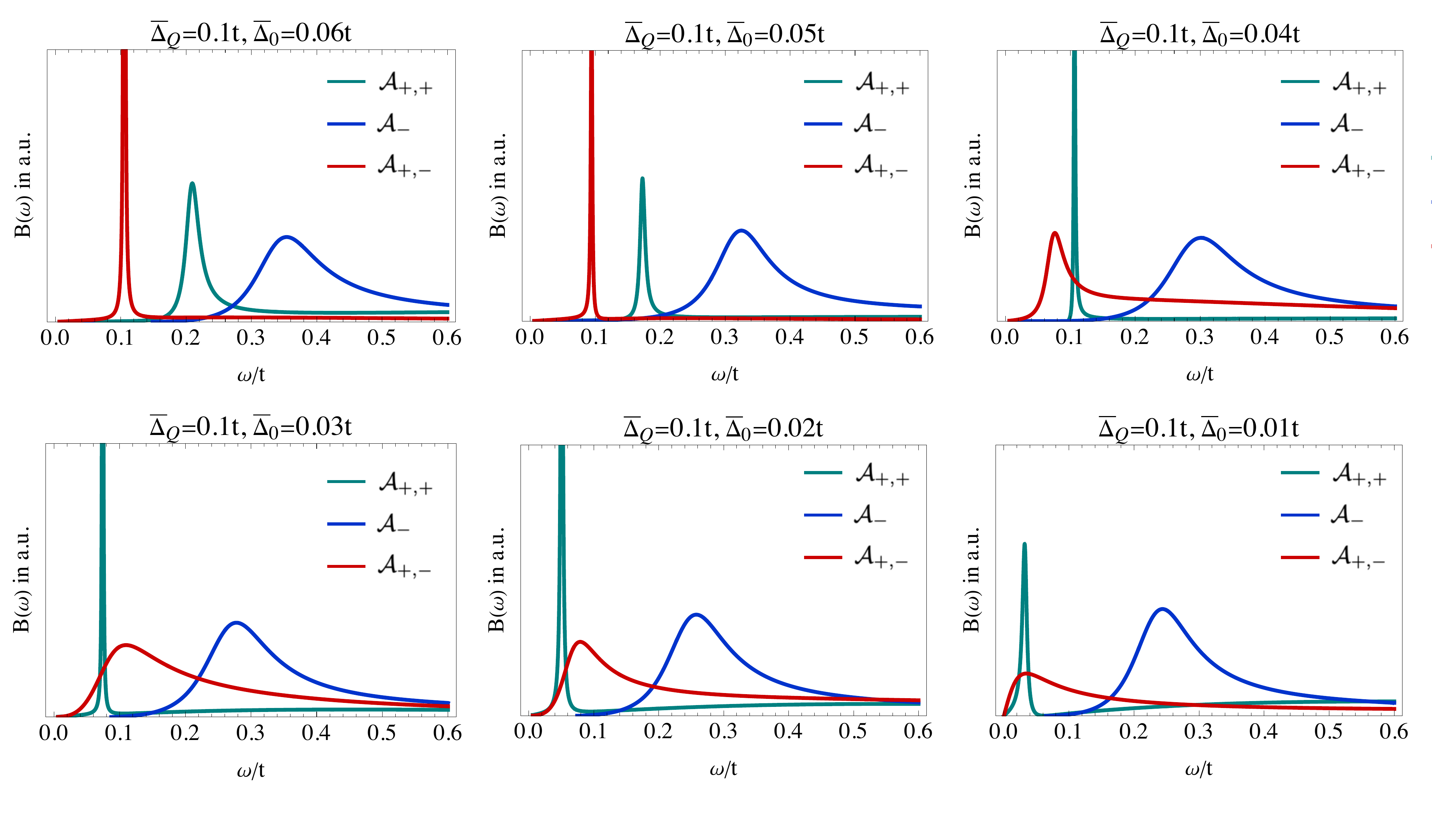}
    \caption{Variation of the Higgs mode spectral functions by fixing $|\overline{\Delta}_{\bQ}|$ and tuning $|\overline{\Delta}_0|$.}\label{fig:ex}
\end{figure}

In Fig.\ref{fig:ex} we present more numerical data showing the evolution of the Higgs spectral functions when $|\overline{\Delta}_0|$ varies while $|\overline{\Delta}_{\bQ}|$ is fixed. Clearly that as $|\overline{\Delta}_0|$ decreases, the spectral weights of $\mathcal{A}_{+,+}$ and $\mathcal{A}_{+,-}$ get interchanged. The long-lived Higgs boson mode is $\mathcal{A}_{+,+}$ in the limit of $|\overline{\Delta}_0|\ll|\overline{\Delta}_{\bQ}|$. In all cases, the peaks of $\mathcal{A}_{+,+}$ and $\mathcal{A}_{+,-}$ are located at a lower energy scale compared to the $\mathcal{A}_{-}$ mode. 

\section{III. Energy hierarchy of Higgs modes: PDW state vs PDW+SC state} 
\label{sec:iii_energy_hierarchy_of_higgs_modes_pdw_state_vs_pdw_sc_state}
In general, 
  analytical interpretation  
 of numerical results  for the positions and the widths of collective modes in a state with PDW order parameters $\Delta_{\pm Q}$ and
SC order parameter $\Delta_0$ is challenging, even at $T=0$,
 because in analytical calculations one has to deal with 
 multi-component 
 fermionic Green's functions.
 However, as we argue below, one prominent feature of numerical results can be understood analytically. 
 Namely, the numerics shows that in a pure PDW state, where there are two longitudinal modes --- symmetric and antisymmetric ($(\mathcal  A_+)$ and $\mathcal  A_-$, respectively) the frequency of the (damped) $\mathcal  A_+$ mode is higher than that of the $\mathcal  A_-$ mode
  (Fig.~2(a)).  In the mixed PDW/SC state, when $\Delta_{\pm Q}$ and $\Delta_0$ are all present,  the antisymmetric mode $\mathcal  A_-$ does not move, but the $A_+$ mode couples with longitudinal fluctuations of $\Delta_0$ into mixed $\mathcal A_{++}$  and $\mathcal A_{+-}$ modes.  When the hybridization coefficient, which we specified by $C_{\mathcal{A}}$, is either large or small,  one of these two modes is predominantly the $\mathcal  A_+$ mode.   The numerical results in Fig.~2(c) clearly show that this mode is now located at a smaller frequency than the $\mathcal  A_-$ mode, and in additional numerical results the distance between the two increases with increasing  $\Delta_0$.
 
 We demonstrate below that this result can be understood by analyzing the collective modes within 
 Ginzburg-Landau expansion, whose coefficients are obtained by evaluating proper fermionic loops.
 
\subsection{1. PDW State}
For a pure PDW state, the GL expansion is justified for $|\Delta_{\pm Q}|\ll T$, and we have, neglecting spatial fluctuations and nonanalytic $\partial_\tau$ dependence that 
 accounts for damping,
    \begin{align}
\mathcal{L}(\Delta_{\pm Q}) = \kappa_0(|\partial_\tau \Delta_Q|^2+|\partial_\tau \Delta_{-Q}|^2 ) 
+ r(|\Delta_Q|^2+|\Delta_{-Q}|^2) + u(|\Delta_{Q}|^4+|\Delta_{-Q}|^4) + \gamma (|\Delta_Q|^2|\Delta_{-Q}|^2).\label{eq:1}
\end{align}

By minimizing this action, one can obtain the mean-field values  $\overline\Delta_{\pm Q}$ and  frequencies of the longitudinal (Higgs) modes $\delta_{\pm Q}(\tau) = \Delta_{\pm Q}(\tau) - \overline\Delta_{\pm Q}$.
We use the 
 gauge
   $\overline\Delta_Q = \overline\Delta_{-Q}\in \mathbb{R}$ and define
    the Higgs modes $A_{\pm}$ as even and odd combinations via
\be
\mathcal{A}_{\pm} = \frac{\delta_Q\pm \delta_{-Q}}{\sqrt{2}}.
\ee
Expanding in $\delta_{\pm Q}(\tau)$, using the mean-field values for $\overline\Delta_{\pm Q}$ we obtain the energies of $A_+$ and $A_-$ modes as
\be
\epsilon_{\mathcal{A}_+}= 2\sqrt{1+ \frac{\gamma}{2\kappa_0}} \overline\Delta_Q,~~~
\epsilon_{\mathcal{A}_-}= 2\sqrt{1-\frac{\gamma}{2\kappa_0}} \overline\Delta_Q.
\label{eq:E}
\ee
The prefactors $\gamma$ and $\kappa_0$  in (\ref{eq:E}) along with $u$ in (\ref{eq:1}) have been obtained in Ref.~\cite{soto-garrido-wang-fradkin-cooper} and found to be positive (and  in fact $\kappa_0=u$).

We see that 
the $\mathcal{A}_-$ mode has a 
  smaller energy than the $\mathcal{A}_+$ mode.  This is consistent with the numerics (Fig. 2(a))  From the physics perspective, a smaller energy for the $\mathcal{A}_-$ mode is the consequence of an effective repulsion (for $\gamma>0$)
between $\Delta_Q$ and $\Delta_{-Q}$, such that it is energetically advantageous for them to fluctuate in opposite directions.

\subsection{2. PDW+SC State}

In the presence of a uniform SC order $\Delta_0$, there are three Higgs modes. By symmetry, $\mathcal{A}_-$ 
 is not affected,
 while $\mathcal{A}_+$ and the Higgs mode of the uniform order $\mathcal{A}_0$
hybridize, 
 leading to two eigenmodes $\mathcal{A}_{++}$ and $\mathcal{A}_{+-}$.   For our consideration, we assume that 
 $\overline\Delta_0$ is larger than $\overline\Delta_{Q}$ (= $\overline\Delta_{-Q}$), expand in $\overline\Delta_{\pm Q}$ in Ginzburg-Landau functional 
  and compute the coefficients  of various $\Delta^2_{Q}$ and $\Delta^4_{Q}$ terms using fermionic propagators for a SC state with $\Delta_0$.  In other words,  we treat $\Delta_0$ as a background field and ignore its fluctuations, which is a separate mode.
   For $\overline\Delta_Q=0.05t, \overline\Delta_0 = 0.1t$, our numerical analysis
   shows that in this situation the $\mathcal A_{++}$ mode predominantly describes fluctuations of a SC order parameter, while 
   $\mathcal A_{+-}$ mode predominantly describes $A_+$ fluctuations of PDW order, namely in this case
\begin{equation}
    \mathcal{A}_{+,+}\approx \mathcal{A}_0, ~~~ \mathcal{A}_{+,-}\approx \mathcal{A}_+.
\end{equation}
Like we said, the numerical data clearly show that in a broad range of $\Delta_Q^0/\Delta_0 = O(1)$,  the energy hierarchy between $\mathcal{A}_{+}$ and $\mathcal{A}_-$ modes is \emph{reversed} compared with a pure PDW state (i.e., the $\mathcal A_{+}$ mode is at a smaller frequency), and the distance between the two modes increases with decreasing  $\Delta_Q^0/\Delta_0 = O(1)$.



In the presence of $\Delta_0\neq 0$, the Ginzburg-Landau expansion for PDW order parameters  reads
\begin{align}
\mathcal{L}(\Delta_{\pm Q}) =& \kappa_0(|\partial_\tau \Delta_Q|^2+|\partial_\tau \Delta_{-Q}|^2 ) + \tilde \kappa \partial_\tau\Delta_{Q}\Delta_{-Q} \non\\
&+ r\(|\Delta_Q|^2+|\Delta_{-Q}|^2\) + \(\frac{\tilde r}2 \Delta_Q \Delta_{-Q} + \frac{\tilde r^*}2 \Delta_Q^* \Delta_{-Q}^*  \)\non\\
&+  u\(|\Delta_Q|^4+|\Delta_{-Q}|^4\) + \gamma  |\Delta_Q|^2 |\Delta_{-Q}|^2 + \(\frac{\bar\gamma}{2} \Delta_Q^2\Delta_{-Q}^2 + \frac{\bar\gamma^*}{2} \Delta_Q^{*2}\Delta_{-Q}^{*2} \) \non\\
&+\frac{\tilde\gamma}2\(|\Delta_Q|^2+ |\Delta_{-Q}|^2\)\(\Delta_Q\Delta_{-Q} +\Delta_Q^*\Delta_{-Q}^*\),
\label{eq:2}
\end{align}
where several $U(1)$ symmetry breaking terms are now allowed since this symmetry is explicitly broken by $\Delta_0$.
  We have used the same notation for 
   the 
    prefactors of the terms which are present with and without $\Delta_0$, but their values certainly change because of $\Delta_0$.

The  values of the prefactors in (\ref{eq:2})  can be obtained by integrating out fermion propagators in the effective vertices for $\Delta_{\pm Q}$. In the presence of $\Delta_0$, there are
 two kinds of fermion propagators, normal ($G(\omega, k)$) and anomalous ($F(\omega,k)$). 
  Evaluating the integrals, we 
   find 
   that
\begin{align}
&r\sim \frac{1}{v_F^2} {|\Delta_0|},~\tilde r \sim \frac{1}{v_F^2} \frac{\Delta_0^{*2}}{|\Delta_0|} \non \\
2\kappa_0\pm \tilde\kappa\sim u\sim \gamma& \sim \frac 1{v_F^2} \frac{1}{|\Delta_0|},  \tilde\gamma \sim -\frac 1{v_F^2} \frac{\Delta_0^{*2}}{|\Delta_0|^3} ,\bar\gamma \sim \frac 1{v_F^2} \frac{\Delta_0^{*4}}{|\Delta_0|^5}
\label{eq:dim}
\end{align}
where all proportionality coefficients, which we didn't write explicitly, 
are positive. A simple 
  analysis shows that the GL expansion is justified even at $T=0$ as long as
\be
|\Delta_Q|\lesssim |\Delta_0|.
\label{eq:7}
\ee

We now minimize the action in (\ref{eq:2}).  A simple experimentation shows that  
 the minimum is reached when  the phases of $\Delta_{Q}$ and $\Delta_{-Q}$ are the same, and the phase of $\Delta_0$ 
 is shifted by $\pi/2$, e.g., when 
\be
\overline\Delta_Q =\overline\Delta_{-Q}\in \mathbb{R}, i\Delta_0 \in \mathbb{R},
\label{eq:3}
\ee
 For such an order,  
   time-reversal is broken. This is the 
    same effect as captured by the $\beta_5 \Delta_0^{*2} \Delta_Q \Delta_{-Q}$ term in the main text. 
     This is not the only choice for the phases, but one can easily make sure that other choices 
       correspond to different phase selections for the sliding mode of  $\rho_Q \sim \Delta_Q \Delta_0^*$. Due to the translation symmetry, all such choices are equivalent  to (\ref{eq:3}).
      Using (\ref{eq:3}) we find that for $2r+\tilde r<0$, the  mean-field values of $\Delta_{\pm Q}$ are
\be
\overline\Delta_Q =\overline\Delta_{-Q}= \sqrt{\frac{-(2r+\tilde r)}{4u+ 2\gamma + 2\bar\gamma + 4\tilde \gamma}}.
\label{eq:mf}
\ee

Expanding around these mean-field values as $\Delta_{\pm Q} = \overline\Delta_{\pm Q}  + \delta_{\pm Q}$, we find the 
 Lagrangian for the Higgs modes $\mathcal{A}_{\pm }$ (fluctuations of $\delta_Q = \pm \delta_{-Q}$) in the form
\begin{align}
\mathcal{L}(\mathcal{A}_+,\mathcal{A}_-) =& (2\kappa_0 + \tilde \kappa ) (\partial_\tau \mathcal{A}_+)^2 + (2r - |\tilde r|) \mathcal{A}_+^2 + (12 u + 6\gamma + 6\bar\gamma + 12 \tilde \gamma) \(\overline\Delta_Q\)^2 \mathcal{A}_+^2 \non\\
&+ (2\kappa_0 - \tilde \kappa ) (\partial_\tau \mathcal{A}_-)^2 + (2r + |\tilde r|) \mathcal{A}_-^2 + (12 u-2\gamma - 2\bar\gamma) \(\overline\Delta_Q\)^2 \mathcal{A}_-^2.
\end{align}
Using \eqref{eq:mf} 
 and rotating from Matsubara to real frequencies, 
  we obtain
\be
\epsilon_{\mathcal{A}_+}= 2\sqrt{\frac{2u+\gamma + \bar\gamma + 2\tilde \gamma}{2\kappa_0 + \tilde \kappa}} \overline\Delta_Q,~~~
\epsilon_{\mathcal{A}_-}= 2\sqrt{\frac{2u-\gamma -\bar\gamma -\tilde \gamma}{2\kappa_0 + \tilde \kappa} \(\overline\Delta_Q\)^2 + \frac12\frac{|\tilde r|}{2\kappa_0 - \tilde \kappa}}.
\label{eq:5}
\ee
For $\Delta_0 =0$, 
$\bar\gamma=\tilde \gamma = \tilde \kappa = \tilde r =0$ by symmetry. Substituting into (\ref{eq:5}), we recover \eqref{eq:E}. 
In the presence of $\Delta_0$, however, the result is different 
as 
the energy of the $\mathcal{A}_-$ mode is enhanced by  the term $\propto\tilde r$.  Using Eq. (\ref{eq:dim}) for estimates, we see that
\be
\epsilon_{\mathcal{A}_+}\sim \overline\Delta_Q \ll \epsilon_{\mathcal{A}_-}\sim \Delta_0,
\ee
i.e., the  $\mathcal{A}_+$ mode is now located at an energy far smaller than the 
$\mathcal{A}_-$ mode.   This fully agrees with the numerics.

Just like the pure PDW case, this result can be understood intuitively. The presence of a uniform order $\Delta_0$ enables an effective attractive ``interaction" (although it is at quadratic level in $\Delta_{\pm Q}$) $\sim - |\tilde r| \Delta_Q \Delta_{-Q}$. Thus it is energetically more favorable for $\Delta_Q$ and $\Delta_{-Q}$ to fluctuate in the same direction. Since this attraction is at quadratic level, it dominates over the repulsive interaction at quartic level we previously discussed.


\subsection{3. Mode-mode coupling in the PDW+SC state: composition of the longitudinal modes and the locations of the peaks in the spectral function}

In this subsection we apply a simple mode-mode coupling theory to get analytical understanding of the composition of the
 coupled symmetric  fluctuations of the d-wave PDW order, $\mathcal{A}_+ (q)= (\mathcal{A}_{\bQ} (q) + \mathcal{A}_{-\bQ} (q))/\sqrt{2}$, and fluctuations of the uniform d-wave SC order $\mathcal{A}_0 (q)$, see Eq. (9) in the main text, $q = (0, \omega)$.  The two coupled modes are labeled in the main text as $\mathcal{A}_{+,+} (q)$ and $\mathcal{A}_{+,-} (q)$.  In the numerical analysis, we found that when the
     mean-field value of the SC order parameter is larger than that of the PDW order, $|\overline{\Delta}_0| > |\overline{\Delta}_{\bQ}|$,  the mode $\mathcal{A}_{+,+}$ nearly coincides with $\mathcal{A}_0 (q)$ while the mode  $\mathcal{A}_{+,-}$ nearly coincides with $\mathcal{A}_+ (q)$.
    For this analysis we assume that both $\mathcal{A}_+ (q) = \mathcal{A}_+ (\omega)$ and $\mathcal{A}_0 (q) = \mathcal{A}_0 (\omega)$ can be described as damped
  oscillators, as they are in the pure PDW and SC states [see panel (a) in Fig. 2 in the main text],
  i.e., prior to mode-mode coupling $\chi^{-1}_{\mathcal{A}_+} =  4 \Delta^2_Q - (\omega + i \gamma_Q)^2 $ and $\chi^{-1}_{\mathcal{A}_0} =  4 \Delta^2_0 - (\omega + i \gamma_0)^2$, where  $\Delta_Q$ and $\Delta_0$ include the renormalizations which we discussed in the previous subsections, but generally  $\Delta_Q$ is comparable to $|{\overline \Delta}_{\bQ}|$ and $\Delta_0$ is comparable to $|{\overline \Delta}_{0}|$.
   We further assume that the coupling between the two modes can be approximated by frequency independent $\lambda \propto (\Delta_0 \Delta_Q)$.  The two coupled modes
   $\mathcal{A}_{+,+}$ and $\mathcal{A}_{+,-}$ are the eigenfunctions of
    \begin{equation}
    {\hat L} = \begin{pmatrix}
        a & \lambda \\
        \lambda & b\\
       \end{pmatrix}
       \label{eq:m1}
       \end{equation}
where $a = \chi^{-1}_{\mathcal{A}_+}$ and $b = \chi^{-1}_{\mathcal{A}_0}$. The spectral functions of the two modes  ($B(\omega)$) are Im $\chi_{\mathcal{A}_{++}}$ and Im $\chi_{\mathcal{A}_{+-}}$, where
$\chi^{-1}_{\mathcal{A}_{++}}$ and $\chi^{-1}_{\mathcal{A}_{+-}}$ are the eigenvalues of the matrix ${\hat L }$.
The eigenfunctions
\begin{equation}
    {\Phi} = \begin{pmatrix}
        \mathcal{A}_{+,+} \\
         \mathcal{A}_{+,-}\\
       \end{pmatrix}
       \label{eq:m2}
       \end{equation}
are obtained from
\begin{equation}
    {\Phi}_0 = \begin{pmatrix}
         \mathcal{A}_{+} \\
         \mathcal{A}_{0}\\
       \end{pmatrix}
       \label{eq:m3}
       \end{equation}
by a conventional rotation:   $\Phi = {\hat K} \Phi_0$, where
  \begin{equation}
    {\hat K} = \begin{pmatrix}
        \cos{\theta} & \sin{\theta} \\
        - \sin{\theta}& \cos{\theta}\\
       \end{pmatrix}
       \label{eq:m4}
       \end{equation}
The value of $\theta$ is determined from the condition that ${\hat K}^{-1} {\hat L} {\hat K}$ is a diagonal matrix:
  \begin{equation}
    {\hat K}^{-1} {\hat L} {\hat K}  = \begin{pmatrix}
        \chi^{-1}_{ \mathcal{A}_{++}} & 0 \\
        0& \chi^{-1}_{ \mathcal{A}_{+-}}\\
       \end{pmatrix}
       \label{eq:m5}
       \end{equation}

 The computation is elementary  and yields
 \bea
 &&  \mathcal{A}_{+,+} = \cos{\theta} \left(  \mathcal{A}_{+} + \tan{\theta}  \mathcal{A}_0\right) \nonumber \\
 &&  \mathcal{A}_{+,-} = \cos{\theta} \left(  \mathcal{A}_{0} - \tan{\theta}  \mathcal{A}_{+}\right)
 \label{eq:m6}
\eea
\bea
&&\cos{2\theta} = \frac{\frac{a-b}{2}}{\sqrt{\left(\frac{a-b}{2}\right)^2 + \lambda^2}} \nonumber \\
&& \tan{\theta} = \frac{1}{\lambda} \left(\frac{a-b}{2} - \sqrt{\left(\frac{a-b}{2}\right)^2 + \lambda^2}\right)
 \label{eq:m6}
\eea
(we have chosen  one of the two sign options and verified that the choice does not affect the results)
Then
\bea
&&\chi_{ \mathcal{A}_{++}} = \frac{1}{\frac{a+b}{2}  +\sqrt{\left(\frac{a-b}{2}\right)^2 + \lambda^2}} \nonumber \\
&&\chi_{ \mathcal{A}_{+-}} = \frac{\frac{a+b}{2}  +\sqrt{\left(\frac{a-b}{2}\right)^2 + \lambda^2}}{ab-\lambda^2}
 \label{eq:m7}
\eea
\\
Using our definitions of $a$ and $b$, we find $a-b = 4 (\Delta^2_Q - \Delta^2_0)$.  For $\Delta_0 \gg \Delta_Q$,
$a-b \approx -4 \Delta^2_0$, and $|a-b|/2 >> \lambda$.   In this situation, $\theta \approx \pi/2$, which implies
 that  $ \mathcal{A}_{++}$ mode is made predominantly out of SC fluctuations and $ \mathcal{A}_{+-}$ mode is made out of PDW fluctuations.
The susceptibility of the $ \mathcal{A}_{++}$ mode is featureless at $\omega \sim \Delta_Q$, but that of the $ \mathcal{A}_{+-}$ mode is peaked at a frequency where  $ab=\lambda^2$, i.e., at positive $\omega$, at  $\omega = \omega_p \approx 2 \Delta_Q (1 - (\lambda/(4 \Delta_0 \Delta_Q)^2)^{1/2} \leq 2\Delta_Q$.  Like we said in the main text, the damping $\delta_Q$ nearly vanishes in the presence of strong SC order and also because even for a pure PDW state the damping is vanishingly small at low frequencies.  In this situation,  the spectral function of the $ \mathcal{A}_{+-}$ mode has a near-$\delta$-functional peak at
$\omega = \omega_p$.
The composition of the $ \mathcal{A}_{+-}$ mode and the presence of a sharp peak in the spectral function of this mode fully agree with the results of numerical calculations.

In the opposite limit $\Delta_0 \ll \Delta_Q$, the situation is more tricky.  Now $a-b \approx 4 \Delta_Q$ is large and positive, hence the solution of (\ref{eq:m6}) is $\theta \approx 0$.  Then $ \mathcal{A}_{+,+}$ mode is now composed largely from
PDW fluctuations and $ \mathcal{A}_{+,-}$ mode is mainly out of SC fluctuations.  This is consistent with the numerical solution as it shows that at  frequencies well above $\Delta_0$  the spectral function of the $\mathcal{A}_{+,+}$ mode  shows the behavior similar to that of $\mathcal{A}_+$ in Fig. 2a of the main text, and the spectral function of the $\mathcal{A}_{+,-}$ mode shows a long tail consistent with the behavior of $B(\omega)$ for a pure SC state in Fig. 2a of the main text.  At the same time, mode-mode coupling analysis yields a sharp peak in the $\mathcal{A}_{+,-}$ mode at $\omega \leq 2 \Delta_0$ and featureless $B(\omega)$ for the $\mathcal{A}_{+,+}$ mode at these frequencies.  Numerical results, however, show that these exists a sharp peak in the spectral function for the $\mathcal{A}_{+,+}$ mode at small $\omega \sim 2 |{\overline \Delta}_0|$.   To reproduce this result in the analytical calculation one has to go beyond the point of departure for our mode-mode coupling analysis. Indeed, we assumed that $\chi^{-1}_{\mathcal{A}_+}$ preserves the same quadratic frequency dependence as for a pure PDW state, i.e., that SC only renormalizes  $\Delta_Q$ compared to $|{\overline \Delta}_{\bQ}|$ (see previous subsection).   It is likely, however, that SC does affect the frequency dependence of $\chi^{-1}_{\mathcal{A}_+}$ such that it develops a peak at $\omega = O(\Delta_0)$   (e.g., $\omega^2$ term gets replaced by $\omega^4/(\omega^2 - (\Delta^*_0)^2)$, where $\Delta^*_0$ is comparable to $\Delta_0$). The analytical verification of this conjecture requires more work and we leave it for further study.

\section{IV. SC coexists with CDW} 
\label{sec:iii_sc_coexists_with_cdw}
Here we consider the case when a uniform SC coexists with a {\it real} CDW order. Such a state looks similar to a PDW state in STM measurement. 
The CDW order can arise from fluctuations of a parent PDW order through the fusion rule $\rho_{2\bQ}\sim\Delta_{\bQ}\Delta_{-\bQ}^*$ so the resultant CDW momentum is $\bm{P}=2\bQ$. 
Alternatively, it can be viewed as an independent order due to some effective interaction which we name as $g_3$.
At mean field level, we assume the CDW order parameter is $\overline\rho_{\bm{P}}=|\overline\rho_{\bm{P}}|e^{i\varphi_{\bP}}$ which obeys $\overline\rho_{\bm{P}}=\overline\rho_{-\bm{P}}^*$. Taking into account fluctuations the total order parameter is the local field 
\begin{equation}
    \begin{aligned}
        \rho(\br)&=\overline\rho_{\bm{P}}(1+\mathcal{A}_{\bP}(\br))e^{i\theta_{\bP}(\br)}e^{i\bm{P}\cdot\br}\\
        &+\overline\rho_{-\bm{P}}(1+\mathcal{A}_{-\bm{P}}(\br))e^{i\theta_{-\bm{P}}(\br)}e^{-i\bm{P}\cdot\br}
    \end{aligned}
\end{equation}
where we use $\mathcal{A}_{\pm\bm{P}}$ and $\theta_{\pm\bm{P}}$ to denote  amplitude and phase fluctuations. 
If we impose that the CDW order parameter is real even in the presence of fluctuations and include only the leading fluctuating effects, we will have the following constraint
\begin{equation}
     \begin{aligned}
         \mathcal{A}_{\bm{P}}(\br)&=\mathcal{A}_{-\bm{P}}(\br)\equiv\mathcal{A}_\rho(\br)\\
         \theta_{\bm{P}}(\br) &=-\theta_{-\bm{P}}(\br)\equiv \theta_\rho(\br)
     \end{aligned}
 \end{equation} 
 and the CDW field becomes\cite{PhysRevLett.31.462,PhysRevB.26.4883,RevModPhys.60.1129}
\begin{equation}
    \rho(\br)=2|\overline\rho_{\bm{P}}|(1+\mathcal{A}_{\rho}(\br))\cos(\bm{P}\cdot\br+\varphi_{\bP}+\theta_\rho(\br)).
\end{equation}
If $\bm{P}$ is incommensurate, $\varphi_{\bP}$ can be arbitrary, a constant shift on $\theta_\rho$ does not cause any energy change and hence $\theta_\rho(\br)$ corresponds to a gapless mode. If instead $\bm{P}$ is commensurate as considered here, $\varphi_{\bP}$ will be locked to discrete values, hence $\theta_\rho$ becomes a massive mode. 
From now on we express the Fourier components of $\rho(\br)$ as
\begin{equation}
    \rho_{\pm\bP}(q)=\overline\rho_{\pm\bP}\left[\delta_{q,0}+\mathcal{A}_\rho(q)\pm i\theta_\rho(q)\right].\label{eq:fluctuatingrho}
\end{equation}
For later convenience we define $s_{\pm\bP}(q)=\mathcal{A}_\rho(q)\pm i\theta_\rho(q)$.

To obtain the effective action for the $\bP=2\bQ$ CDW order parameter, 
we use  the new basis
\begin{equation}
    {\Psi'}_{\sigma, k}^\dagger=[\psi_{\sigma}^\dagger(k), \psi_{\bm{P},\sigma}^\dagger(k),\psi_{2\bm{P},\sigma}^\dagger(k),\psi_{3\bm{P},\sigma}^\dagger(k)],\label{eq:newfermion}
\end{equation}
such that the action under consideration can be written as 
\begin{equation}
    \begin{aligned}
        S=&\frac{1}{g_2}\sum_q|\Delta_0(q)|^2+\frac{1}{g_3}\sum_q \rho_{\bm{P}}(q)\rho_{-\bm{P}}(-q)\label{eq:action11}\\
    &+\frac{1}{4}\int_{k,k'} ({\Psi'}_{\up,k}^\dagger,{\Psi'}^\text{T}_{\down,-k})\hat{\mathcal{K}'}_{k,k'}\begin{pmatrix}
        {\Psi'}_{\up,k'}\\
        {{\Psi'}^\dagger}^\text{T}_{\down,-k'}
    \end{pmatrix}
    \end{aligned}
\end{equation}
where 
\begin{equation}
    \hat{\mathcal{K}'}_{k,k'}=\begin{pmatrix}
        {{\hat{G'}}_p}^{-1}(k)\delta_{k,k'}+\hat\rho_{k-k'} & \hat{\Phi'}_{k-k'}\\
        \hat{\Phi'}^\dagger_{k'-k} & \hat {G'}_h^{-1}(k)\delta_{k,k'}-\hat\rho^\text{T}_{k-k'}.\label{eq:mathcalKp}
    \end{pmatrix}
\end{equation}
The matrix representation for the particle Green's function $\hat{G'}_p$, the hole Green's function $\hat{G'}_h$, the pairing order parameter $\hat{\Phi'}_{k-k'}$ and the CDW order parameter $\hat\rho_{k-k'}$ are 
\begin{equation}
    \begin{aligned}
        &\hat{G'}_p(k)=\begin{pmatrix}
        G_{p,0}(k) & 0 & 0 & 0\\
        0 & G_{p,2}(k) & 0 & 0 \\
        0 & 0 & G_{p,4}(k) & 0 \\
        0 & 0 & 0 & G_{p,6}(k) \\
    \end{pmatrix}, ~~\hat{G'}_h(k)=\begin{pmatrix}
        G_{h,0}(k) & 0 & 0 & 0\\
        0 & G_{h,2}(k) & 0 & 0 \\
        0 & 0 & G_{h,4}(k) & 0 \\
        0 & 0 & 0 & G_{h,6}(k) \\
    \end{pmatrix}\\
    &\hat\rho_{q}=\begin{pmatrix}
        0 & \rho_{-\bm{P}}(q) & 0 & \rho_{\bm{P}}(q)\\
        \rho_{\bm{P}}(q) & 0 & \rho_{-\bm{P}}(q) & 0 \\
        0 & \rho_{\bm{P}}(q) & 0 & \rho_{-\bm{P}}(q) \\
        \rho_{-\bm{P}}(q) & 0 & \rho_{\bm{P}}(q) & 0\\
    \end{pmatrix},~~~\hat{\Phi'}_{q}=\Delta_0(q)\begin{pmatrix}
        f_{\bk} & 0 & 0 & 0\\
        0 & 0 & 0 &  f_{\bk+\bP}\\
        0 & 0 & f_{\bk+2\bP} & 0\\
        0 & f_{\bk+3\bP} & 0 & 0 
    \end{pmatrix}
    \end{aligned}\label{eq:matrixGpp}
\end{equation}
where $q=k-k'$. The diagonal entries in $\hat{G'}_p$ and $\hat{G'}_k$ are the elementary Green's functions in Eq.\eqref{eq:phGreen}, and $\rho_{\pm\bP}(q)$ are defined in Eq.\eqref{eq:fluctuatingrho}.


After integrating out the fermions, we obtain the following mean field and fluctuation action,
\begin{equation}
    \begin{aligned}
        S_\text{MF}&=\frac{1}{g_2}|\overline\Delta_0|^2+\frac{1}{2g_3}\overline\rho_{\bP}\overline\rho_{-\bP}-\frac{1}{4}\text{Tr}\ln\left(1+\mathcal{G}’_0\overline\Sigma'\right),\\
        S_\text{FL}&=\frac{1}{g_2}\sum_q |\overline\Delta_0h_0(q)|^2+\frac{1}{g_3}\overline\rho_{\bP}\overline\rho_{-\bP}\sum_qs_{\bP}(q)s_{-\bP}(-q)+\frac{1}{4}\text{Tr}\sum_{n=1}^\infty\frac{(-\overline{\mathcal{G}'}\tilde\Sigma')^n}{n},
    \end{aligned}
\end{equation}
where 
\begin{equation}
    \begin{aligned}
     &[\mathcal{G}'_0]_{k,k'}=\delta_{k,k'}\begin{pmatrix}
         \hat{G}'_p(k) & 0\\
         0 & \hat{G}'_h(k)
     \end{pmatrix},~~~ [\overline\Sigma']_{k,k'}=\delta_{k,k'}\begin{pmatrix}
        \overline\rho & \overline{\Phi'}\\
        \overline{\Phi'}^\dagger & - \overline\rho^T\\
    \end{pmatrix}, \\
    & \overline{\mathcal{G}'}=\left({{\mathcal{G}'}^{-1}_0}+\overline{\Sigma}'\right)^{-1},[\tilde\Sigma']_{k,k'}=\begin{pmatrix}
        \tilde\rho_{k-k'} & \tilde{\Phi}'_{k-k'}\\
        {{{\tilde{\Phi}}{'}}^\dagger}_{k-k'} & - \tilde\rho_{k-k'}^\text{T}\\
    \end{pmatrix}.
    \end{aligned}\label{eq:CDWFL}
\end{equation}
Here as before, we use $\overline\rho$ and $\overline\Phi'$ to denote the $q=0$ component of $\hat\rho_q$ and $\hat\Phi'_q$, which include mean field order parameters only, and use $\tilde\rho_q$ and $\tilde\Phi'_q$ to denote the $q\neq0$ component of $\hat\rho_q$ and $\hat\Phi'_q$.

\begin{figure}
    \includegraphics[width=8cm]{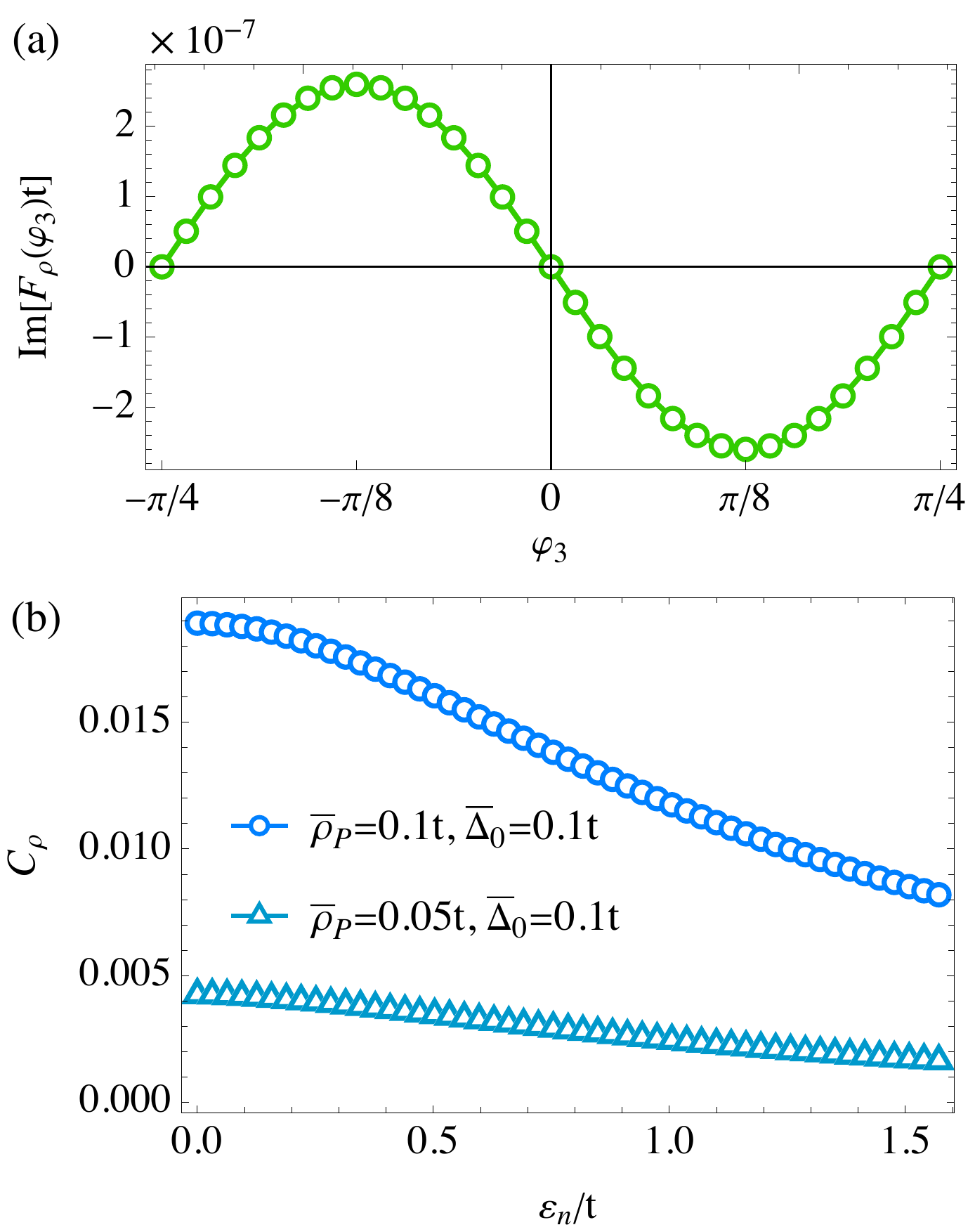}
    \caption{(a) Im$F_\rho(\varphi_3)$ for $\overline\rho_{\bP}=0.05t$ and $\overline\Delta_0=0.1t$. (b) Plot of the coefficient $C_\rho$ as a function of the Matsubara frequency $\varepsilon_n$ at $\bq\to0$ limit. }\label{fig:CDWphi}
\end{figure}

The coupled mean field gap equations are obtained in a similar way Eq.\eqref{eq:gap2} is obtained, i.e. by treating $\overline\Delta_0$ and $\overline\rho_{\bP}$ as complex variational parameters and taking the saddle point equations 
\begin{equation}
    \begin{aligned}
        \frac{\overline\Delta_0}{g_2}=\frac{1}{4}\text{Tr}\left\{\overline{\mathcal{G}'}\frac{\delta\overline\Sigma'}{\delta \overline\Delta_0^*}\right\}, ~ \frac{\overline\rho_{\bP}}{g_3}=\frac{1}{4}\text{Tr}\left\{\overline{\mathcal{G}'}\frac{\delta\overline\Sigma'}{\delta \overline\rho_{\bP}^*}\right\}.\label{eq:gap3}
    \end{aligned}
\end{equation}
For the purpose of convenience, we choose some fixed $|\overline\Delta_0|$ and $|\overline\rho_{\bP}|$, and solve $\varphi_{\bm{0},\bP}$ and $g_{2,3}$ from Eq.\eqref{eq:gap3}, with the constraint that $g_{2,3}$ both are real. To see this, we define the following function 
\begin{equation}
    F_{\rho}(\varphi_{\bP})=\frac{1}{4 \overline\rho_{\bP}}\text{Tr}\left.\left\{\overline{\mathcal{G}'}\frac{\delta\overline\Sigma'}{\delta \overline\rho_{\bP}^*}\right\}\right|_{|\overline\rho_{\bP}|,|\overline\Delta_0|\text{ fixed}},
\end{equation}
and show the plot of $\text{Im}[F_\rho(\varphi_{\bP})]$ in Fig.\ref{fig:CDWphi}(a).
From this we have confirmed that the mean field gap equation has solution only for $\varphi_{\bP}=0,\pi/4,...$. Moreover, the have confirmed that only the $\varphi_{\bP}=0$ is a stable solution. 

Again, this can be easily understood by  expanding $S_\text{MF}$ to the quartic order, the resultant GL free energy density is 
\begin{equation}
    \begin{aligned}
        \mathcal{F}_\text{GL}&=\alpha_1'|\overline\Delta_0|^2+\alpha_2'\overline\rho_{\bP}\overline\rho_{-\bP}+\beta_1'|\overline\Delta_0|^4+\beta_2'\overline\rho_{\bP}^2\overline\rho_{-\bP}^2\\
        &+\beta_3'\left(\overline\rho_{\bP}^4+\overline\rho_{-\bP}^4\right)+\beta_4'|\overline\Delta_0|^2\overline\rho_{\bP}\overline\rho_{-\bP}+...
    \end{aligned}
\end{equation}
Here we found that $\beta_3'<0$ so the system prefers $\varphi_{\bP}=0$ in its ground state, meaning there is some CDW phase locking with the commensurate $\bP$.  Note that the SC phase is absence in the expansion, so one usually does not expect TRS breaking in this case.

\begin{figure*}
    \includegraphics[width=15.5cm]{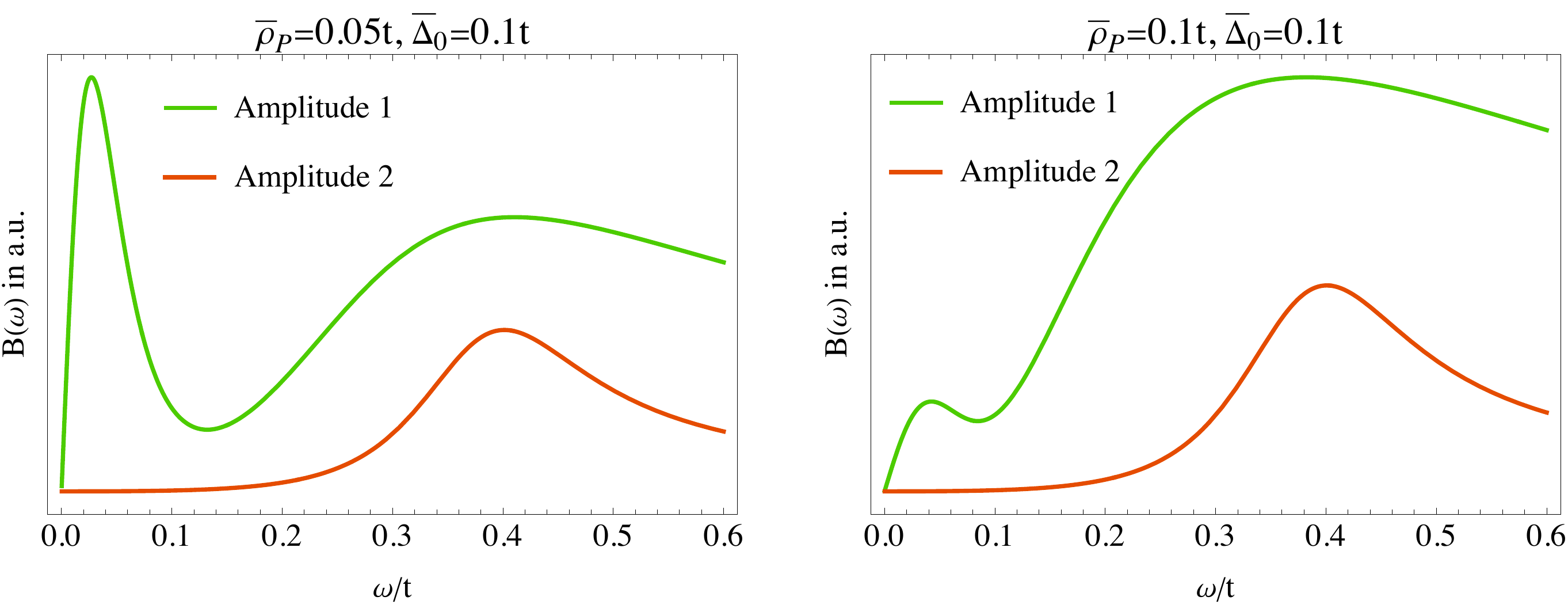}
    \caption{Spectral function of the two collective amplitude modes in a CDW+SC system. See Eq.\eqref{eq:amplitude23} for details. Both of these plots are obtained at $T=5\times10^{-3}t$.}\label{fig:CDWSC_Higgs}
\end{figure*}

For the fluctuating effects, we follow the same procedures out lined in Sec. and truncate $S_\text{FL}$ to Gaussian level. To this end, we use the basis
\begin{equation}
    \zeta''(q)=[\mathcal{A}_\rho(q),\mathcal{A}_0(q),\theta_\rho(q),\theta_0(q)]^T\label{eq:flucbasis3}
\end{equation}
such that the fluctuating $\tilde\Sigma'$ can be compacted expressed as
\begin{equation}
    [\tilde\Sigma']_{k,k'}=\left(\hat{O}_{1},\hat{O}_{2},\hat{O}_{3},\hat{O}_{4}\right)\cdot{\zeta''}(q).\label{eq:fluct3}
\end{equation}
The explicit forms of these $\hat O$-matrices are 
\begin{equation}
    \hat O_1=\begin{pmatrix}
        \overline\rho_{\mathcal{A}} & 0 \\
        0  & -\overline\rho_{\mathcal{A}}^\text{T}
    \end{pmatrix}, ~~\hat O_2=\begin{pmatrix}
         0 & \overline{\Phi'}_{0}\\
         \overline{\Phi'}_{0}^\dagger & 0\\
     \end{pmatrix}, ~~ \hat O_3=\begin{pmatrix}
        \overline\rho_\theta & 0 \\
        0  & -\overline\rho_\theta^\text{T}
    \end{pmatrix}, ~~ \hat O_4=\begin{pmatrix}
         0 & i\overline{\Phi'}_{0}\\
         -i\overline{\Phi'}_{0}^\dagger & 0\\
     \end{pmatrix}\label{eq:matrixO}
\end{equation}
in which 
\begin{equation}
    \overline\rho_{\mathcal{A}}=\begin{pmatrix}
        0 & \overline\rho_{-\bm{P}} & 0 & \overline\rho_{\bm{P}}\\
        \overline\rho_{\bm{P}} & 0 & \overline\rho_{-\bm{P}} & 0 \\
        0 & \overline\rho_{\bm{P}} & 0 & \overline\rho_{-\bm{P}} \\
        \overline\rho_{-\bm{P}} & 0 & \overline\rho_{\bm{P}} & 0\\
    \end{pmatrix}, ~ \overline\rho_\phi=i\begin{pmatrix}
        0 & -\overline\rho_{-\bm{P}} & 0 & \overline\rho_{\bm{P}}\\
        \overline\rho_{\bm{P}} & 0 & -\overline\rho_{-\bm{P}} & 0 \\
        0 & \overline\rho_{\bm{P}} & 0 & -\overline\rho_{-\bm{P}} \\
        \overline\rho_{-\bm{P}} & 0 & \overline\rho_{\bm{P}} & 0\\
    \end{pmatrix}, ~ \overline{\Phi'_0}=\overline\Delta_0\begin{pmatrix}
        f_{\bk} & 0 & 0 & 0\\
        0 & 0 & 0 &  f_{\bk+\bP}\\
        0 & 0 & f_{\bk+2\bP} & 0\\
        0 & f_{\bk+3\bP} & 0 & 0 
    \end{pmatrix}.
\end{equation}

The fluctuating part of the action, truncated at Gaussian level, is 
\begin{equation}
    \begin{aligned}
        S^{(2)}_\text{FL}
        &=\sum_q \zeta''^T(q) \hat\Gamma^{-1}(q) \zeta''(-q).
    \end{aligned}
\end{equation}
Here
\begin{equation}
     \hat\Gamma^{-1}(q)=\hat U^{-1}+\hat\Pi(q)\label{eq:invGamma3}
 \end{equation} 
is the inverse RPA type interaction dressed by collective modes. The bare interaction in this case becomes 
 \begin{equation}
     \hat U^{-1}=\begin{pmatrix}
         \frac{\overline\rho_{\bP}\overline\rho_{-\bP}}{g_3} & 0 & 0 & 0\\
         0 & \frac{|\overline\Delta_0|^2}{g^2} & 0 & 0\\
         0 & 0 &  \frac{\overline\rho_{\bP}\overline\rho_{-\bP}}{g_3} & 0\\
         0 & 0 & 0 & \frac{|\overline\Delta_0|^2}{g^2}\\
     \end{pmatrix}.
 \end{equation}
and the four-by-four matrix $\hat\Pi(q)$ becomes
 \begin{equation}
     M_{i,j}(q)=\frac{1}{8}\sum_{k}\text{Tr}\left[\overline{\mathcal{G}}_{k+\frac{q}{2}}\hat O_i\overline{\mathcal{G}}_{k-\frac{q}{2}}\hat O_j\right].\label{eq:M3}
 \end{equation}
 This matrix has the same symmetry properties as those in Eq.\eqref{eq:M1} and \eqref{eq:M2}, i.e. the matrix elements are odd function of $q$ if they are couplings between amplitude and phase; otherwise they are even in $q$. 
 We find from our calculations that the CDW phase mode $\theta_\rho(q)$ does not couple to any of the other modes. 

 At small $q$, we neglect the amplitude-phase coupling. The four eigen modes in this system are:  the two original phase modes $\theta_\rho(q)$ and $\theta_0(q)$, and two amplitude modes resulting from the coupling between $\mathcal{A}_\rho(q)$ and $\mathcal{A}_0(q)$:
 \begin{equation}
    \mathcal{A}_{\pm}\propto \mathcal{A}_{0}\pm [C_{\rho}(q)]^{\pm1}\mathcal{A}_{\rho}\label{eq:amplitude23}
\end{equation}
where the parameter $C_\rho(q)$ can be found from the eigen vectors of $\hat\Gamma^{-1}_A(q)$. In Fig.\ref{fig:CDWphi} (b) we show a plot of $C_\rho$ as a function of Matsubara frequency $\varepsilon_n$ at $\bq=0$, $T=5\times10^{-3}$, $|\overline\Delta_0|=0.1t$ and two different $|\overline\rho_{\bP}|$. We see that in both cases, even if $|\overline\Delta_0|$ and $|\overline\rho_{\bP}|$ are comparable, $C_\rho(q)\ll1$ [for comparison see the plot of $C_{\mathcal{A}}$ in the main text]. 
This means that the coupling between the two amplitude modes are very small: $\mathcal{A}_+$ in Eq.\eqref{eq:amplitude23} is mainly dominated by $\mathcal{A}_0(q)$ and $\mathcal{A}_-$ is mainly dominated by $\mathcal{A}_\rho(q)$. 
In Fig.\ref{fig:CDWSC_Higgs} we present the numerical results for the spectral functions for $|\overline\rho_{\bP}|=|\overline\Delta_0|=0.1t$.


\section{V. Derivation of the Raman susceptibility} 
\label{sec:iii_derivation_of_the_raman_susceptibility}
To probe the collective modes discussed above, we study the dressed Raman intensity for different cases. 
The dressed Raman susceptibility is calculated in parallel to the RPA summation of polarization, such that various collective modes are encoded through a series bubbles containing particle-hole excitations\cite{PhysRevB.29.4976,PhysRevB.82.014507,RevModPhys.79.175}. Note that long-range Coulomb interactions and vertex corrections are neglected in our calculations, while the latter are important to restore the correct results constraint by conservation laws if a constant Raman vertex is used\cite{PhysRevB.96.014503}. 
The Raman susceptibility is defined as 
\begin{equation}
    \chi_R(\bq,\tau)=\braket{\mathcal{T}_\tau \tilde \rho (\bq,\tau)\tilde \rho (-\bq,0)}
\end{equation}
where $\mathcal{T}_\tau$ is the time-ordering and the Raman density is 
\begin{equation}
    \tilde \rho (\bq,\tau)=\sum_{\bk,\sigma}\gamma(\bk)\psi^\dagger_\sigma(\bk+\bq/2,\tau)\psi_\sigma(\bk-\bq/2,\tau).
\end{equation}
To compute this response function from an effective action, we introduce the source term coupling to $\tilde \rho(\bq,\tau)$, 
\begin{equation}
    S_\eta=\int d\tau \sum_{\bq}\eta(\bq,\tau)\tilde \rho(-\bq,\tau)=\sum_q\eta(q)\tilde\rho(-q),\label{eq:source}
\end{equation}
and then explicitly integrate out both the fermion fields and the fluctuating fields to obtain a generating functional  $Z[\eta]$.  This allows for the calculation of $\chi_{R}$ through functional derivatives, namely
\begin{equation}
    \chi_R=-\left.\frac{\delta^2\ln Z[\eta]}{\delta \eta\delta \eta}\right|_{\eta=0}.
\end{equation}

In the presence of some unidirectional commensurate density wave order (e.g. PDW or CDW), the unit cell is extended to $N$ times larger than the original one, and the ordering vector $\overline\bq$ must satisfy $N\overline\bq=0$. For the PDW order considered in this paper, the ordering vector is $\bQ=(\frac{\pi}{4},0)$ and hence $N=8$; while for the CDW order with momentum $\bP=2\bQ$, we have $N=4$. It is then convenient to introduce the following fermion basis
\begin{equation}
    \Psi_{\sigma, k}^\dagger=[\psi_{\sigma}^\dagger(k), \psi_{\overline\bq,\sigma}^\dagger(k),...,\psi_{(N-1)\overline\bq,\sigma}^\dagger(k)].
\end{equation}
Using this basis, the source action $S_\eta$ defined in Eq.\eqref{eq:source} now takes the form
\begin{equation}
    S_\eta=\frac{1}{N}\sum_{k,k'} (\Psi_{\up,k}^\dagger,\Psi^\text{T}_{\down,-k})\mathcal{G}^{-1}_\eta(k,k')\begin{pmatrix}
        \Psi_{\up,k'}\\
        {\Psi^\dagger}^\text{T}_{\down,-k'}
    \end{pmatrix}
\end{equation}
where 
\begin{equation}
    \mathcal{G}^{-1}_\eta(k,k')=\eta(k'-k)\hat\gamma_{\frac{\bk+\bk'}{2}}.
\end{equation}
The matrix representation of the Raman vertex can be found easily. For instance, when the ordering vector is $\bQ=(\frac{\pi}{4},0)$, we have
\begin{equation}
   \hat\gamma_{\bk}=\begin{pmatrix}
        \hat\gamma_{p,\bk} & 0\\
        0 & \hat\gamma_{h,\bk}
    \end{pmatrix},~
    \hat\gamma_{p,\bk}=\begin{pmatrix}
         \gamma_{\bk} & 0 & 0 & 0 & 0 & 0 & 0 & 0\\
         0 & \gamma_{\bQ+\bk} & 0 & 0 & 0 & 0 & 0 & 0\\
         0 & 0 & \gamma_{2\bQ+\bk} & 0 & 0 & 0 & 0 & 0\\
         0 & 0 & 0 & \gamma_{3\bQ+\bk} & 0 & 0 & 0 & 0\\
         0 & 0 & 0 & 0 & \gamma_{4\bQ+\bk} & 0 & 0 & 0\\
         0 & 0 & 0 & 0 & 0 & \gamma_{5\bQ+\bk} & 0 & 0\\
         0 & 0 & 0 & 0 & 0 & 0 & \gamma_{6\bQ+\bk} & 0\\
         0 & 0 & 0 & 0 & 0 & 0 & 0 & \gamma_{7\bQ+\bk}\\
     \end{pmatrix}, ~ \hat\gamma_{h,\bk}=-\hat\gamma_{p,-\bk}.
\end{equation} 
The diagonal entries $\gamma_{\bk}$ are various Raman vertex considered in this paper.
Combining this $S_\eta$ with the fluctuation action $S_\text{FL}$, and integrating out fermions, we obtain
\begin{equation}
    \begin{aligned}
         &-\frac{1}{N}\text{Tr}\ln\left(\bar{\mathcal{G}}^{-1}+\mathcal{G}_\eta^{-1}+\tilde\Sigma\right)\\
         =&-\frac{1}{N}\text{Tr}\ln\bar{\mathcal{G}}^{-1}-\frac{1}{N}\text{Tr}\ln\left[1+\bar{\mathcal{G}}\left(\mathcal{G}_\eta^{-1}+\tilde\Sigma\right)\right]\\
         =&-\frac{1}{N}\text{Tr}\ln\bar{\mathcal{G}}^{-1}+\frac{1}{N}\text{Tr}\sum_{n=1}^\infty\frac{\left[-\bar{\mathcal{G}}\left(\mathcal{G}_\eta^{-1}+\tilde\Sigma\right)\right]^n}{n},
    \end{aligned}
 \end{equation} 
where we remind that $\bar{\mathcal{G}}$ and $\tilde\Sigma$ 
are mean field full Green's function and fluctuating self energy respectively. Examples considered in this paper are given in 
 Eq.\eqref{eq:Gh} and Eq.\eqref{eq:CDWFL}.

At Gaussian level, the additional terms arising from the source $\eta$ are, 
\begin{equation}
    \begin{aligned}
        &\frac{1}{2N}\sum_{k,k'}\text{Tr}\left(\overline{\mathcal{G}}_k[\mathcal{G}_\eta^{-1}]_{k,k'}\overline{\mathcal{G}}_{k'}[\tilde\Sigma]_{k',k}\right)+ \frac{1}{2N}\sum_{k,k'}\text{Tr}\left(\overline{\mathcal{G}}_k[\tilde\Sigma]_{k,k'}\overline{\mathcal{G}}_{k'}[\mathcal{G}_\eta^{-1}]_{k',k}\right)+\frac{1}{2N}\sum_{k,k'}\text{Tr}\left(\overline{\mathcal{G}}_k[\mathcal{G}_\eta^{-1}]_{k,k'}\overline{\mathcal{G}}_{k'}[\mathcal{G}_\eta^{-1}]_{k',k}\right).
    \end{aligned}
\end{equation}
The first two terms are identical by cyclic permutations under the trace. 
After some manipulation, the total fluctuating action incorporating the source term gets modified into
\begin{equation}
    \begin{aligned}
        S_\text{FL}[\eta]=&\sum_q \zeta^T(q) \hat\Gamma^{-1}(q) \zeta(-q)+2\sum_{q}\eta(-q)\Lambda^T(q)\cdot\zeta(q)+\sum_{q}\eta(q)\eta(-q)K(q)\\
        =&\sum_q \left[\zeta^T(q)+\eta(q)\Lambda^T(-q)\hat\Gamma(q)\right]\hat\Gamma^{-1}(q)\left[\zeta(-q)+\eta(-q)\hat\Gamma(q)\Lambda(q)\right]+\sum_q\eta(q)\eta(-q)\left[K(q)-\Lambda^T(-q)\Gamma(q)\Lambda(q)\right]\label{eq:action_source}
    \end{aligned}
\end{equation}
where $\Lambda^T(q)=[..., \Lambda_i(q), ...]$, $\zeta(q)$ is the basis for fluctuating fields, which, depending on different situations,  can be Eq.\eqref{eq:flucbasis1}, \eqref{eq:flucbasis2} or \eqref{eq:flucbasis3},    and 
 \begin{equation}
     \begin{aligned}
         \Lambda_{i}(q)&=\frac{1}{2N}\sum_{k}\text{Tr}\left[\overline{\mathcal{G}}_{k+\frac{q}{2}}\hat X_i\overline{\mathcal{G}}_{k-\frac{q}{2}}\hat\gamma_{\bk}\right],\\
     K(q)&=\frac{1}{2N}\sum_{k}\text{Tr}\left[\overline{\mathcal{G}}_{k+\frac{q}{2}}\hat\gamma_{\bk}\overline{\mathcal{G}}_{k-\frac{q}{2}}\hat\gamma_{\bk}\right]
     \end{aligned}
 \end{equation}
 Here $\hat X$ can be $\hat D$, $\hat E$ or $\hat O$ depending on different cases. 
 We have explicitly checked that $\Lambda_i(q)$ is even in $q$ if it belongs to the amplitude fluctuation subspace, and is odd in $q$ if it belongs to the phase fluctuation subspace. By definition, $K(q)$ is even in $q$.

From Eq.\eqref{eq:action_source} we can integrating our the fluctuation field $\zeta(q)$ to obtain $Z[\eta]$, which results in the following response function
\begin{equation}
    \chi_R=-\left.\frac{\delta^2\ln Z[\eta]}{\delta \eta\delta \eta}\right|_{\eta=0}=K(q)-\Lambda^T(-q)\Gamma(q)\Lambda(q).
\end{equation}


\section{VI. The role of disorder} 
\label{sec:the_role_of_disorder}
Here we present a quantitative discussion about the role of weak disorder, aiming to examine whether or not in the case of PDW+SC case the spontaneous TRSB is robust against disorder. We assume the following form of weak disorder that couples to the composite $1\bQ$ CDW order $\rho_{\bQ}\sim \Delta_{0}^*\Delta_{\bQ}+\Delta_0 \Delta_{-\bQ}^*$ via 
\begin{equation}
    H_\text{dis}=\int d\br V_\text{dis}(\br)\left\{[\Delta_{0}^*(\br)\Delta_{\bQ}(\br)+\Delta_0(\br) \Delta_{-\bQ}^*(\br)]e^{i\bQ\cdot \br}+c.c.\right\}\label{eq:Sdis}
\end{equation}
where we reminds that $\Delta_{\overline{\bq}}(\br)$ ($\overline{\bq}=\bm{0},\pm\bQ$) is the order parameter fields which depends on $\br$ slowly. Here we consider the disorder effect on phase modes only, so $\Delta_{\overline{\bq}}(\br)$ can be parametrized as
\begin{equation}
    \Delta_{\overline{\bq}}(\br)=|\overline{\Delta}_{\overline{\bq}}|e^{i\varphi_{\overline{\bq}}(\br)}
\end{equation}
where 
\begin{equation}
    \varphi_{\overline{\bq}}(\br)={\varphi}_{\overline{\bq}}+\theta_{\overline{\bq}}(\br),\label{eq:phifluctuation}
\end{equation}
and we remind that $\varphi_{\overline{\bq}}$ is a uniform phase configuration determined in the clean limit saddle point solutions. This gives
\begin{equation}
    H_\text{dis}=\int  d\br V_\text{dis}(\br)|\overline{\Delta}_0||\overline{\Delta}_{\bQ}|\left\{ e^{-i[\varphi_0(\br)-\varphi_{\bQ}(\br)]+i\bQ\cdot\br}+e^{i[\varphi_0(\br)-\varphi_{-\bQ}(\br)]+i\bQ\cdot\br}+c.c.\right\}.
\end{equation}
Let's first assume as in the clean limit that the fluctuating part $\theta_{\overline{\bq}}(\br)$ is small, and approximate Eq.\eqref{eq:Sdis} as
\begin{equation}
    \begin{aligned}
        H_\text{dis}&\approx\int  d\br V(\br)\left\{ e^{-i(\varphi_0-\varphi_{\bQ})+i\bQ\cdot\br}i[-\theta_0(\br)+\theta_{+}(\br)] +e^{i(\varphi_0-\varphi_{-\bQ})+i\bQ\cdot\br}i[\theta_0(\br)-\theta_{-}(\br)]+c.c.\right\}\\
        &=i \int\frac{d\bk}{(2\pi)^2}\left\{\left[V_{-\bk-\bQ}e^{i(\varphi_0-\varphi_{-\bQ})}-V_{-\bk+\bQ}e^{-i(\varphi_0-\varphi_{-\bQ})}\right][\theta_0(\bk)-\theta_{-}(\bk)]\right.\\
        & ~~~~~~~~~~~~~~~~~~~\left.\left[V_{-\bk+\bQ}e^{i(\varphi_0-\varphi_{\bQ})}-V_{-\bk-\bQ}e^{-i(\varphi_0-\varphi_{\bQ})}\right][\theta_0(\bk)-\theta_{+}(\bk)]\right\}.\label{eq:Hdis}
    \end{aligned}
\end{equation}
Here we have defined $V(\br)=V_\text{dis}(\br)|\overline{\Delta}_0||\overline{\Delta}_{\bQ}|$ and used $\theta_{\pm}=\theta_{\pm\bQ}$ for brevity.

We next find the fluctuating fields $\theta_{\overline{\bq}}(\bk)$ in the presence of disorder. For the clean limit, the Hamiltonian for the phase part is 
\begin{equation}
    H_p=\int  d\br \left[\frac{\kappa}{2}[\nabla\theta_{\bm{0}}(\br)]^2+\frac{\kappa'}{2}\left([\nabla\theta_{+}(\br)]^2+[\nabla\theta_{-}(\br)]^2\right)\right]
\end{equation}
The saddle point solutions for $\varphi_{\overline{\bq}}$ are given by minimizing $H_\text{dis}+H_p$ with respect to $\varphi_{\overline{\bq}}$, which yields 
\begin{equation}
    \begin{aligned}
        \kappa \nabla^2\theta_{\bm{0}}(\br)&=iV(\br)\left[e^{i[\varphi_{0}(\br)-\varphi_{-\bQ}(\br)]+i\bQ\cdot\br}+e^{i[\varphi_{0}(\br)-\varphi_{\bQ}(\br)]-i\bQ\cdot\br}-c.c.\right],\\
         \kappa' \nabla^2\theta_{\pm}(\br)&=iV(\br)\left[e^{-i[\varphi_0(\br)-\varphi_{\pm\bQ}(\br)]\pm i\bQ\cdot\br}-c.c.\right],
    \end{aligned}\label{eq:phasesaddle}
\end{equation}
Substituting Eq.\eqref{eq:phifluctuation} into Eq.\eqref{eq:phasesaddle}, and assuming ${\varphi}_{\overline{\bq}}$ is a constant, we have to linear order in $V$ (which amounts to replacing $\varphi_{\overline{\bq}}(\br)$ with the constant profile $\varphi_{\overline{\bq}}$)
\begin{equation}
    \begin{aligned}
        \kappa \nabla^2\theta_{\bm{0}}(\br)&=iV(\br)\left[e^{i[\varphi_{0}-\varphi_{-\bQ}]+i\bQ\cdot\br}+e^{i[\varphi_{0}-\varphi_{\bQ}]-i\bQ\cdot\br}-c.c.\right],\\
         \kappa' \nabla^2\theta_{\pm}(\br)&=iV(\br)\left[e^{-i[\varphi_0-\varphi_{\pm\bQ}]\pm i\bQ\cdot\br}-c.c.\right],
    \end{aligned}\label{eq:phasesaddle2}
\end{equation}
or in momentum space
\begin{equation}
    \begin{aligned}
        \kappa k^2\theta_0(\bk)&=i V_{\bk-\bQ}e^{-i[\varphi_0-\varphi_{\bQ}]}-iV_{\bk+\bQ}e^{i[\varphi_0-\varphi_{\bQ}]}+i V_{\bk+\bQ}e^{-i[\varphi_0-\varphi_{-\bQ}]}-iV_{\bk-\bQ}e^{i[\varphi_0-\varphi_{-\bQ}]},\\
        \kappa' k^2 \theta_{+}(\bk)&=-i V_{\bk-\bQ}e^{-i[\varphi_0-\varphi_{\bQ}]}+iV_{\bk+\bQ}e^{i[\varphi_0-\varphi_{\bQ}]},\\
        \kappa' k^2 \theta_{-}(\bk)&=-i V_{\bk+\bQ}e^{-i[\varphi_0-\varphi_{-\bQ}]}+iV_{\bk-\bQ}e^{i[\varphi_0-\varphi_{-\bQ}]}.
    \end{aligned}\label{eq:phasesaddle3}
\end{equation}
They apparently satisfy the following relation
\begin{equation}
     \kappa\theta_0(\bk)=-\kappa'[\theta_{+}(\bk)+\theta_{-}(\bk)].
\end{equation}

To find the disorder averaged energy change perturbatively, we assume $\overline{V_{\bk}}=0$ and the following disorder correlations 
\begin{equation}
    \begin{aligned}
        \overline{V_{\bk}V_{\bk'}}&=\delta_{\bk,-\bk'}F ~~\text{for}~ \bk\approx \bQ\\
        \overline{V_{\bk}V_{\bk'}}&\approx 0 ~~\text{overwise.}
    \end{aligned}\label{eq:disav}
\end{equation}
It is then straightforward to obtain [by substituting $\theta$ from Eq.\eqref{eq:phasesaddle3} into Eq.\eqref{eq:Hdis} and then averaging over disorder using Eq.\eqref{eq:disav}] the energy shift after disorder average using Eq.\eqref{eq:disav},
\begin{equation}
    \delta\mathcal{F}=-\int\frac{d^2\bk}{(2\pi)^2}\frac{2F}{\kappa k^2}\left[1+\frac{\kappa}{\kappa'}-\cos\left(2\bar\varphi_0-\bar\varphi_{\bQ}-\bar\varphi_{-\bQ}\right)\right].
\end{equation}
We first notice that this expression is logarithmically divergent in the infra-red (IR) 
in 2D.  Specifically, it means that at long-distances the phase fluctuations are large compared to $2\pi$ - thus violating the conditions under which these expressions were derived. This divergence is cut off in 3D, even if the system is quasi-2D.  In strictly 2D (as discussed in Ref. \cite{PhysRevB.108.L100505}), the logarithm is cutoff at something analogous to the Larkin length, which diverges as $F \to 0$.  However,  since the divergence is only logarithmic, the results are not very sensitive to the exact nature or value of the IR cutoff.

The most significant point is that the energy shift is negative, meaning that the presence of weak disorder tends to reduce the free energy compared to the clean case. 
Moreover, $\delta\mathcal{F}$ is minimized by the condition  
\begin{equation}
    2\bar\varphi_0-\bar\varphi_{\bQ}-\bar\varphi_{-\bQ}=\pi+2m\pi,~ m\in\mathbb{Z}\label{eq:disTRSB}
\end{equation}
which is the same as the TRSB condition we obtained in the clean limit. Therefore, we can conclude from the above analysis that the presence of disorder, at least in the weak limit, also favors TRSB. 

The breaking of TRS in our case also indicate the 1$\bQ$ CDW is not a long range order, even in the clean case. However, it is interesting to look at the 1$\bQ$ CDW correlation in the presence of disorder. First we note that, the $1\bQ$ order in terms of the fluctuating phases is given by
\begin{equation}
    \rho_{\bQ}(\br)\sim|\rho|e^{-i(\varphi_0-\varphi_{\bQ})}\left(e^{-i[\theta_0(\br)-\theta_+(\br)]}+e^{i[\theta_0(\br)-\theta_-(\br)]}e^{i(2\bar\varphi_0-\bar\varphi_{\bQ}-\bar\varphi_{-\bQ})}\right).
\end{equation}
As we pointed out above, $2\bar\varphi_0-\bar\varphi_{\bQ}-\bar\varphi_{-\bQ}=\pi ~\text{mod} ~2\pi$ even in the presence of disorder, so
\begin{equation}
    \rho_{\bQ}(\br)\sim|\rho|e^{-i(\varphi_0-\varphi_{\bQ})}\left(e^{-i[\theta_0(\br)-\theta_+(\br)]}-e^{i[\theta_0(\br)-\theta_-(\br)]}\right).
\end{equation}
Therefore, correlation averaged over disorder is 
\begin{equation}
    \begin{aligned}
        \overline{\rho_{\bQ}(\br)\rho_{\bQ}^*(0)}&\sim |\rho|^2\left(\overline{e^{-i[\theta_0(\br)-\theta_+(\br)]}e^{i[\theta_0(0)-\theta_+(0)]}}+\overline{e^{i[\theta_0(\br)-\theta_-(\br)]}e^{-i[\theta_0(0)-\theta_-(0)]}}\right)\\
        &=|\rho|^2\left[\exp\left(-\frac{1}{2}
        \overline{[\theta_0(\br)-\theta_+(\br)-\theta_0(0)+\theta_+(0)]^2}\right)+\exp\left(-\frac{1}{2}
        \overline{[\theta_0(\br)-\theta_-(\br)-\theta_0(0)+\theta_-(0)]^2}\right)\right]
    \end{aligned}
\end{equation}
Note that
\begin{equation}
    \begin{aligned}
        [\theta_0(\br)-\theta_+(\br)-\theta_0(0)+\theta_+(0)]^2=\int \frac{d^2\bk d^2\bk'}{(2\pi)^4}&\left[\theta_0(\bk)\left(e^{i\bk\cdot\br}-1\right)-\theta_+(\bk)\left(e^{i\bk\cdot\br}-1\right)\right]\times\\
        &\times\left[\theta_0(\bk')\left(e^{i\bk'\cdot\br}-1\right)-\theta_+(\bk')\left(e^{i\bk'\cdot\br}-1\right)\right]
    \end{aligned}
\end{equation}
and use the following relations [which can be easily deduced from combining Eq.\eqref{eq:phasesaddle3}, \eqref{eq:disav} and \eqref{eq:disTRSB}],
\begin{equation}
    \begin{aligned}
        \overline{\theta_0(\bk)\theta_0(\bk')}&=\delta_{\bk,-\bk'}\frac{8F}{\kappa^2 k^4}\\
        \overline{\theta_0(\bk)\theta_+(\bk')}&=\overline{\theta_0(\bk)\theta_-(\bk')}=-\delta_{\bk,-\bk'}\frac{4F}{\kappa\kappa' k^4}\\
        \overline{\theta_+(\bk)\theta_+(\bk')}&=\overline{\theta_-(\bk)\theta_-(\bk')}=\overline{\theta_+(\bk)\theta_-(\bk')}=\delta_{\bk,-\bk'}\frac{2F}{\kappa'^2 k^4}
    \end{aligned}
\end{equation}
one can obtain
\begin{equation}
    \frac{1}{2}\overline{[\theta_0(\br)-\theta_+(\br)-\theta_0(0)+\theta_+(0)]^2}=F \int\frac{d^2\bk}{(2\pi)^4}\left(\frac{2}{\kappa}+\frac{1}{\kappa'}\right)^2\frac{(e^{i\bk\cdot\br}-1)(e^{-i\bk\cdot\br}-1)}{k^4}\label{eq:discorrelator}
\end{equation}
and 
\begin{equation}
    \overline{[\theta_0(\br)-\theta_+(\br)-\theta_0(0)+\theta_+(0)]^2}=\overline{[\theta_0(\br)-\theta_-(\br)-\theta_0(0)+\theta_-(0)]^2}.
\end{equation}
Eq.\eqref{eq:discorrelator} apparently has IR log-divergence if no regularization is imposed. Such a divergence can be rescued by performing a more careful self-consistent treatment\cite{PhysRevB.108.L100505}. Here, if we assume there is an IR momentum cutoff $k_0$, carrying out the integral in Eq.\eqref{eq:discorrelator} gives
\begin{equation}
    \frac{1}{2}\overline{[\theta_0(\br)-\theta_+(\br)-\theta_0(0)+\theta_+(0)]^2}=|\br|^2/\xi^2
\end{equation}
where 
\begin{equation}
    \xi^{-2} = \frac{F}{16\pi^3}\left(\frac{2}{\kappa}+\frac{1}{\kappa'}\right)^2\ln\frac{1}{k_0}.
\end{equation}
Therefore, the disorder averaged 1$\bQ$ order correlation is given by
\begin{equation}
    \overline{\rho_{\bQ}(\br)\rho_{\bQ}^*(0)}\sim 2|\rho|^2 e^{-|\br|^2/\xi^2}.
\end{equation}
The correlation length $\xi$ becomes large when the disorder variance $F\to 0$.


\bibliography{Ramanv2}